 \documentclass[final,authoryear,5p]{elsarticle}

\usepackage{epsfig}

\usepackage{amssymb}

\usepackage[ps2pdf,%
a4paper=true,%
breaklinks=true,%
colorlinks=true,%
pdfauthor={First Author et al.},%
pdftitle={Template for manuscripts in Advances in Space Research}%
]{hyperref}

\journal{Advances in Space Research}

\def\referee#1{#1}

\begin{document}

\begin{frontmatter}



\title{Understanding space weather to shield society:\\  A global road map for 2015-2025 commissioned by COSPAR and ILWS}

\author[label1]{Carolus J. Schrijver\corref{cor}}
\address[label1]{Lockheed Martin Solar and Astrophysics Laboratory, 3251 Hanover Street, Palo Alto, CA94304, USA}
\cortext[cor]{Corresponding authors}
\ead{schrijver@lmsal.com}

\author[label2]{Kirsti Kauristie\corref{cor}}
\address[label2]{Finnish Meteorological Institute, Finland}

\author[label3]{Alan D. Aylward}
\address[label3]{University College London, Dept. of physics and astronomy, Gower Street, London WC1E 6BT, UK}

\author[label18]{Clezio M. Denardini}
\address[label18]{Instituto Nacional de Pesquisas Espaciais, Brazil}

\author[label27]{Sarah E. Gibson}
\address[label27]{HAO/NCAR, P.O. Box 3000, Boulder, CO 80307-3000, USA}

\author[label4]{Alexi Glover}
\address[label4]{RHEA System and ESA SSA Programme Office, Darmstadt, Germany}

\author[label26]{Nat Gopalswamy}
\address[label26]{NASA Goddard Space Flight Center, Greenbelt, MD, USA}

\author[label5]{Manuel Grande}
\address[label5]{Univ. of Aberystwyth, Penglais STY23 3B, UK}

\author[label6]{Mike Hapgood}
\address[label6]{RAL Space and STFC Rutherford Appleton Laboratory, Harwell Oxford, Didcot, UK}

\author[label7]{Daniel Heynderickx}
\address[label7]{DH Consultancy BVBA, Diestsestraat 133/3, 3000 Leuven, Belgium}

\author[label8]{Norbert Jakowski}
\address[label8]{German Aerospace Center, Kalkhorstweg 53, 17235 Neustrelitz, Germany}

\author[label9]{Vladimir V. Kalegaev}
\address[label9]{Skobeltsyn Institute of Nuclear Physics, Lomonosov Moscow State University, Moscow, Russia}

\author[label10]{Giovanni Lapenta}
\address[label10]{KU Leuven, Celestijnenlaan 200B, Leuven 3001, Belgium}

\author[label11]{Jon A. Linker}
\address[label11]{Predictive Science Inc., San Diego, CA, USA}

\author[label12]{Siqing Liu}
\address[label12]{National Space Science Center, Chinese Academy of Sciences, Haidian District, Beijing 100190, China}

\author[label13]{Cristina H. Mandrini}
\address[label13]{Instituto de Astronomia y Fisica del Espacio, Buenos Aires, Argentina}

\author[label14]{Ian R. Mann}
\address[label14]{Dept. of physics, Univ. Alberta, Edmonton, AB, T6G 2J1, Canada}

\author[label16]{Tsutomu Nagatsuma}
\address[label16]{Space Weather and Environment Informatics Lab.,
National Inst. of Information and Communications Techn., Tokyo 184-8795, JAPAN}

\author[label17]{Dibyendu Nandi}
\address[label17]{Center for Excellence in Space Sciences and Indian Institute of Science, Education and Research, Kolkata, Mohanpur 74125, India}

\author[label19]{Takahiro Obara}
\address[label19]{Planetary plasma and atmospheric research center, Tohoku University, 6-3 Aoba, Aramaki, Aoba, Sendai 980-8578, Japan}

\author[label20]{T. Paul O'Brien}
\address[label20]{Space science department/Chantilly, Aerospace Corporation, Chantilly, VA 20151, USA}

\author[label21]{Terrance Onsager}
\address[label21]{NOAA Space Weather Prediction Center, USA}

\author[label22]{Hermann J. Opgenoorth}
\address[label22]{Swedish Institute of Space Physics, 75121 Uppsala, Sweden}

\author[label23]{Michael Terkildsen}
\address[label23]{Space Weather Services, Bureau of Meteorology, Australia}

\author[label24]{Cesar E. Valladares}
\address[label24]{Institute for scientific research, Boston College, Newton, MA 02459, USA}

\author[label25]{Nicole Vilmer}
\address[label25]{LESIA, Observatoire de Paris, CNRS, UPMC, Universit{\'e} Paris-Diderot, 5 place Jules Janssen, 92195 Meudon, France}





\begin{abstract}

  There is a growing appreciation that the environmental conditions
  that we call space weather impact the technological infrastructure
  that powers the coupled economies around the world. With that comes
  the need to better shield society against space weather by improving
  forecasts, environmental specifications, and infrastructure
  design. We recognize that much progress has been made and continues
  to be made with a powerful suite of research observatories on the
  ground and in space, forming the basis of a Sun-Earth system
  observatory. But the domain of space weather is vast - extending
  from deep within the Sun to far outside the planetary orbits - and
  the physics complex - including couplings between various types of
  physical processes that link scales and domains from the microscopic
  to large parts of the solar system. Consequently, advanced
  understanding of space weather requires a coordinated international
  approach to effectively provide awareness of the processes within
  the Sun-Earth system through observation-driven models. This roadmap
  prioritizes the scientific focus areas and research infrastructure
  that are needed to significantly advance our understanding of space
  weather of all intensities and of its implications for
  society. Advancement of the existing system observatory through the
  addition of small to moderate state-of-the-art capabilities designed
  to fill observational gaps will enable significant advances. Such a
  strategy requires urgent action: key instrumentation needs to be
  sustained, and action needs to be taken before core capabilities are
  lost in the aging ensemble. We recommend advances through priority
  focus (1) on observation-based modeling throughout the Sun-Earth
  system, (2) on forecasts more than 12\,hrs ahead of the magnetic
  structure of incoming coronal mass ejections, (3) on understanding
  the geospace response to variable solar-wind stresses that lead to
  intense geomagnetically-induced currents and ionospheric and
  radiation storms, and (4) on developing a comprehensive
  specification of space climate, including the characterization of
  extreme space storms to guide resilient and robust engineering of
  technological infrastructures. The roadmap clusters its
  implementation recommendations by formulating three action pathways,
  and outlines needed instrumentation and research programs and
  infrastructure for each of these. An executive summary provides an
  overview of all recommendations.

\end{abstract}


\end{frontmatter}

\parindent=0.5 cm

\tableofcontents

\section*{Executive Summary}\label{sec:exsum}

Space weather is driven by changes in the Sun's magnetic field and by the consequences of that variability in Earth's magnetic field and upper atmosphere. This results in a variety of manifestations, including geomagnetic variability, energetic particles, and changes in Earth's uppermost atmosphere. All of these can affect society's technological infrastructures in different ways.

{\bf Space weather is generally mild but some times extreme.} Mild space weather storms can degrade electric power quality, perturb precision navigation systems, interrupt satellite functions, and are hazardous to astronaut \referee{health}. Severe space storms have resulted in perturbations in the electric power system and have caused loss of satellites through damaged electronics or increased orbital drag. For rare extreme solar events the effects could be catastrophic with severe consequences for millions of people.

{\bf Societal interest in space weather grows rapidly:} As science and society increasingly recognize the impacts of space weather on the infrastructure of the global economy, interest in, and dependence on, space weather information and services grows rapidly. Apart from having societal relevance, understanding space weather is an exciting science revealing how the universe around us works.

{\bf Space weather is an international challenge:} Significant scientific problems require substantial resources, with observations having to cover the terrestrial globe and span the vast reaches of the heliosphere between Earth and the Sun.

{\bf Mitigating against the impacts of space weather can be improved} by designing less susceptible, more resilient technologies, combined with better environmental knowledge and more reliable forecasts. This roadmap outlines how we can achieve deeper understanding and better forecasts, recognizing that the expectations for space weather information differ between societal sectors, and that capabilities to observe or model space weather phenomena depend on available and anticipated technologies.

{\bf The existing observatories that cover much of the Sun-Earth system provide a unique starting point:}  Moderate investments now that fill key capability gaps can enable scientific advances that could not be otherwise achieved, while at the same time providing a powerful base to meet many operational needs. 
Improving understanding and forecasts of space weather requires addressing scientific challenges within the network of physical processes that connect the Sun to society. The roadmap team identified the highest-priority areas within the Sun-Earth space-weather system whose advanced scientific understanding is urgently needed to address current space weather service user requirements. The roadmap recommends actions towards such advanced understanding, focusing on the general infrastructure to support research as well as on specific concepts for instrumentation to meet scientific needs.

{\bf Roadmap recommendations:}

\noindent {\em Research: observational, computational, \&\ theoretical needs:}
\begin{enumerate}
\item	Advance the international Sun-Earth system observatory along with models to improve forecasts based on understanding of real-world events through the development of innovative approaches to data incorporation, including data-driving, data assimilation, and ensemble modeling;
\item	Understand space weather origins at the Sun and their propagation in the heliosphere, initially prioritizing post-event solar eruption modeling to develop multi-day forecasts of geomagnetic disturbance times and strengths, after propagation through the heliosphere;
\item	Understand the factors that control the generation of geomagnetically-induced currents (GICs) and of harsh radiation in geospace, involving the coupling of the solar wind disturbances to internal magnetospheric processes and the ionosphere below; 
\item	Develop a comprehensive space environment specification, first to aid scientific research and engineering designs, later to support forecasts.
\end{enumerate}

\noindent {\em Teaming: coordinated collaborative research environment:}
\begin{enumerate}
\item[I] Quantify vulnerability of humans and of society's infrastructure for space weather by partnering with user groups; 
\item[II]	Build test beds in which coordinated observing supports model development;
\item[III]	Standardize (meta-)data and product metrics, and harmonize access to data and model archives;
\item[IV]	Optimize observational coverage of the Sun-society system.
\end{enumerate}

\noindent {\em Collaboration between agencies and communities:}
\begin{enumerate}
\item[A]	Implement an open space-weather data and information policy;
\item[B]	Provide access to quality education and information materials;
\item[C]	Execute an international, inter-agency assessment of the state of the field on a   5-yr basis to adjust priorities and to guide international coordination;  
\item[D]	Develop settings to transition research models to operations;
\item[E]	Partner with the weather and solid-Earth communities to share lessons learned.
\end{enumerate}

The roadmap's research recommendations are expanded in three pathways that reflect a blend of the magnitude of societal impact, scientific need, technological feasibility, and likelihood of near-term success. Each pathway needs recommendations of any preceding it implemented to achieve full success, but can be initiated in parallel.  The pathways are designed to meet the variety of differential needs of the user communities working with different types of impacts. Recommendations within each pathway are grouped into actions that can be taken now, soon, or on a few-year timescale, each listed in priority order within such group.

\underline{Pathway I recommendations:} to obtain forecasts more than 12 hours ahead of the magnetic structure of incoming coronal mass ejections and their impact in geospace to improve alerts for geomagnetic disturbances and strong GICs, related ionospheric variability, and geospace energetic particles:

\noindent {\em Maintain existing essential capabilities:}
\begin{itemize}
\item	solar magnetic maps (GBO, SDO) and EUV/X-ray images at arcsec and few-second res. (SDO; Hinode), and solar spectral irradiance observations;
\item	solar coronagraphy, best from multiple perspectives (Earth's view and L1: GBO and SoHO; and well off Sun-Earth line: STEREO);
\item	in-situ measurements of solar-wind plasma and magnetic field at, or upstream of, Sun-Earth L1 (ACE, SoHO; DSCOVR);
\item	for several years, continue to measure the interaction across the bowshock-magnetopause (as now with Cluster/ARTEMIS/THEMIS; soon with MMS), to better understand wind-magnetosphere coupling;
\item	satellite measurements of magnetospheric magnetic and electric fields, plasma parameters, soft auroral and trapped energetic particle fluxes (e.g., Van-Allen Probes, LANL satellites, GOES, ELECTRO-L, POES, DMSP);
\item	ground-based sensors for solar, heliospheric, magnetospheric, and iono-/thermo-/mesospheric data to complement satellite data.
\end{itemize}

\noindent {\em Model capability, archival research, or data infrastructure:}
\begin{itemize}
\item	near-real time, observation-driven 3D solar active-region models of the magnetic field to assess destabilization and to estimate energies; 
\item	data-driven models for the global solar surface-coronal field;  
\item	data-driven ensemble models for the magnetized solar wind;
\item	data assimilation techniques for the global ionosphere-magnetosphere-atmosphere system using ground and space data for nowcasts and near-term forecasts of geomagnetic and ionospheric variability, making optimal use of selected locations where laboratory-like test beds exist or can be efficiently developed;
\item	coordinated system-level research into large-scale rapid morphological changes in Earth's magnetotail and embedded energetic particle populations and their linkage to the ionosphere;
\item	system-level study of the mechanisms of the particle transport/acceleration/losses driving currents and pressure profiles in the inner magnetosphere;
\item	stimulate research to improve global geospace modeling beyond the MHD approximation (e.g., kinetic and hybrid approaches);
\item	develop the ability to use solar chromospheric and coronal polarimetry to guide full-Sun corona-to-heliosphere field models.    
\end{itemize}

\noindent {\em Deployment of new/additional instrumentation:}
\begin{itemize}
\item	binocular imaging of the solar corona at $\sim$1-arcsecond and at least 1-min. resolution, with about 10$^\circ$ to 20$^\circ$ separation between perspectives;
\item	observe the solar vector-field at and near the surface and the overlying corona at better than 200-km resolution to quantify ejection of compact and low-lying current systems from solar active regions;
\item	(define criteria for) expanded in-situ coverage of the auroral particle acceleration region and the dipole-tail field transition region (building on MMS) to determine the magnetospheric state in current (THEMIS, Cluster) and future high-apogee constellations, using hosted payloads and cubesats where appropriate;
\item	(define needs, then) increase ground- and space-based instrumentation to complement satellite data of the magnetospheric and ionospheric variability to cover gaps (e.g., in latitude coverage and over oceans); 
\item	an observatory to expand solar-surface magnetography to all latitudes and off the Sun-Earth line [starting with Solar Orbiter];
\item	large ground-based solar telescopes (incl.\ DKIST) to perform multi-wavelength spectro-polarimetry to probe magnetized structures at a range of heights in the solar atmosphere, and from sub-active-region to global-corona spatial scales;
\item	optical monitors to measure global particle precipitation to be used in assimilation models for geomagnetic disturbances and ionospheric variability.                                                       
\end{itemize}

\underline{Pathway II recommendations:} to understand the particle environments of (aero)space assets leading to improved environmental specification and near-real-time conditions.

With the Pathway-I requirements implemented: 

\noindent {\em Maintain existing essential capabilities:}
\begin{itemize}
\item	LEO to GEO observations of electron and ion populations (hard/$\sim$MeV and soft/$\sim$keV; e.g., GOES, \ldots), and of the magnetospheric field, to support improved particle-environment nowcasts;
\item	maintain a complement of spacecraft with high resolution particle and field measurements (such as the Van Allen Probes).
\end{itemize}

\noindent {\em Model capability, archival research, or data infrastructure:}
\begin{itemize}
\item	specify the frequency distributions for fluences of energetic particle populations [SEP, RB, GCR] for a variety of orbital conditions, and maintain archives of past conditions;
\item	develop, and experiment with, assimilative integrated models for radiation-belt particles towards forecast development including data from ionosphere, thermosphere and magnetosphere and below, and validate these based on archival data.
\end{itemize}

\noindent {\em Deployment of new/additional instrumentation:}
\begin{itemize}
\item	deploy high- and low-energy particle and electro-magnetic field instruments to ensure dense spatial coverage from LEO to GEO and long-term coverage of environment variability (incl., e.g.,  JAXA's ERG).
\end{itemize}

\underline{Pathway III recommendations:} to enable pre-event forecasts of solar flares and coronal mass ejections, and related solar energetic particle, X-ray, EUV and radio wave eruptions for near-Earth satellites, astronauts, ionospheric storm forecasts , and polar-route aviation, including all-clear conditions.

\noindent {\em Maintain existing essential capabilities (in addition to Path\-way-I list):}
\begin{itemize}
\item	solar X-ray observations (GOES);
\item	observe the inner heliosphere at radio wavelengths to study shocks and electron beams in the corona and inner heliosphere;
\item	maintain for some years multi-point in-situ observations of SEPs on- and off Sun-Earth line throughout the inner heliosphere (e.g., L1, STEREO; including ground-based neutron monitors);
\item	maintain measurements of heavy ion composition (L1: ACE; STEREO; near-future: GOES-R). 
\end{itemize}

\noindent {\em Model capability, archival research, or data infrastructure:}
\begin{itemize}
\item	develop data-driven predictive modeling capability for field eruptions from the Sun through the inner heliosphere;
\item	investigate energetic particle energization and propagation in the inner heliosphere, aiming to develop at least probabilistic forecasting of SEP properties [cf. Pathway-I for heliospheric data-driven modeling];
\item	ensemble modeling of unstable active regions to understand energy conversions into bulk kinetic motion, photons, and particles. 
\end{itemize}

\noindent {\em Deployment of new/additional instrumentation:}
\begin{itemize}
\item	new multi-point in-situ observations of SEPs off Sun-Earth line throughout the inner heliosphere to understand population evolutions en route to Earth (e.g., Solar Orbiter, Solar Probe Plus).
\end{itemize}

{\bf Concepts for new priority instrumentation:}

\noindent {\em Pathway I:}
\begin{enumerate}
\item	Quantify the magnetic structure involved in nascent coronal ejections though binocular vision of the source-region EUV corona, combined with 3D mapping of the solar field involved in eruptions through (near-) surface vector-field measurements and high-resolution atmospheric imaging;
\item	Understand the development of strong geomagnetically-induced currents through magnetotail-to-ionosphere in-situ probes, complemented with coordinated ground-based networks for geomagnetic and ionospheric variability;
\item	Map the global solar field, and use models and observations to determine the foundation of the heliospheric field, to drive models for the solar-wind plasma and magnetic field;
\item	Image the aurorae as tracers supporting the quantification of the state of the magnetosphere-ionosphere system;
\end{enumerate}

\noindent {\em Pathway II:}
\begin{enumerate}
\item[5]	Combined ground- and space-based observations for the modeling of the dynamic radiation-belt populations;
\end{enumerate}

\noindent {\em Pathway III:}
\begin{enumerate}
\item[6]	In-situ multi-point measurements to understand solar energetic particles in the Sun-Earth system.
\end{enumerate}

\section{Introduction}
As technological capabilities grow, society constructs a rapidly deepening insight into the forces that shape the environment of our home planet. With that advancing understanding comes a growing appreciation of our vulnerability to the various attributes of space weather. The variable conditions in what we think of as "space" drive society to deal with the hazards associated with living in close proximity to a star that sustains life on Earth even as it threatens humanity's technologies: the dynamic magnetism manifested by the Sun powers a sustained yet variable solar wind punctuated by explosive eruptions that at times envelop the planets, including Earth, in space storms with multiple potentially hazardous types of conditions. Powerful magnetic storms driven by solar eruptions endanger our all-pervasive electric power grid and disrupt the many operational radio signals passing through our planet's upper atmosphere (including the satellite navigation signals that is now vital to society). Energetic particle populations can lead to malfunctions of satellites and put astronauts at risk. 

These solar-powered effects in Earth's environment, collectively known as space weather (with the shorthand notation of SWx), pose serious threats to the safe and efficient functioning of society. In recognition of the magnitude of the hazards, governments around the world are investing in capabilities to increase our awareness of space weather, to advance our understanding of the processes involved, and to increase our ability to reliably forecast, prepare for, design to, and respond to space weather. Scientists with a wide variety of expertise are exploring the magnetism of the Sun from its deep interior to the outermost reaches of the planetary system, and its impacts on planetary environments. Great strides forward have been made towards a comprehensive study of all that happens between the magnetized interior of the Sun and mankind's technological infrastructure. An ensemble of in-situ and remote-sensing instruments on the Earth and in space is uncovering and quantifying many aspects of evolving space weather from Sun to Earth. These instruments monitor the progress of space weather across the vast interplanetary volume known as the heliosphere and observe how space weather impacts pass through Earth's magnetic cocoon and into Earth's upper atmosphere. But this ensemble is subject to major observational gaps and needs to be strengthened by improved coordination and integration. Scientific research and observations are increasingly combined in computer models and analyses that describe and forecast space weather conditions, aiming to protect society from the dangers of space storms, but still lacking essential capabilities in many parts of the overall Sun-to-Earth chain. 

Similar to terrestrial weather, space weather occurs all the time and it frequently affects our technological infrastructure in ways that are not catastrophic, but that are costly in their aggregate value to the global economy. Impacts on satellite-based navigation and timing capabilities or satellite communications affect users of the global navigation satellite system (GNSS) from precision agriculture to national security. Space weather is a serious constraint on the resilience of GNSS and thus a vital issue for a huge and growing range of economic activities: impacts are seen throughout the multitude of industry sectors that use satellite-based navigation systems, including on the aviation industry as it seeks to exploit GNSS as a way to optimize aircraft routing and hence airspace capacity; on the development of self-driving cars that rely on GNSS; on the accuracy of maritime drilling operations; on the timing systems that synchronize power and cell-phone networks in some countries, and on the finance industry as it moves to microsecond-resolution time stamping of financial transactions. Satellite anomalies and failures due to radiation storms impact the flow of information from satellite phones to weather monitoring, and from scientific exploration to intelligence monitoring. Space-weather impacts on the all-pervasive electric power grids have an even larger reach, with impacts ranging from rather frequent power-quality variations (such as voltage variations, frequency drifts and harmonics, and very short interruptions) to potential infrequent but major power interruptions.

Also similar to terrestrial weather, extreme space storms are expected to have major impacts. With our sensitive electrical, electronic, and space-based technologies expanding rapidly into a tightly woven network of applications we are not certain about the magnitudes of the impacts of extreme space weather, but studies suggest that we should be seriously concerned. For example, a single unusually severe geomagnetic storm has been hypothesized to cause long-term power outages to tens of millions of citizens of multiple countries and is thus listed among national security risks (e.g., the UK national risk register). Permanent loss of multiple satellites by severe energetic-particle storms has significant consequences for communications, surveillance, navigation, and national security.  

Although small compared to the risks deemed to be involved, the resources required to understand space weather and its impacts are substantial: the domain to be covered for successful space-weather forecasting and preparedness is vast, the physics of the intricately coupled processes involved is complex, and the diversity of the rapidly-growing user communities crosses all aspects of society and its globally-connected economy. Moreover, the impacts of space weather span the globe. For those reasons, an international approach is paramount to successfully advancing our scientific understanding of space weather. This realization prompted the Committee on Space Research (COSPAR) of the International Council for Science (ICSU) and the International Living With a Star (ILWS) Steering Committee to commission a strategic assessment of how to advance the science of space weather with the explicit aim of better meeting the user needs around the globe.  This report is the outcome of that activity.

In the spring of 2013, the leadership of COSPAR and ILWS appointed a team of experts charged to create this roadmap (see Appendix A for the process followed). The mission statement by COSPAR's Panel on Space Weather (PSW) and the ILWS steering committee asks the team to ``[R]eview current space weather capabilities and identify research and development priorities in the near, mid and long term which will provide demonstrable improvements to current information provision to space weather service users'', \referee{thereby expressing a focus on the terrestrial environment. The charge continued with} the expectation that ``the roadmap would cover as minimum:
\begin{itemize}
\item	Currently available data, and upcoming gaps
\item	Agency plans for space-based space weather data (national and international): treating both scientific and monitoring aspects of these missions.
\item	Space and ground based data access: where current data is either proprietary or where the geographic location of the measurement makes data access difficult
\item	Current capability gaps which would provide a marked improvement in space weather service capability.
\end{itemize}
The outcome should centre on a recommended approach to future developments, including coordination and addressing at least:
\begin{enumerate}
\item	 Key science challenges
\item	 Data needs, space and ground based
\item	 Smooth and organised transition of scientific developments into reliable services''.
\end{enumerate}

For the purpose of this roadmap we use the following definition: {\em Space weather refers to the variable state of the coupled space environment related to changing conditions on the Sun and in the terrestrial atmosphere, specifically those conditions that can influence the performance and reliability of space-borne and ground-based technological systems, and that can directly or indirectly endanger human well-being.} Aspects of space situational awareness such as space debris in Earth orbit and asteroids or other near-Earth objects are not considered in this roadmap. The structure under which the roadmap team operates implies that it focus exclusively on civilian needs, although many of its conclusions are likely directly pertinent to the security and military sectors of society. 

This roadmap identifies high-priority challenges in key areas of research that are expected to lead to a better understanding of the space environment and an improvement in the provision of timely, reliable information pertinent to effects on space-based and ground-based systems. Among those is the realization that we cannot at present use observations of the Sun to successfully model the magnetic field in coronal mass ejections (CMEs) en route to Earth, and thus we cannot forecast the strength of the perturbation of the magnetospheric field that will occur. Another example is that we understand too little of magnetic instabilities to forecast the timing and energy release in large solar flares or in intense (sub)storms in geospace. Advances in these areas will strengthen our ability to understand the entire web of physical phenomena that connect Sun and Earth, working towards a knowledge level to enable forecasts of these phenomena at high skill scores. 

The roadmap prioritizes those advances that can be made on short, intermediate, and decadal time scales, identifying gaps and opportunities from a predominantly, but not exclusively, geocentric perspective.
This roadmap does not formulate requirements for operational forecast or real-time environmental specification systems, nor does it address in detail the effort required to utilize scientific advances in the improvement of operational services. This roadmap does, however, recognize that forecasts (whether in near-real time or retrospectively) can help uncover gaps in scientific understanding or in modeling capabilities, and that test beds for forecast tools serve both to enable quantitative comparison of competing models and as staging environment for the phased transition of research tools to an operational application. 

The main body of this roadmap is complemented by Appendices that contain more detailed or supplemental information, describe the roadmap process, summarize scientific advances and needs, and outline instrumentation concepts. Appendix G provides a list of the acronyms and abbreviations used in this document.

\section{Space weather: society and science}

The task defined by COSPAR and ILWS is founded on the realization that space weather is a real and permanent hazard to society that needs to be, and can be, addressed by combining scientific research with engineering ingenuity: protecting society from space weather requires that we adequately understand the physical processes of space weather, that we characterize the conditions to which technological infrastructures need to be designed, that we learn to effectively forecast space weather, and that the consequences of acting on such forecasts are accepted as necessary for the protection of societal infrastructure. 

Societal use of, and dependence on, ground-based electrical systems and space-based assets has grown tremendously over the past decades, by far outpacing population growth as society continues to grow its electrical/electronic and space-based technologies.  Global electricity use has increased by a factor of about 1.6 over the 15-year period between 1997 and 2012 (International Energy Agency, 2013).  The global satellite industry revenue has multiplied by a factor of about 4.2 over that period (to US{\$}190 billion per year for 2012, part of a total value of US{\$}304 billion for the overall space industry, with over 1,000 operating satellites from over 50 countries; Satellite Industry Association, 2013). In contrast, the global population grew by approximately 20\%\ over that period (Population reference bureau, 2013), demonstrating our increasing use of electrical power and satellite-based information per capita. 

With that growth in electrical/electronic and space-based technologies comes increasing vulnerability to space weather: where a century ago the main risk was associated with the telegraph systems we now see impacts in the electric power grid, in satellite functionality, in the accuracy of navigation and timing information, and in long-range high-frequency (HF) radio communication. We see an increasing interest in understanding space weather impacts and the threats these pose are spread over a variety of civilian sectors (and non-civilian sectors that lie beyond the scope of this Roadmap). Selected reports on these impacts (that themselves provide information on more literature on the subject) are compiled in an on-line resource list\footnote{http://www.lmsal.com/$\sim$schryver/COSPARrm/SWlibrary.html} that accompanies this report; that resource list also includes a glossary of solar-terrestrial terms\footnote{http://www.swpc.noaa.gov/content/space-weather-glossary}, and links to a National Geographic introduction to space weather accessible via YouTube\footnote{http://www.lmsal.com/$\sim$schryver/COSPARrm/SWlibrary.html{\#}youtube}, and lectures related to space weather, its impacts, and its science in the NASA Heliophysics Summer School\footnote{http://www.vsp.ucar.edu/Heliophysics/}.

\begin{figure}
\includegraphics[width=8.8cm]{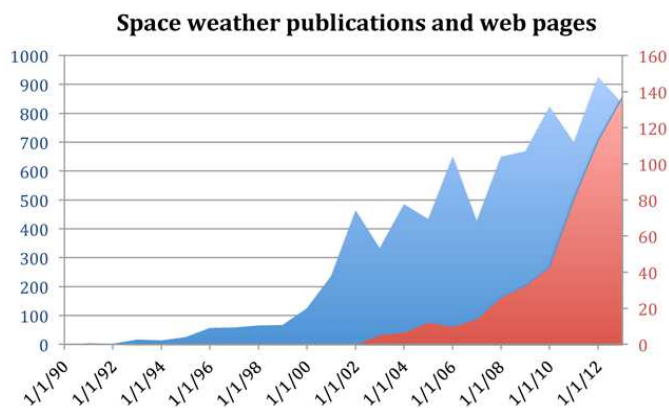}
\caption{Number of publications per year with ``space weather'' in the abstract in NASA/ADS (blue; left axis), and the number of web sites returned by a Google search for ``space weather'' within calendar years (since 2003) in thousands (red; right axis).  \label{fig1}}
\end{figure}
The reality of the threat to society posed by space weather is increasingly acknowledged - reflected, for example, in the exponential growth of the number of web pages on space weather (totaling over 130,000 new entries in 2013; see Figure 1) and in the number of customers subscribing to alert and forecast services (exceeding, for example, 44,000 for the US Space Weather Prediction Center). A core difficulty facing any study that attempts a cost-benefit analysis for space weather is inadequate knowledge of the technological and economic impacts of ongoing space weather and of the risk posed by extreme space storms. This hampers the quantitative identification of the most significant impacts of space weather and consequently the prioritization of the research areas and the deployment of infrastructure to protect against space weather. For our roadmap, existing assessments of threats and impacts compiled by organizations currently engaged in providing space weather infromation to affectd sectors proved adequate for a prioritization of recommendations, but achieving a quantification of the SWx impact on societal technologies is important for the process of allocating the required resources for research and forecasting, and of determining what sectors of society should be involved in appropriating which resources. For example, although the threat posed by geomagnetic storms is broadly recognized as real (e.g., Krausmann, 2011; Langhoff and Straume, 2012), establishing the vulnerability and consequences of such an event has proven to be very difficult (Space Studies Board, 2008; DHS Office of Risk Management and Analysis, 2011; JASON, 2011), hampering a cost-benefit assessment of investments that could make impacted systems less vulnerable by suitable engineering or by improved forecasting (DHS Office of Risk Management and Analysis, 2011).

Like terrestrial weather, space weather manifests itself as a variety
of distinct phenomena, and like terrestrial weather, it ranges from
benign to extremely severe. Most frequently, space weather is very
weak in intensity with apparently little impact on technology. Strong,
severe, or extreme geomagnetic conditions (as measured by the Kp
index, on NOAA's G
scale\footnote{http://www.swpc.noaa.gov/NOAAscales/} occur only 5\%\
of days through a solar magnetic cycle. Even though there are no
reports of catastrophic failures in the US high-voltage power grid,
\referee{one study finds that there appears to be} an increase by
40\%{$\pm$}20\%\ in insurance claims for industrial electrical and
electronic equipment on the 5\%\ most geomagnetically active days (as
measured by the rate of change in the geomagnetic field strength)
relative to quiet days, and there is an increase of 30\%{$\pm$}10\%\
in the occurrence frequency of substantial disturbances in the US
high-voltage power grid (Schrijver et al., 2014). \referee{Another
  study suggests that, overall, approximately 4\%\ of the disturbances
  in the US high-voltage power grid reported to the US Department of
  Energy appear to be} attributable to strong but not extreme
geomagnetic activity and its associated geomagnetically induced
currents (GICs; Schrijver and Mitchell, 2013). \referee{More such
  studies are needed to validate these findings and particularly to
  help uncover the pathways by which these impacts occur (a point to
  which we return in our recommendations below).}

Other aspects of space weather can adversely affect satellites.  Severe solar energetic-particle (SEP) storms, for example, impact satellites directly, while expansion of the terrestrial upper atmosphere by magnetospheric variability (through Joule heating) may affect low-orbiting satellites by modifying their orbits through increased drag which is an issue for on-orbit operations, collision avoidance, and could eventually lead to early re-entry (e.g., Rodgers et al., 1998). A series of such storms during the 2003 October-November time frame, for example, saw considerable impacts on satellites through electronic single-event upsets (SEUs), solar-array degradation, modified orbit dynamics for spacecraft in low-Earth orbits, and noise on both housekeeping data and instrument data. Another manifestation of space weather is the enhancement of radiation belt (RB) particles and of magnetospheric plasma that cause charging/discharging phenomena or state upsets in satellite electronics. For 34 Earth and space science missions from NASA's Science Mission Directorate, for example, 59\%\ of the spacecraft experienced such effects (Barbieri and Mahmot, 2004). A graphic laboratory demonstration of discharging inside dielectric materials (a critical space weather impact for satellites in geosynchronous and middle Earth orbit) is available on YouTube\footnote{http://youtu.be/-EKdxzZ52zU}.

A third impact for space weather occurs via severe modification of trans-ionospheric signals by highly variable plasma density in space and time, thus affecting customers of GNSS services. The economic impact of this type of space weather has yet to be investigated, being complicated by the fact that it will mostly occur well downstream of the immediate service providers and also by the fact that GNSS technology is rapidly evolving even as the total numbers of users and uses increases, increasingly in layered applications that may hide just how GNSS-dependent a system is.

The threat posed to society by the most severe space storms that occur a few times per century is largely unknown and the magnitude of such a threat is consequently highly uncertain: the technological landscape evolves so rapidly that our modern-day highly-interconnected societal infrastructure has not been subjected to the worst space storms that can occur. Some reports put the threat by the most severe space storms among the significant threats faced by our technology-dependent society. Geomagnetic disturbances (GMDs) on electrical systems have been known to impact technology for over 150 years, starting with the telegraph systems, and showing a clear correlation with the sunspot cycle (Boteler et al., 1998). Among the insurance-industry reports that review the space weather risk landscape from the industry perspective (e.g., Hapgood, 2011), one (Lloyd's, 2013) concludes for the US in particular that ``the total population at risk of extended power outage from a Carrington-level storm [i.e., unusually strong but likely to occur approximately once per century] is between 20-40 million, with durations of 16 days to 1-2 years'', while recognizing that even for weaker storms ``the potential damage to densely populated regions along the Atlantic coast is significant.'' The World Economic Forum (2013) includes vulnerability to geomagnetic storms explicitly in its listing of top environmental risks deemed to be able to significantly impact the global economy. The US National Intelligence Council (2013) noted that ``[u]ntil ‘cures' are implemented, solar super-storms will pose a large-scale threat to the world's social and economic fabric''. A report on a risk analysis for space weather impacts in the UK (Royal Academy of Engineering, 2013) stated that ``the reasonable worst case scenario would have a significant impact on the national electricity grid.'' Impacts of geomagnetic storms are not limited to grids at high latitudes: although geomagnetic latitude is an important factor in GMD strengths, ``geological conditions tend to override the effect of latitude'' (NERC, 1989). In addition, lower-latitude regions can experience GICs arising from fluctuations in the magnetospheric ring current as demonstrated by the serious impact of the Halloween 2003 storms on the power grid in South Africa (e.g., Gaunt, 2013). 

Even if we disregard the uncertain impact of extreme space weather, we
know that impacts of moderate to extreme space storms that occur a few
hundred times per 11-year solar cycle have consequences that merit
substantial attention and investment. \referee{One study of the
  overall cost to the US economy alone of non-catastrophic
  disturbances in the US power grid attributable to geomagnetically
  induced currents suggests that the impacts may be as large as
  US{\$}5-10 billion/year (Schrijver et al., 2014; we note that
  more detailed follow-up studies are needed to validate this result,
  because that study had to combine its significant findings on
  insurance-claim statistics with other sources of information to
  estimate the financial impact).} Economic impacts of space weather
through other technological infrastructures have yet to be
established, but threat assessments suggest ``that space weather is
the largest contributor to single-frequency GPS errors and a
significant factor for differential GPS'' (American Meteorological
Society, 2011), for an industry that is worth of order US{\$}100
billion/year worldwide (American Meteorological Society, 2011; Pham,
2011). A recent study (Schulte in den Baumen et al., 2014) made a
first attempt to couple a GIC impact model with an economic model of
global trade showing how GIC impact in three different regions (China,
Europe and North America) would drive impacts across the world
economy. It reinforces the message that space weather is a global
problem - that a physical impact in one region can damage economies
far from the impact site.

Given the persistent presence of the threat, society's increasing
exposure to space weather, and the likely low-frequency but
high-impact extreme-storm scenarios, it is not surprising that calls
for the preparation for, forecasting of, mitigation against, and
vulnerability assessment for space weather impacts by the
international community echo in various studies over the past decade,
including reports from academia (Hapgood, 2011), from the US National
Research Council (2008), the UN Committee on the Peaceful Use of Outer
Space (2013), and from the Organisation for Economic Co-operation and
Development (OECD, 2011). \referee{Recognition of the societal hazard
  posed by space weather is also reflected in reports by, for example,
  the North American Electric Reliability Corporation (NERC, 2012b,
  20131, 2013b) and in the order by the US Federal Energy Regulatory
  Commission (FERC, 2013) to develop reliability standards for geomagnetic
  disturbances.}

The user base interested in forecasts of space weather is growing rapidly with the increased awareness of space-weather threats and impacts. The official US space-weather forecast center (the Space Weather Prediction Center of the US National Oceanographic and Atmospheric Administration), for example, sees a continuing rapid growth in subscribers to its ``Product Subscription Service''\footnote{http://pss.swpc.noaa.gov} that was initiated in 2005 and that exceeded 40,000 individual subscribers early in 2014. A survey of the subscribers to the SWPC service in 2013 enabled an assessment of the interests from the user side (Schrijver and Rabanal, 2013), which concluded that ``[s]pace weather information is most commonly obtained for reasons of [indirect impacts through interruptions of power or communications on] human safety and continuity or reliability of operations. The information is primarily used for situational awareness, as aid to understand anomalies, to avoid impacts on current and near-future operations by implementing mitigating strategies, and to prepare for potential near-future impacts that might occur in conjunction with contingencies that include electric power outages or GPS perturbations. Interest in, anticipated impacts from, and responses to the three main categories of space weather [- geomagnetic, radiation, and ionospheric storms -] are quite uniform across societal sectors. Approximately 40\%\ of the respondents expect serious to very serious impacts from space weather events if no action were taken to mitigate or in the absence of adequate space weather information. The impacts of space weather are deemed to be substantially reduced because of the availability of, and the response to, space weather forecasts and alerts.'' It appears that many users of space weather forecasts apply the forecast information to avoid impacts on their systems and operations, either by increased monitoring given situational awareness or by taking preventive mitigating actions. As with terrestrial weather, this means that the economic value of space weather forecasts likely significantly exceeds the total costs of detrimental impacts, and that this value would increase as forecast accuracy and specificity would increase. Other valuable uses of space weather information as indicated by the subscribers to space-weather information lie in anomaly analysis and system design specification (Schrijver and Rabanal, 2013).

The study of space weather is important because of its societal relevance and much headway has been made in recent years (see a brief discussion in Appendix C). Space weather also teaches us about the physical processes of the local cosmos that is our home within the Galaxy. More commonly known as the field of Sun-Earth connections, or as heliophysics particularly within the US, this is the science of the astrophysical processes that occur in the deep solar interior, in the vast reaches of the solar atmosphere that extend beyond the furthest planet out to the interstellar medium, and that includes the variety of coupling processes to the planets and natural satellites of our own solar system. Understanding all of these processes and interactions between diverse environments is our stepping stone to understanding what happens elsewhere in the universe in similar environments, as much as to understanding the distant past and future of our own planetary system and its host star.   

\begin{figure*}[ht]
\includegraphics[width=\textwidth]{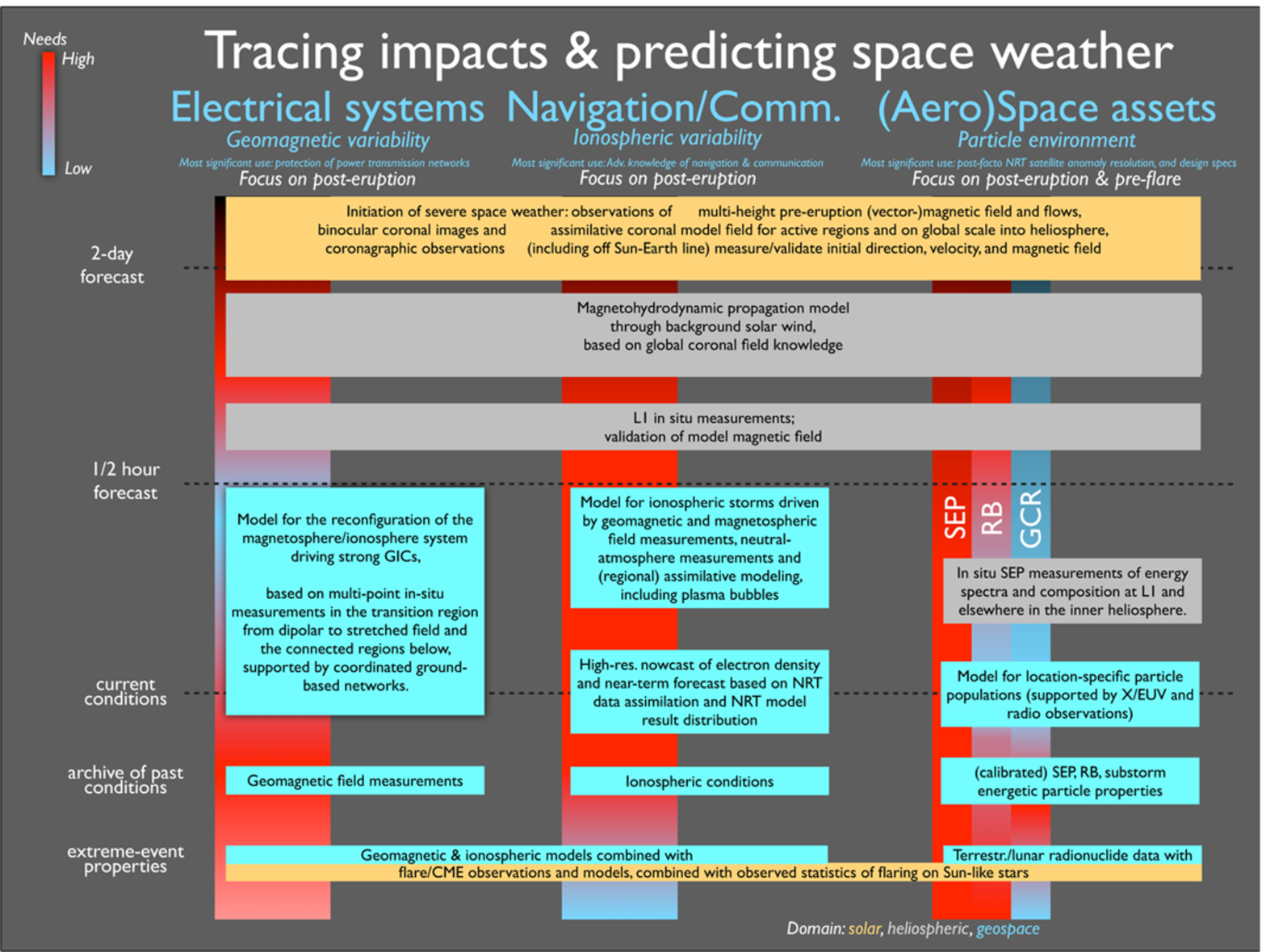}
\caption{Overview of the primary impacts and their societal sectors of space weather.  The red shading in the background indicates the priority needs for the user communities behind each of the impacts, differentiated by time scale for forecast or for archival information as shown on the left. Text boxes identify the primary needed observations, archival measurements, and models to complete the impact chain, differentiated (using color, see legend) by solar, heliospheric, and geospace domains.  \label{fig2}}
\end{figure*}
\section{User needs}
The complexity of the interconnected system of physical processes involved in space weather precludes a single, comprehensive, yet understandable and concise exploration of the network as a whole. To effectively and efficiently address its charge, the roadmap team therefore developed a strategy that focuses on three largely distinct space weather phenomena with largely complementary impact chains into societal technologies: a) geomagnetic disturbances that drive electric currents through the electric power infrastructure, b) variability in the ionospheric electron density that impacts positioning and navigation systems, and c) energetic particles for space assets, astronauts and stratospheric air traffic (with possible impacts on passengers, crews, and avionics). Many other systems are susceptible to these and other manifestations of space weather, whether directly or indirectly, but together these three chains encompass the most significant impacts and together they cover most of the research and forecast needs that would be needed for the other types of impacts. Figure 2 lays out these three chains side by side, with an approximate vertical time axis to map the user needs and the physical domains involved. Some details of the impact tracing exercise are described in Appendix B.  

The societal needs for space weather information, including forecasts, are rapidly growing as witnessed by the exponential growth in the US NOAA/SWPC product customer base, the growth in space-weather research publications (Fig. 1), and in the variety of strategic studies and reports (see Section 2). Of the subscribers to the SWPC electronic alert and forecast services customers, only about 20\%\ deem the quality and content of the available information adequate (Schrijver and Rabanal, 2013). Even though about two-thirds of these customers considers the information ``generally adequate'', the forecasts leave much to be desired owing to a lack of scientific understanding in multiple sectors of the Sun-Earth connected system and to a poor knowledge of the magnitude of impacts. 

When provided with warnings of significant space weather storms, the subscribers to space weather information services typically either increase monitoring or implement preventive action in roughly comparable fractions. Increased availability of real-time information and increased specificity and accuracy of the forecasts should enable more targeted and effective actions, while lowering the costs associated with false alarms. Improved quality of forecasts and improved quantification of the system vulnerability can put society on the desirable path of either efficiently acting on space-weather forecasts as one would on terrestrial weather forecasts, or by improving system design and operations that reduce vulnerability as done for other natural forces, at least up to some value that is ideally based on a well-founded cost-benefit risk assessment. 

In the three technological infrastructures of our impact tracing exercise different approaches to mitigate the SWx risk are used and consequently also the needs of SWx  services supporting the societal sectors using these infrastructures are different:

\subsection{Electric power sector}

The electric power sector is primarily affected by geo\-mag\-netically-induced currents (GICs), which can push transformers off their linear domains leading to dissipative heat that can damage transformers, and to generation of harmonics of the primary 60-Hz or 50-Hz wave. These harmonics can cause protective systems to trip and can affect systems operated by power customers downstream in the distribution network. Power-grid operators can reduce or prevent damage by optimizing design, by introducing protective equipment, or by redistributing and changing the power-generation resources so that fewer long-distance transfers are needed and more near-locally-generated power is available to counter frequency and voltage modulations. Other solutions, such as that adopted in the UK with its relatively compact power system, are to bring all available grid links into operation in order to maximize redundancy and to spread GICs over the whole system, reducing the impact on individual system elements.  To optimize system protection, improved knowledge of the impacts is needed, including extreme-event scenarios, which requires historical and pre-historical information. 

In order to effectively plan the power grid operations subject to strong GMDs, the sector has expressed a need for a reliable warning at least half a day ahead of coming storms, preferably with a reliable forecast of the magnitude and duration of the geomagnetically induced currents (GICs). The propagation time for CMEs from Sun to Earth is typically of order 2-4 days, and even for the fastest known events exceeds 0.75 days. Hence, the need for a forecast for CME arrivals about 12 hours ahead requires the formulation of a forecast well after a solar eruption has occurred (and should have been observed), but long before it approaches geospace (here defined loosely as a domain encompassing everything from Earth's middle atmosphere out to the upstream magnetopause and downstream beyond the magnetotail), or any sensors currently available upstream of Earth. 

In Figure 2, the highest-priority user needs for the power sector are indicated by red-shaded areas in the bar in the background of the left-hand column. The deepest red areas for GIC service needs are in the forecasts with half-day lead times and in the specifications of extreme conditions based on archival information on past conditions. Interests in 12hrs-ahead warnings of GMDs are also expressed by industries using the geomagnetic field directly, including mining, drilling, and surveying. We note that addressing the needs of the GMD/GIC and ionospheric communities also provides information on modulations of satellite drag caused by current dissipation within the ITM domain of Earth's upper atmosphere.

\subsection{Positioning, navigation, and communication}
Positioning and navigation services such as provided by the GNSS are most affected by ionospheric electron-density variability via strong coupling with magnetospheric and thermospheric changes driven by CMEs and high-speed solar wind streams or related to X-ray and extreme ultraviolet (EUV) eruptions on the solar surface or (of particular importance for the low-latitude ionosphere) upward propagating waves originating in the lower Earth atmosphere. Ionospheric range errors of up to 100m in single frequency GNSS applications can be largely mitigated by proper ionospheric models, or more accurately by regional and/or global maps of the total electron content (TEC) or by 3D electron density reconstructions provided in near-real time (i.e., within minutes, or even seconds, of measurements being made). Precise and safety-critical positioning and navigation, as required, for instance, in aviation, suffer in particular from steep spatial gradients and temporal variability of the plasma density. Thus, plasma bubbles and small-scale electron density irregularities (known as plasma turbulence) may cause strong fluctuations in signal amplitude and phase that are called radio scintillations; severe phase scintillation may even cause loss of phase lock of the signal. 

In professional GNSS services position accuracy is supported by augmentation systems. These include (a) wide area satellite-based augmentation systems (SBAS) such as EGNOS in Europe, WAAS in North America, and MSAS in Asia), and (b) ground-based augmentation systems that provide local services such as around sea- and air-ports. These systems monitor the accuracy and reliability of the positioning signal and provide an integrity flag that can warn GNSS users when back-up solutions for navigation must be used (well demonstrated by the WAAS system during the Halloween storms of 2003). SBAS systems bring a further vulnerability to space weather in that their messages must cross the ionosphere as both uplink to, and downlink from, a satellite hosting an SBAS relay system. Both uplink and downlink are vulnerable to ionospheric radio scintillation. Similarly the increasing use of satellite links for communication with, and tracking of, aircraft opens up a vulnerability to scintillation. On the other hand, forecasts of plasma perturbations have a great potential for mitigating space weather impact on radio systems by alerting operators and customers of telecommunication and navigation systems and remote sensing radars as well. Thus, for example, airplanes whose HF and satellite communication may be disturbed heavily typically require - as is the case for weather conditions - a forecast of ionospheric electron density variability a good fraction of a day ahead of time. Correcting for, or accommodating, these impacts on HF telecommunication and GNSS signals requires an extensive network of ionospheric measurement devices, rapid modeling, and rapid dissemination of correction information. The needs of most of the GNSS customers thus lands in the time domain from current state (nowcast) to of order an hour ahead (see middle column of Figure 2), but somewhat longer-term forecasts are valuable, for example, for planning purposes of operations and emergency response activities. These needs will evolve as dual-frequency and/or differential GNSS usage become more widespread.

Another aspect of ionospheric electron density variability driven by space weather is radio wave absorption in the frequency range below about 30 MHz. This arises when high-energy inputs to the atmosphere generate additional ionization at altitudes around 90 km altitude and below known as D region in the ionosphere. In this region the neutral density is high and therefore also the collision frequency between charged particles and neutrals. Thus, charged particles that are excited by electromagnetic waves in the frequency range mentioned above lose their energy very fast, i.e. the radio wave will be absorbed or at least heavily damped. D region ionization can arise from a variety of high-energy space weather phenomena: intense bursts of X-rays from solar flares and the precipitation of solar energetic particles (SEP). These have the possibility to disrupt HF communications that are used by civil aviation in remote regions, especially over the oceans and poles. The impact of flare X-rays appears to be a nuisance rather than a serious problem, since they produce only short-lived disruptions (of order an hour) that can be mitigated by well-established procedures and by use of satcom applications as an increasingly common backup. However, the impact of SEPs is very significant as their impact is greatest in the polar regions (where satcom backup is not currently available) and is long-lived (up to days) so that procedural measures are insufficient. Thus polar flights are diverted to other less vulnerable routes during SEP events, imposing significant extra costs on airlines (around {\$}100k per diversion; National Research Council, 2008). This problem could be mitigated in the future with increased use of satcom the polar regions, including use of the potential Canadian PCW (Polar Communications and Weather) satellite system.

The presence of a strong D region also impacts low-frequency navigation systems by advancing the arrival of skywave, i.e., the LF signal reflected from the ionosphere that interferes with ground-wave, which is the LF signal propagating as an interface wave along the surface of the Earth. Thus older LF systems such as LORAN and DECCA were prone to major position errors, for example during REP events at mid-latitudes. However, most of these old systems have now been taken out of service in favor of more accurate GNSS. New LF systems, such as the eLORAN system now deployed operationally in the UK, can match the accuracy of GNSS, but like GNSS include integrity checks to warn when location data are not accurate.

\subsection{(Aero)space assets}
The impacts of high-energy solar-energ\-etic and radiation-belt particles are strongest in space-based assets, including satellites and astronauts, but extend to stratospheric aircraft as these rely increasingly on ever-smaller electronics systems distributed throughout the vehicle. Such effects are caused by energetic particles (those of galactic origin all the time, and SEP where these have strong fluxes at energies $>$500\,MeV) that penetrate to the Earth's atmosphere and interact with atmospheric neutral particles. Such collisions produce cascades of neutrons and ions that can interact with aircraft to produce single event effects (SEE) and increased dose rate for passengers and crew. During SEP events radiation effects at aircraft trajectories are strongest at high latitudes where solar particles can more easily penetrate. Whether there are detrimental effects on the health of airline crews and passengers remains to be established, but assessing the risk, regardless of the outcome of the studies, is clearly of importance. 

The requirements from the involved sectors separate somewhat by type of radiation, as indicated in Figure 2.  In the background, with mostly longer-term and moderate variability, is the galactic cosmic ray (GCR) population, for which archival data spanning centuries is important and in principle available in natural records such as ice sheets, rocks, sediments, and the biosphere. Radiation-belt (RB) particles need to be characterized for system design purposes based on archival and extreme-event information, and their nowcast and forecast is important for planning of spacecraft special operations and maneuvers. SEPs are highly variable and dangerous during high-intensity, large-fluence events not only for satellites, but also for astronaut activities (which puts particular value on the all-clear forecasts for 1-3 days). Archival and near-real time information is needed for design and anomaly resolution, while forecasts on time scales of up to a day or so are, as for the RB particles, important for planning of spacecraft and astronaut operations. The time-scale needs for the three different populations of energetic particles are shown separately in color-coded background bars in Figure 2. 

The customer requirements for each infrastructure described above are based on present-day needs. We anticipate that the requirements for GMDs and energetic particles will remain in place for many years to come, perhaps even with more urgency as the power-grid reach and the power-dependence of society grow and as we depend increasingly on space-based assets. For navigation and positioning services the requirements appear to be already changing as technologies advance: whereas satellite-only, single-frequency navigation systems are sensitive to ionospheric electron density variability, use of dual-frequency solutions and GNSS augmentation systems are reducing the impact of large scale electron density variations. The mitigation of harm from ionospheric scintillation is probably now the key challenge for future GNSS systems.

\section{Promising opportunities and some challenges}
\subsection{The opportunity of improved CME forecasts}

One conclusion from the exercise with three sample impact chains (c.f., Appendix B) is that during recent years research in solar-terrestrial physics has advanced so much (compare Appendix C) that within the time span of next 5-10 years we anticipate far more accurate and specific forecasts of incoming solar-wind perturbations that will have lead times from some hours to a couple of days. Such forecasts (pertaining to the heliospheric evolution of post-eruption solar activity prior to their arrival at Earth) would improve our capabilities to respond to the user needs described in Section 3 in several ways.  Therefore, we have dedicated one branch in our recommendations (which we refer to as Pathway I) particularly to such research and collaboration activities that can be coupled with the anticipated advancements in post-eruption forecasts.

Ongoing solar missions have given us guidance on optimal solar surface observations to support modeling so that improved estimates on the CME magnetic structure and energy content, as well as the propagation in the heliosphere can be achieved. With this information not only the arrival timing of CMEs but in particular their geo-effectiveness could be provided more reliably than today for lead times well beyond 0.5-1 hour (which is the characteristic travel time for heliospheric perturbations from a sentinel placed in the solar wind upstream of Earth at the so-called L1 point where a spacecraft can readily remain at a nearly fixed point essentially on the Sun-Earth line). At the same time, our capabilities to monitor and model the near-Earth energetic particle and radiation environment have advanced so much that forecasts of similar lead times can be generated to support also operations of (aero-)space assets.  

The improved CME forecasts would be beneficial also for electrical power systems because the most severe incidences of geomagnetic disturbances are almost always associated with magnetospheric evolution in response to the stresses imparted by the variations in the solar wind. In many cases support would come also for the technologies in communication and navigation systems, although these applications would still need to tolerate the direct disturbances from solar eruptions which come in the form of light and particle bursts. 

\begin{figure*}[ht]
\includegraphics[width=\textwidth]{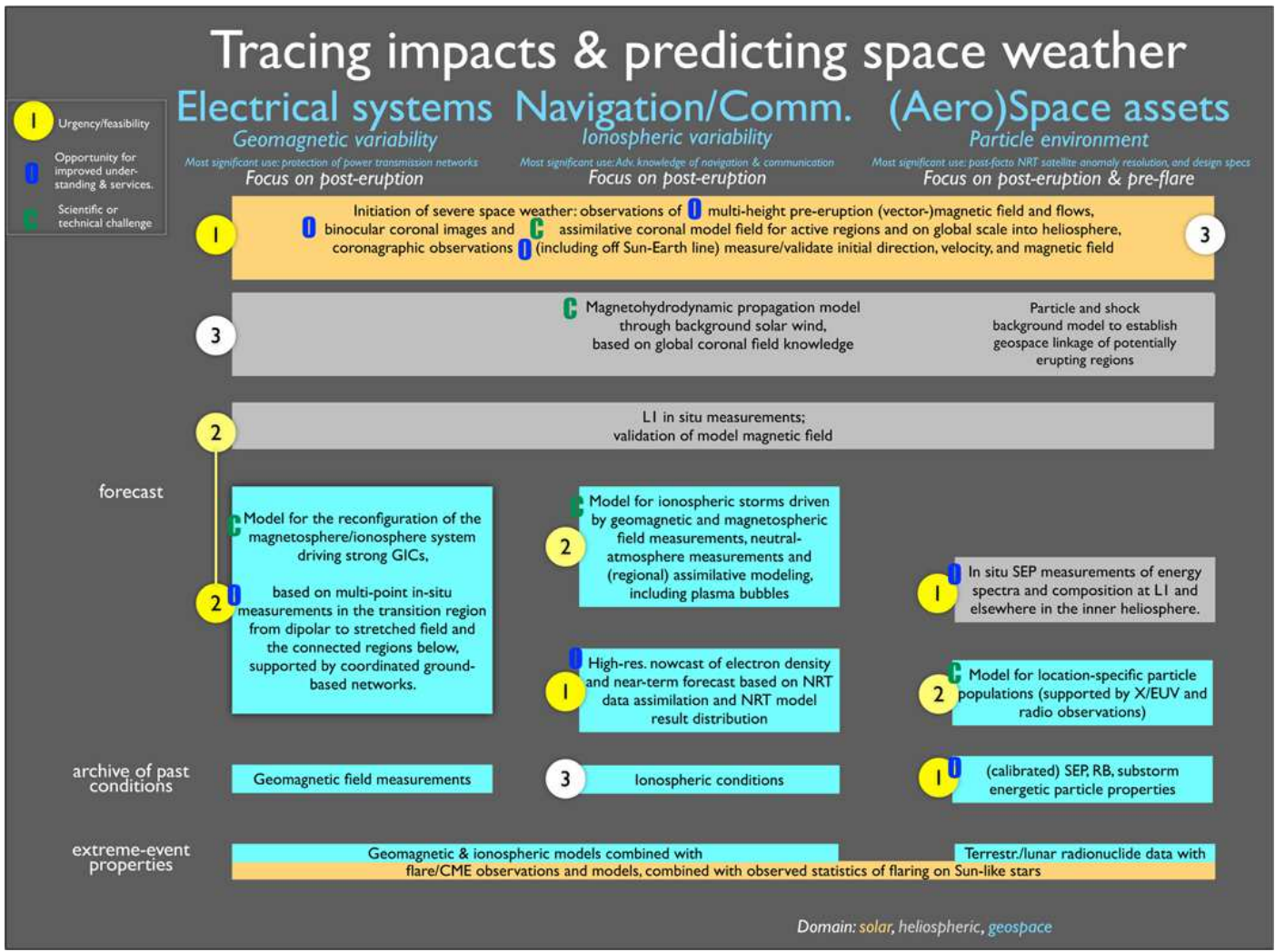}
\caption{Identification of the top-priority needs to advance understanding of space weather to better meet the user needs, differentiated by impact. Within each chain, letters identify opportunities for immediate advance (O), and the largest challenges (C) that need to be addressed.   \label{fig3}}
\end{figure*}
Below, we first describe some promising opportunities in solar and heliospheric research that support the efforts for CME forecasts with lead times well in excess of 1\,hr. Harvesting these opportunities will naturally generate also some new challenges, not just in solar/heliospheric research but also in geospace research as there additional work is needed for efficient utilization of the improved characterization of solar driving. Luckily, there are some opportunities also on the geospace side that we should now seize in order to gain better understanding of certain processes in the magneto\-sphere-ionosphere-thermosphere system that control the severity of SWx disturbances in the three impact areas. Challenges and opportunities in the research on factors controlling large GMD and GIC are addressed after the discussion of solar and heliospheric matters (Sections 4.1 and 4.2). In Sections 4.3 and 4.4 we discuss the tasks for improved forecasts and environment specification for navigation and positioning and (aero-)space assets. Main points of our reasoning are presented in the diagrams of Figures 2 and 3. 

The diagram in Figure 3 does not only address the case of CME prediction (post-eruption), it also describes some tasks associated with the attempts to identify active areas on the solar surface even before their eruption in order to achieve forecasts with lead times of several days. Although our current scientific understanding does not provide adequate support for such predictions with good enough confidence levels, it is good to keep also this need in the picture as a long-term goal. The track for forecasts with lead times beyond a few days is described below in Section 4.5. Specification of extreme conditions and solar-cycle forecasts are discussed in Section 4.6.

\subsubsection{Solar surface}
Once space weather phenomena need to be forecast with lead times that have the solar-wind driver beyond the Sun-Earth L1 site that lies one million miles upwind from our planet, the observations required to make such forecasts shift to the origins of the space weather that lie within the solar corona and the innermost domains of the heliosphere. The reason for that shift all the way towards the Sun is that we do not presently have the technological means to station a spacecraft on the Sun-Earth line somewhere halfway the Sun and the Earth (solar sails are being considered, but have yet to be demonstrated to be technologically feasible), nor do we have the financial resources to maintain a large fleet or near-Sun orbiters with at least one always near the Sun-Earth line (although constellations of spacecraft have been considered for this). Given this situation, we conclude that forecasts beyond one hour for solar-wind-driven magnetospheric and ionospheric variability are necessarily based on observations of the domain near the Sun and on heliospheric propagation models from there out to the planets, unless novel observational techniques or instrumentation can be developed and deployed. 

The timing of the impacts of heliospheric storms on geospace is currently generally forecast based on coronagraphic observations from which propagation speeds and directions are estimated. Observations from the L1 sentinel point are nowadays provided by the LASCO coronagraph onboard the aging ESA/NASA SoHO mission (launched in 1996), aided by the STEREO coronagraphs and heliospheric imagers at least when in appropriate phases of their orbits around the Sun. These observations enable estimation of arrival times that have recently become accurate to within a quarter to half a day, while ongoing modeling of the background solar wind enables a rough estimate of the densities and shock strengths of the incoming storm. Coronagraphic imaging (now routinely in use, but generally not extending to close to the solar surface) or alternatively high-sensitivity high-altitude coronal EUV imaging (which appears technologically feasible) is therefore critical to forecasts of solar-wind driven geomagnetic disturbances and related phenomena in the geospace on time scales of a few hours to a few days. The potential loss of L1 coronagraphy as now performed by SoHO's LASCO is thus a weak link in the currently available inventory of space-weather inputs, until a replacement coronagraph is launched or until alternative means of obtaining heliospheric observations become routinely available. 

When looking to forecast CME arrival times, one long-standing technique (Hewish, 1955) that is now maturing for space weather purposes is Interplanetary Scintillation (IPS; Hewish, Scott, and Wills, 1964). This uses the scintillation patterns observed in the signals received from astronomical radio sources. This scintillation comes from the scattering of radio waves by small-scale plasma density irregularities propagating with the bulk solar-wind outflow. Thus, IPS provides another means of remotely sensing the solar wind. As a complement to, say, white-light heliospheric imagers similar to those currently aboard the STEREO spacecraft well off the Sun-Earth line, IPS can provide information on CME speeds and thus on CME arrival times. When combined with modeling techniques and/or in-situ data, other parameters such as CME masses can also be obtained, along with CME propagation directions and arrival times (e.g., Bisi et al., 2010, and references therein).

Another technique that is being developed to map the approach of the
CME is based on loss-cone anisotropy of ground-based muon
observations: behind the shock (if present) and inside the CME there
is a cosmic-ray density depleted region (associated with a Forbush
decrease) that some times results in precursory signatures observable upstream of the shock (1 to 7 hours ahead). A detailed study has recently been undertaken by a group of Canadian experts (Trichtchenko et al., 2013) who have delivered a roadmap for development of a pre-operational system.

Successful prediction of the arrival times of heliospheric perturbations is but one factor in the forecast of the magnitude or duration of geomagnetic perturbations. Critical for the forecasting of magnitude and duration of GMDs is the knowledge of the magnetic field structure within approaching solar-wind perturbations that none of the above methods can provide. Most important in determining the magnitude of the geospace response (be it immediate or delayed) is commonly said to be the direction of the magnetic field at the leading edge, while the trailing structure affects the storm duration. There is, however, as yet only a poor understanding of the coupling of the heliospheric field into geomagnetic activity and the data show that more characteristics of the solar-wind field are definitely required (e.g., Newell et al., 2007).

As a prospect for the future, a radio technique known as Faraday rotation (FR) may offer a means of remotely sensing the heliospheric magnetic field. The FR technique is already used by astronomers to remotely sense galactic magnetic fields and there are now growing efforts to apply the technique to study the heliospheric field. This is a significant research challenge but may be within reach with cutting edge radio telescope technologies, such as LOFAR and those being developed for the Square Kilometer Array (SKA). 

Despite some potential for the future, for the near-term, however, the only path towards successfully forecasting the intensity and duration of strong magnetospheric storms, of related ionospheric current and electron density variability, and of GICs beyond the next hour or so, requires that we can model the field that left the Sun. Specifically, we need to develop the means to model the internal properties of the configuration of twisted magnetic field that was injected into the heliosphere in the earliest phases of a CME and the subsequent evolution of that field configuration in its interaction with the high-coronal and inner-heliospheric magnetic field into which it is thrust.

Establishing the detailed geometry and strength of the coronal
magnetic field prior to eruptions (c.f. the yellow box at the top of
Figure 3), be it from strong-field compact active regions or from
weak-field extended quiet-Sun regions, is an area of active
research. For active regions, the key ingredients to this are
high-resolution magnetic field observations of the solar surface
layers and the near-surface (chromospheric) layers, and high-resolution EUV imaging in narrow thermal regimes of the overlying corona (flagged as ``opportunities'' in Figure 3). Polarimetric imaging is used to derive the vector-magnetic field, while coronal imaging provides the pathways of coronal loops that outline the magnetic field. At present, field extrapolations based on photospheric vector-magnetic maps are limited to phases of slow evolution, often using the approximation that all forces within the magnetic field are balanced (in the so-called non-linear force-free approximation in which curvature and pressure-gradient Lorentz forces cancel against each other). But even when limiting this modeling to phases of slow evolution, models commonly fail to reproduce the observed coronal loop geometry, revealing the sensitivity of the field to boundary conditions above, below, and in the vicinity of an erupting solar region, or to the assumed initial conditions, or both (e.g., DeRosa et al., 2008). For weak-field regions, vector fields at the surface and near-surface are difficult to establish, so X-ray or EUV imaging and coronal polarimetric measurements sensitive to field direction (see, e.g., {B{\c a}k-St{\c e}{\'s}licka} et al., 2013) may be used to constrain the magnetic field and, as called out as an opportunity for large telescopes (such as DKIST) polarimetric measurements sensitive to coronal magnetic field strength may also be used in future.  In general, it appears that combined measurements of the surface magnetic field and of the coronal field as traced by its plasma are needed to significantly advance beyond present-day limited and often unsuccessful field modeling (for example, Malanushenko et al., 2014).

The most urgent challenge (labeled accordingly ``C'' in the top yellow box in Figure 3) to be tackled to achieve reliable forecasts of intensities and durations of geomagnetic storms on time scales of a few hours to a few days is thus the realization of algorithms and observations needed for coronal field modeling. These algorithms require at least full-Sun magnetograms combined with active-region scale vector-magnetic data and coronal high-resolution narrow-band imaging to provide crisp images of the thermal plasma glow: that combination provides information on the solar surface magnetic field and, though the tracing of the coronal field in the coronal emission patterns, also on the field in the coronal volume above. Tracing the coronal structures requires not only good angular resolution, but also separation of images into narrow thermal bands so that structures stand out well from their surroundings. Single-perspective imaging may be shown to be adequate, but binocular imaging, necessarily involving an observatory off the Sun-Earth line by approximately 10-20 degrees, would substantially aid in correctly interpreting the 3-dimensional coronal field structure that is otherwise seen only in projection against the sky. Thus a second point of view for coronal imaging substantially off the Sun-Earth line is identified as offering the potential of a big leap forward in space-weather forecasting on the hours to days time scale. Surface and near-surface magnetic field measurements and EUV coronal imaging need to be continued, because longer-term forecasts are not feasible without these.

\subsubsection{Heliosphere}
Intermediate to the initial phases of solar eruptions and the geospace response lies the vast expanse of the surrounding corona coupled with the inner heliosphere (gray boxes in Figure 3): first the CMEs propagate through the surrounding coronal field that is strong enough to modify the primary direction of the initiating eruption up to a few to some ten solar radii, then they interact with the pre-existing magnetized solar wind into which the eruptions plow on their way to the outer heliosphere. Here lies what is presently a challenge in the form of the development of a data-driven heliospheric propagation model that must include the properties of the magnetic field and that is, ideally, modified by ingestion of information on the heliospheric state by the STEREO spacecraft or by tomographic methods relying on interplanetary radio scintillation. The development and validation of such codes emerged as challenge in our impact tracings, that can be successfully dealt with after local coronal field modeling of the state before and after eruptions has been developed to provide input to such models, and if the global coronal field is known well enough to know how the nascent CME may have been modified and deflected even before entering the heliosphere proper. Only once such propagation models are available based on the inner-boundary initial condition of what has left the solar domain can the heliospheric model be fully tested against the observables obtained at L1 or, for the purpose of model testing and validation, at any other point within the extended heliosphere, be it closer or further from the Sun, be it by planetary or heliospheric missions with in-situ instrumentation. 

Along with the development of this heliospheric MHD modeling
capability must come the test of whether it is adequate for these
few-day forecasts to measure only the Earth-facing side of the solar
magnetic field or whether more extended surface coverage of
line-of-sight or vector-magnetic field is required to adequately model
the foundation of the solar wind based on the lowest-order components
of the coronal field that are determined by field on all sides of the
Sun, including its hard-to-observe polar regions. Similarly, it should
be established how important observations are of the inner heliosphere
off the Sun-Earth line as data to be assimilated: are coronagraphic
observations from well off the Sun-Earth line (such as from around the
Sun-Earth L5 or L4 regions that trail and lead Earth in its orbit
around the Sun by some 40 to 80 degrees where spacecraft can maintain
a relatively stable position with respect to Sun and Earth) critically
important, or useful as guides, or are they mostly superfluous provided that
Earth-perspective data and advanced coronal and heliospheric modeling
are available? Depending on the outcome of the study of that question,
coronal and inner-heliospheric imaging, such as done by the STEREO
spacecraft in some phases of their orbits, should become structural
ingredients of the space-weather science and forecasting
infrastructure. We note that the implementation of a mission for such
a perspective would also help address the 180-degree ambiguity in the
direction of the vector-magnetic component transverse to the line of
sight that currently is an intrinsic problem of even the most advanced
solar spectro-polarimetric instrumentation.

\subsection{Challenges and opportunities for geomagnetic disturbances}

\subsubsection{L1 observations: Validation of $>$1\,hr forecasts and interaction with the magnetosphere}
On a par with establishing the structure of the erupting solar field as it begins to propagate into the heliosphere is establishing how incoming solar-wind field structures interact with the magnetospheric field to drive geomagnetic variability either immediately or with delay after storing energy in the magnetotail. L1 in-situ particle and field measurements are critical in providing input to models for near-term forecasts, and for validation of longer-term forecasts of storms advancing through the heliosphere. Therefore, the continuity of L1 measurements should be ensured so that besides the needs of operational services also the needs from the research community are taken into account.

\subsubsection{Reconfigurations in the magneto\-sphere-iono\-sphere system and strong GICs}
The coupled magnetosphere-ionosphere (MI) system responds to the solar wind through processes of energy input at the magnetopause, followed by transport and storage in the magnetosphere, and finally release of such stored magnetic energy from the magnetospheric tail either to the inner magnetosphere, to the ionosphere/thermosphere, or ejection into the solar wind outward through the distant magnetotail. Recent discoveries from the THEMIS and Cluster missions reveal the tantalizing and complex physics of the release of such stored energy via the development and penetration of earthward propagating azimuthally narrow channels of high speed plasma (bursty bulk flows, BBF) that transport magnetic flux and embedded plasma earthward. On the other hand, nearer-Earth dynamics that couple these flows to the inner edge of the plasma sheet and to the flow-braking region to the ionosphere produce large field-aligned currents and hence GICs that are still inadequately understood. In particular, the conditions leading to, and the exact physical processes responsible for, large field aligned currents (FACs) to reach the ionosphere and drive large GICs are not known.

At times, most of the magnetic energy in the magnetosphere is stored for one to several hours, to be released in one drastic and intense substorm, leading to harmful GIC effects. At other times, a large portion of the energy is transported earthward and immediately, but gradually, released through a more steady process causing quite moderate GIC effects. For major solar wind drivers there will always be a resulting magnetic storm, but even such a magnetic storm is made up of several interacting and overlapping release mechanisms including extreme convection, recurrent substorms, and recurrent plasma injections to near-Earth region in a variety of combinations, thus making the prediction of exactly when GIC effects will occur, and exactly how intense they will be, very difficult. 

It has been recognized that the exact path of energy dissipation in the MI system is dependent on details in the temporal development and other characteristics of the solar wind driver disturbance, the pre-history of solar wind conditions before the arrival of a major event, and the pre-conditioning of the ionosphere, thermosphere, plasmasphere, radiation belts and, indeed, the magnetotail itself. For example, the composition of low-ionization heavy (ionospheric) ions versus high-ionization low-mass (solar-wind) ions in the magnetotail depends entirely on the pre-event history of magnetospheric activity, but plays an important role in the release of stored solar-wind energy in the form of substorms, through various instabilities.  Likewise, the readiness of the ionosphere (and the magnetic field lines connecting the ionosphere to the magnetosphere) to redirect magnetospheric current through ionospheric channels is pre-conditioned by auroral precipitation that modifies conductance during the substorm growth phase, when the energy stored in the magnetotail is at first only partially released. 

The challenge for understanding large GICs hence requires a deeper understanding of the dynamic evolution of the system level state of the coupled near-Earth magneto\-sphere-iono\-sphere-thermo\-sphere (MIT) system. Especially important is the ability to understand - and to distinguish between - more gradual dissipation of energy and those events associated with explosive release of stored energy. However, the physical processes that distinguish between extremely rapid and more gradual energy release are very poorly understood. A number of important processes for energy transport and dissipation are known, but their inter-relationships and the pre-existing state of the system which is the most likely to lead to large GICs is also not known.

Almost certainly the main driver of near-Earth tail dynamics is the
conversion and release of stored magnetic energy in that region. The
NASA magnetospheric multi-scale (MMS) is targeted to address details
of the kinetic physics leading to fast magnetic reconnection and the
release of earthward flows mainly in the distant tail at 25\,R$_{\rm
  E}$ (R$_{\rm E}$ is the symbol used for a distance equal to one
Earth radius of 6371\,km). However, the nature of the dynamical
processes operating at the critical transition region between
dipole-like (closed) and tail-like (stretched) magnetic fields
(between about 6\,$R_{\rm E}$ and 12\,$R_{\rm E}$) remains to be
properly explored, because this region has been markedly undersampled
by previous missions. We have seen glimpses of some of the processes
that couple the tail to the ionosphere; these indicate the importance
of active MIT coupling in determining the overall ability of the
system to drive large GICs: some times localized intensifications in auroral arcs are seen prior to substorm onset, which indicate changes in FACs close to the location of where the substorm onset occurs later. The exact character of pressure gradients at the earthward edge of the plasma sheet certainly also plays an important role in this coupling. Perhaps most significant is the fact that it is currently impossible to distinguish between events which only result in small-scale expansion of bright and dynamic aurorae, often called pseudo-breakups, and those which lead to large and rapid reconfigurations of the tail and full substorm development with bright aurorae covering the whole midnight sector of the auroral oval. Likely plasma sheet kinetic plasma instabilities play an important role, but their ability to communicate with the ionosphere, in particular the overall nature of Alfv{\'e}n wave exchange with the ionosphere, and the balance between Ohmic dissipation and the acceleration of field-aligned electrons, and their consequences for destabilizing the tail and driving large FACs and hence GICs are not known.

From the GIC perspective, whether local instabilities at the inner edge of the plasma sheet precede and communicate with the near-Earth neutral line, or whether global tail reconfigurations begin with reconnection in the mid-tail is hardly relevant. The critical missing understanding lies in the near-Earth MI coupling processes that allow large currents to be driven in the ionosphere, with the consequent production of large GICs. Thus the ionospheric ability to close major FACs in the substorm current wedge (SCW, a current system where the magnetospheric dawn-to-dusk currents from magnetotail are diverted to the midnight sector of the auroral ionosphere) is crucial for the understanding significant GIC effects.

In order to predict the exact route of energy dissipation, or even the timing of the onset of violent energy release after a certain storage period, it is absolutely essential to understand and consequently improve modeling on the principles of pre-conditioning in the coupled MIT system (c.f. ``challenge'' in the blue box of the left column in Figure 3). To this end, satellite- and ground-based data are required from all parts of the coupled system (flagged as ``opportunity'' in Figure 3), both in (and below) the ionosphere, the inner magnetosphere and the magnetotail (including the plasma sheet where the instabilities leading to energy release take place), and the lobe region in which the plasma sheet is embedded, where the energy storage and release can be monitored. Overall, this leads to requirements for multi-point characterization of the plasma in the transition region between dipole and tail-like fields, together with multi-point monitoring of the medium-altitude region (at several thousand kilometers) of auroral acceleration and wave reflection on conjugate field lines, with extensive support from under-lying multi-instrument ground arrays of magnetometers, optical instruments, and HF radars, etc. Monitors of the solar wind and of incoming flows in the tail, perhaps utilizing existing assets are also required.

\subsection{Research for improved forecasts of ionospheric storm evolution}
User communities of GNSS and HF communication systems have expressed needs both for nowcasts/near-term predictions and for longer-term predictions with multi-hour to multi-day lead times (cf., background bars in Fig. 2). The research and development work to fulfill these needs should address both rapidly evolving steep gradients in the ionospheric electron density (global and regional scales) and processes causing ionospheric scintillation (local and micro scales). Forecasts of ionospheric perturbations and of upper-atmospheric variability driven by solar flaring and CMEs rely on the advances described above in the two previous sections. Ionospheric perturbations driven by either solar flares or CMEs and associated geomagnetic storm or substorm phenomena require flare forecasts of strength and onset time of flares and of CME arrival times and of CME field properties. 
Statistical ionosphere models tuned with data from ground-based networks or low Earth orbiting (LEO) satellites can in many cases provide relatively good results for nowcasts or short-term forecasts in regional or global scales (labeled as ``opportunity'' in the blue box of middle column in Figure 3). However, with increasing lead times more comprehensive physics-based modeling with data assimilation is required. We need to understand what factors control electron density variations during ionospheric storms, including neutral atmosphere composition temperature and dynamics, ion-neutral coupling, solar EUV, particle precipitation, plasma transport, and ion chemistry (``challenge'' in Figure 3).  Operational use of data-driven models requires the availability of various actual observation data sets covering the solar energy input including the associated energy spectrum (solar spectral irradiance), the knowledge of magnetospheric key parameters such as FACs and the convection electric field, thermospheric conditions such as composition and winds, and current state of the ionosphere such as the electron density distribution.  

The largest challenges in ionospheric modeling are most probably associated with the physics causing scintillation in radio signals. Theoretical studies addressing the growth and decay of plasma instabilities should be continued to gain better understanding on processes causing scintillation at high latitudes and at equatorial latitudes. As these areas provide different background conditions for the instabilities, for example due to different internal magnetic field orientation and solar forcing, there is no single general approach to address them both. 

Ground-based observations of the Earth's upper atmosphere have been expanded extensively in the last decade, especially with new coherent and incoherent radars, new ionosondes, new all-sky photometers, and new digital imagers from several different generations, mainly set close and around the magnetic equator and at high latitudes to observe and study plasma bubbles and auroral instabilities and their driving mechanisms. Despite such increases in the observations of these space weather phenomena, there are still scientific issues that are not solved, including, for example, the day-to-day forecast of plasma bubbles as well as teh day-to-day neutral wind that drives their development.

\subsection{Steps for improved radiation belt forecasts and specification}
Near-real time solar-wind conditions are the controlling factors for the near-Earth energetic particle population, but the preceding evolution of that system is also critical; the result is an energetic particle environment in which the outer radiation belts are dominated by locally accelerated particles. For example, efficient energetic particle production mechanisms appear to need a seed population of energetic electrons. Relativistic electrons enhancements are an important space weather factor with a strong influence on satellite electronics. Around half of all magnetic storms are followed by a corresponding enhancement of relativistic electron fluxes. However, GCRs- and SEPs can also penetrate into the terrestrial magnetosphere. SEPs can originate either by direct acceleration in the corona, which can result in very prompt fluxes in the case of direct interplanetary linkage, or by acceleration at interplanetary shocks, from where SEPs can propagate towards Earth through direct magnetic-field linkage or by being carried within the CME so that their SEP populations may eventually pass by, and then temporarily envelop, Earth.

SEPs and GCRs originate from within the solar system and beyond, in the Galaxy, respectively, and propagate to the Earth's orbit. The geomagnetic field controls the penetration of SEP and GCR particles into the near Earth environment, such that lower energy particles penetrate only into polar regions (for example, energies below about 1GeV penetrate only poleward of about 55 degrees magnetic latitude), but progressively lower latitudes are penetrated as energies increase (an energy approaching 20 GeV is required to penetrate to the equator; Beer, 2010). Thus the high-energy GCR particles penetrate much more deeply than do most solar particles; only rarely do SEPs have sufficient energy to reach deep into the equatorial regions. The latitude to which particles penetrate is known as the geomagnetic cutoff and depends on both particle properties and on the geomagnetic field strength. This geomagnetic cutoff filtering is important for satellite and aviation operators, as it controls the particle properties to which their systems are exposed. 

GCRs and very high energy SEPs that penetrate deep into the atmosphere collide with atmospheric species to produce neutrons, muons and other secondary particles. Ground-level and aircraft measurements of these secondary particles give useful insights into the very high-energy part of SEP energy spectra. However, the transport of this particle radiation through the magnetosphere and atmosphere needs to be better understood; measurements, especially on aircraft, are needed to challenge and stimulate GCR and SEP radiation transport models.

Trapped energetic particles in the inner magnetosphere form radiation belts. The inner proton belt is fairly stable, but the outer radiation belt comprises a highly variable population of relativistic electrons - with fluxes often increasing abruptly but generally decaying only slowly. Solar-wind changes produce changes to trapped-particle transport, acceleration, and loss. However, the pre-existing magnetospheric state is also important and therefore many of the challenges and opportunities listed above for the GMD/GIC case are relevant also for improved understanding on SWx effects in the radiation environment. Continuous solar-wind and solar observations are needed to predict magnetospheric disturbances and corresponding auroral precipitations, field-aligned current variations, trapped particle variations and any SEP penetration.

Intensification of magnetospheric convection during storm times, local particle acceleration due to substorm activity, and resonant wave-particle interaction are the main fundamental processes that cause particle energization and loss. However, the details of the mechanisms which may be able to contribute to these processes - including a variety of wave particle interactions, enhanced convection and radial diffusion - remain a subject for active research. While progress has been made in the theoretical understanding of these competing processes, there is as yet no consensus on which of these will be significant in particular situations, and no real predictive methods that can give precise fluxes at different local and universal times. There is therefore a pressing need to confront theoretical models with detailed measurements in order to resolve these shortfalls (see right hand column in Figure 3). Data incorporation is needed to take into account local plasma processes affecting the magnetic field variations and corresponding particle accelerations, losses etc. As much as possible data are needed for nowcasting of energetic particle fluxes. GEO, LEO, MEO, GTO data are very important. Data calibration is needed to provide the correct result. It is essential for satellite operators to know the past, current and future condition of the space environment around their satellites, especially in case of satellite anomaly and/or before critical operation.

Finally, there are short-lived populations of ring-current particles produced by substorms and enhanced magnetospheric convection. Currently, our understanding of substorm onset is evolving rapidly, in particular with respect to local dipolarization structures. However, we are still a long way from an ability to predict precise events. Current spacecraft measurement configurations are probably not sufficient to make accurate predictions. Because these effects are the major contributor to spacecraft failures due to charging, it is essential that current investigations are maintained and extended (reflected as highest-priority opportunity in Fig. 3). Magnetic field models (empirical, numerical) with particle tracing codes are needed to nowcast/forecast particle distribution. They require current or predicted solar-wind information for nowcasting/forecasting. Data sources are solar wind monitoring data and solar UV images, models of solar wind propagation from the Sun. 

\subsection{Challenges for forecasts with lead times beyond 2 days}
Forecasts of substantial space weather events that extend beyond a few days require, by definition, forecasts of energetic events well before they occur on the Sun. This requires advancing our understanding of the storage and instability mechanisms for the solar magnetic field. Many of the issues involved in that will be addressed by the science described in Section 4.1.1, including knowledge of the 3D structure of the field and estimations of energy and helicity budgets. With the advancement of that understanding will also come the ability to experiment with the field configurations to deepen insights into the conditions under which such fields become unstable for flares and CMEs (see Appendix D for more discussion and a listing of requirements). Support of the research described in Section 4.1 will thus also support pre-event forecasts for that reach some hours to a day before the event occurs on the sun.

For forecasts of CME arrivals well beyond the few days that it takes to travel through the heliosphere to reach Earth, this means that forecasts need to be made even before the source regions of potentially geo-effective CMEs have rotated from near the east limb towards the Earth-perspective central meridian. In contrast, prompt SEP storms around Earth often originate from solar regions that have rotated several days past central-meridian passage towards the west limb. This adds some complication in that the magnetic maps of such regions suffer from perspective changes, but that can be dealt with as long as the regions are not too close to the limb (i.e., within about 4-5 days of central meridian passage). What is a much larger challenge here is that in many cases such longer-term forecasts for active regions need to be made even before much of the involved field has emerged onto the solar surface.  
It appears that field configurations leading to the most intense space weather originating in active-region flares and eruptions form and decay within a day or two of the emergence of twisted bundles of magnetic flux or the shearing by convective flows of such bundles. Once emerged and interacting with the pre-existing field, the available non-potential energy can be processed either through a rapid conversion in a flare/eruption or can be gradually dissipated, likely contributing somehow to the thermal energy of the solar atmosphere (therein finding an intriguing analogy with the options for magnetospheric relaxation discussed above).

The successful quantification of the subsurface magnetic field that will be involved in possible solar eruption upon emergence by helioseismic techniques currently lies beyond our capabilities, and even beyond a proof of principle in advanced helioseismic magneto-convective simulations. It thus appears premature to invest in instrument opportunities solely for helioseismic investigations of the eastern/leading solar limb for the purpose of exploring the potential of, say, five-day pre-event forecasts. We do recommend, however, continued support for the development and testing of helioseismic methods that may reveal how we can detect and characterize magnetic flux bundles about to emerge onto the solar surface a day or more ahead of time.
The background solar-wind sector structure involved in recurring geomagnetic perturbations is driven mostly by the rotation of the largest-scale field structures on the Sun. The forecasts of the sector structure and the non-CME-related patters of fast and slow solar wind streams will benefit from the advanced knowledge of the global solar field and the associated heliospheric modeling described in Section 4.1.2. The more extensive the observational magnetograph coverage is of the solar active-region belt, the better multi-day forecasts will become for the overall solar wind stream and sector structures. As the requirements for this overlap with those articulated in Section 4.1.2, we do not repeat these here.

\subsection{Specification of extreme conditions and forecasts of the solar cycle}
Quantitative knowledge of low-frequency, high-impact space weather cannot be derived by observing the Sun-Earth system in modern times, simply because too few of the severe events have occurred, and the most extreme may not yet have been observed at all as yet.  Rather, we must gather data that tell us about solar activity over many centuries in order to reach useful, reliable conclusions about events that happen once per century or once per millennium. For energetic particle storms and the GCR background, such information can be obtained extending over tens of millennia by combining terrestrial and lunar radionuclide studies based on rocks, ice cores, and biosphere-modulated radio-nuclide records (including, for example, carbon-14). 
To quantify the frequency spectrum for solar flares we can look at the multitude of Sun-like stars accessible by nighttime ground-based or space-based astronomical telescopes, as demonstrated by, for example, NASA's Kepler mission and by X-ray and EUV space astrophysics missions (Schrijver and Beer, 2014). Current evidence suggests that even for our aging Sun events that are tens of times more intense than the most energetic solar ones observed during the space era may occur on time scales of once per century to millennium. We recommend that the solar-stellar links be strengthened to better quantify the probability distribution of the most intense but infrequent solar flares by using archival data on Sun-like stars already available and by supporting possible future opportunities to add to that statistical information. 

As to energetic particle storms, it is prudent to make the moderate investment needed to harvest the radio-nuclide information stored in ice, biosphere, and the rocks brought back from the Moon during the Apollo era. Parts of this overall project involve improving our knowledge of the distribution of energies over particle spectra and photon spectra, as well as understanding how terrestrial particle spectra can be combined with photon spectra to understand how solar/stellar flares and particle storms are related for the most intense and dangerous events. 

An entirely different class of problems is to be addressed for multi-year to multi-decade variability that is driven by the general patterns of the solar cycle. Information on past solar cycles is, of course, coming from studies of sunspot patterns, of radionuclide studies associated with GCR modulations by the large-scale structure of the solar wind, and even from studies of potential climate impacts in the pre-industrial era on time scales from decades to millennia. The inter-disciplinary work that is required for that should be stimulated as an aid to environmental specification for designs of long-lasting infrastructures, both in space and on the ground. Multi-year to multi-decade forecasts, on the other hand, need to see investments in understanding the solar dynamo. The observational requirements for this involve continued observations of the solar magnetic field and of the flows involved in transporting it across the solar surface and throughout its interior. Here, continued magnetic observations as required for the science described in Section 4.1 supports the need for surface coverage. Continued helioseismic observations to study the variability of deep flow patterns involved in the dynamo are needed in addition. Moreover, strengthening ties with the astrophysical community looking into dynamo activity in Sun-like stars is needed.

Involvement in the study of habitability of exoplanets is a natural stimulus for both the extreme event knowledge and improved dynamo understanding because exoplanet space weather is a dynamic emerging field that can benefit from Sun-Earth connections knowledge as much as its discoveries can inform us about extremes and long-term trends in space weather in our own planetary system. We encourage stimulus of these research fields, but as these lie at the fringe of our central charge, we do not at present make any explicit recommendations for investments in research and its observational infrastructure, other than emphasizing that funding agencies would be well advised to use the synergy between solar and stellar research and between planetary-system and exoplanet habitability to improve our abilities to specify and predict extreme space-weather conditions and long-term trends.

\section{General recommendations}
With so much yet to understand, and so much space (literally) to cover observationally and in computer models, the needs readily outweigh the means we can expect to have at our disposal. A major effort of the roadmap team therefore focused on identifying where investments would make the largest impact in advancing our scientific insights to best meet the needs of the space weather users. We base that analysis on these identified primary needs: 
\begin{itemize}
\item	For the research community: Comprehensive knowledge of conditions throughout the Sun-Earth system in the past based on models guided by observations, and the understanding of the physical mechanisms involved in determining these conditions.
\item	For the space-weather customer community: Knowledge of historical and recent conditions and knowledge of current conditions and forecasts for coming hours to days, and specification of extreme conditions in geospace and throughout interplanetary space. Different user communities require different types of data and distinct levels of accuracy and specificity.
\end{itemize}

From these follow primary requirements: 
\begin{itemize}
\item	Comprehensive, affordable, sustained observational coverage of the space weather system from Sun to society; 
\item	Data archiving, sharing, access, and standardization between the various researcher and customer communities;
\item	Advanced data-driven and experimental models as tools for analysis, interpretation, and visualization of events and their spatio-temporal context, and for forecasting/specification;
\item	Interaction and coordination between national, regional, and international researcher, user, and agency communities for research guidance, prioritization, impact assessment, and funding;
\item	Education of researchers, customers, and general public.
\end{itemize}

These requirements need to be addressed in the context of a changing paradigm in the science of space weather: whereas many studies to date tend to focus on what they refer to as ``space weather events'', there is a rapidly growing realization that in fact significantly longer time sequences need to be studied because there appears to be an important if not crucial impact of what is either described as hysteresis or as pre-conditioning in the system. This is now recognized to reach from the solar environment to deep within geospace: solar active regions emerge preferentially in locations where other such emergence has occurred before; nascent CMEs often plow through high coronal field that is still relaxing from preceding eruptions; interplanetary solar-wind structures are often mergers of multiple sequential CMEs; geo-effectiveness of CMEs is dependent on pre-conditioning of the magnetosphere by activity over the preceding days; substorms appear to release stresses in the magnetotail built-up over some period of time; preceding conditions in the ionosphere control the timing and intensity of substorm onsets; and so forth. Consequently, there must be a shift away from the study of short time intervals that attempt to study space weather as if pre-conditioning were unimportant, moving towards the analysis of multi-day windows of space weather that makes allowance for significant effects of hysteresis in the Sun-Earth system and its components.

The impact tracings discussed in the preceding section, and detailed in Appendices B, D and E, lead us to formulate the following general priorities for actions that guide us to the detailed implementation suggestions for developments of new observational, computational, and theoretical capabilities that are discussed in the subsequent section. We begin with a concise listing of the priorities for three distinct target audiences, followed in the next three subsections (5.1, 5.2, and 5.3) by a point-by-point expansion into the types of actions to be considered or implemented.

\paragraph{Research: observational, computational, and theoretical needs:}
\begin{enumerate}
\item	{\bf Advance the international Sun-Earth system observatory along with models to improve forecasts based on understanding of real-world events through the development of innovative approaches to data incorporation, including data-driving, data assimilation, and ensemble modeling.} The focus needs to be on developing models for the Sun-Earth system, at first as research tools focusing on actual conditions and later to transition to forecast tools, making use of the existing system before components are lost as instrumentation fails or is discontinued. 
\item	{\bf Understand space weather origins at the Sun and their propagation in the heliosphere, initially prioritizing post-event solar eruption modeling to develop multi-day forecasts of geomagnetic disturbance times and strengths:} Advance the ability to forecast solar inputs into geospace at least 12\,hrs ahead based on observations and models of their solar drivers as input into heliospheric propagation models.
\item	{\bf Understand the factors that control the generation of geomagnetically-induced currents (GICs) and of harsh radiation in geospace, involving the coupling of the solar wind disturbances to internal magnetospheric processes and the ionosphere below:} Advance the ability to forecast the response of the geospace system to driving by solar-wind variability. 
\item	{\bf Develop comprehensive space environment specification:} Create a reference specification of conditions and their likelihoods for the local cosmos.
\end{enumerate}

\paragraph{Teaming of research and users: coordinated collaborative environment:}
\begin{enumerate}
\item[I]	{\bf Quantify the vulnerability of technological infrastructure to space weather phenomena} jointly with stakeholder groups.
\item[II]	{\bf Build test beds in which coordinated observing supports model development:} Stimulate the development of (a) state-of-the-art environments for numerical experimentation and (b) focus areas of comprehensive observational coverage as tools to advance understanding of the Sun-Earth system, to validate forecast tools, and to guide requirements for operational forecasting. 
\item[III]	{\bf Standardize (meta-)data and product metrics and harmonize access to data and model archives:} Define standards for observational and model data products, for data dissemination, for archive access, for inter-calibration, and for metrics. Define data sets needed to test physical models and forecast systems.
\item[IV]	{\bf Optimize observational coverage:} Increase coverage of the Sun-Earth system by combining observations with data-driven models, by optimizing use of existing ground-based and space-based resources, by developing affordable new instrumentation and exploring alternative techniques, and through partnerships between scientific and industry sectors. 
\end{enumerate}

\paragraph{Collaboration between agencies and communities:}
\begin{enumerate}
\item[A]	{\bf Implement an open space-weather data and information policy:} Promote data sharing through (1) open data policies, (2) trusted-broker environments for access to space-weather impact data, and (3) partnerships with the private sector.
\item[B]	{\bf Identify, develop, and provide access to quality education and information materials for all stakeholder groups:} Collect and develop educational materials on space weather and its societal impacts, and create and support resource hubs for access to these materials, and similarly for space-weather related data and data products. Stimulate collaboration among universities in order to promote homogenized space-weather education as part of the curricula. 
\item[C]	{\bf Execute an international, inter-agency assessment of the state of the field to evolve priorities subject to scientific, technological, and user-base developments to guide international coordination:} Identify an organizational structure (possibly the COSPAR/ILWS combination leading to this roadmap) to perform comprehensive assessments of the state of the science of space weather on a 5-year basis to ensure sustained development and availability of high-priority data, models, and research infrastructure. This activity can be a foundation to align the plans of research agencies around the world if, as for example for this roadmap, agency representatives are engaged in the discussions.
\item[D]	{\bf Develop settings to transition research tools to operations.} Establish collaborative activities to host, evaluate, and compare numerical models (looking at the Community Coordinated Modeling Center [CCMC] and the Joint Center for Satellite Data Assimilation [JCSDA] as examples, setup, staffed appropriately from research and user communities) and to assess quantitatively their skill at forecasting/specifying parameters of high operational value. Determine the suitability of research models for use in a space weather service center. Foster continuous improvement in operational capabilities by identifying the performance gaps in research and operational models and by encouraging development in high priority areas.
\item[E]	{\bf Partner with the weather and solid-Earth communities to share infrastructure and lessons learned.} To improve understanding of the couplings between weather and space-weather variability and to quantify potential climate impacts by effects related to space weather. Another type of partnership here involves the transfer of knowledge and ``lessons learned'' from the climate/weather communities on techniques for data-driven assimilative ensemble modeling and on the development of forecasts and their standards based on that. For the solid-Earth community: translation of geomagnetic variability into electric fields involved in GICs.
\end{enumerate}

The above ``General Recommendations'' (listed in priority order within each target group) as formulated by this roadmap team align with many found in other studies and workshops (e.g., Committee on Progress and Priorities of U.S. Weather Research and Research-to-Operations Activities, 2010; American Meteorological Society, 2011; EC Joint Research Center by Krausmann, 2011; Committee on a Decadal Strategy for Solar and Space Physics, 2012; United Nations Committee on the Peaceful Uses of Outer Space, 2013). This roadmap advances from the general and system-wide recommendations in those earlier reports by including a rationale for the prioritization of investments for the worldwide community subject to limited resources.  

\subsection{Research: observational, computational, and theoretical needs}
\vskip 2mm
{\em 1)	Advance the international Sun-Earth system observatory along with models to improve forecasts based on understanding of real-world events through the development of innovative approaches to data incorporation, including data-driving, data assimilation, and ensemble modeling.}

The Sun-Earth system is currently observed by an unprecedented array of ground- and space-based instruments that together provide a comprehensive view, but one that is so sparse that the loss (through failure or termination) of any one key asset would have a substantial impact on the overall study of space weather phenomena. Note that our recommendation on access to the data from this system observatory is listed above as item A, and expanded upon in Section 5.3, item A.

{\bf Recommendation:} Funding agencies should place the continuation of existing observational capabilities within the context of the overall Sun-Earth System observatory as a very high priority when reviewing instrumentation for continued funding: the value of any one observable within the entire set of Sun-Earth observables should significantly outweigh whether the instrument has met its own scientific goals or whether it, by itself only, merits continuation based on its standalone scientific potential. The prioritization of international assets should be taken from a community-based assessment, such as this roadmap and its recommended successors. Of particular importance to be continued are: Earth-perspective magnetic maps and X/EUV images, solar spectral irradiance monitors (including X-rays to UV and radio), coronagraphy (ideally from multiple perspectives), solar radio imaging, in-situ (near-)L1 measurements of the solar wind plasma (including composition) and its embedded field and energetic particles, ground-based sensors for geomagnetic field and ionospheric electron density variability and neutral-atmosphere dynamics, LEO to GEO electron and ion populations, in-situ magnetospheric magnetic and electric fields, and geomagnetic field measurements, space-based auroral imagers, and energetic-particle sensors (including ground-based neutron monitors) where available throughout the inner heliosphere.

Most space weather models, whether for research or for forecasts, currently rely on snapshots or trends in observables, or on extrapolations from one location to model conditions in another, with little or no direct guidance by observations of evolving conditions at the boundaries and in the interior of the modeled volume. This is in part because data incorporation strategies are only in their infancy, and in part because data types and model algorithms are not optimized for assimilation processes.

{\bf Recommendation:} Funding agencies should immediately begin to give preference to modeling projects that use, or are specifically designed to eventually enable, direct or indirect guidance by observational data. Funding agencies should require efficient online access to relevant data sets by standardization of interfaces and by the creation of interface hubs and tools that transform data sets into standard formats that are compatible with common practices in different disciplines. Each agency can focus on its own data, but agencies should coordinate in the development of data standards and standardized interface, emphasizing the importance of near-real-time data availability. The ILWS program partnership can be a coordinating hub in this process.
 
\vskip 2mm
{\em 2)	Understand space weather origins at the Sun and their propagation in the heliosphere, initially prioritizing post-event solar eruption modeling to develop multi-day forecasts of geomagnetic disturbance times and strengths, after propagation through the heliosphere.}

This priority is associated with two scientific focus areas. The first is the need to enable a much-improved quantification of the non-potential magnetic field of source regions of solar activity pre- and post-eruption, for which we propose local-area binocular imaging as well as high-resolution comprehensive imaging that includes vector-mag\-neto\-graphy of the solar surface and chromosphere on the scale of the erupting region. The second involves advanced modeling capabilities for the coupling from the localized source-region corona into the overall heliosphere, and the computation of the propagation of plasma and embedded magnetic field through the heliosphere towards geospace, for which we propose increased coverage of the solar surface magnetic field, and advanced modeling capabilities for the global coronal and inner-heliospheric dynamic magnetic field and solar-wind plasma.  The specific recommendations associated with this research focus area are presented in detail in Sections 6, 7a and 7c.

\vskip 2mm
{\em 3)	Understand the factors that control the generation of geomagnetically-induced currents (GICs) and of harsh radiation in geospace, involving the coupling of the solar wind disturbances to internal magnetospheric processes and the ionosphere below.}

This priority is associated with several recommendations. The first involves satellite coverage in the domain between the dipole-tail transition region at the inner edge of the plasma sheet and in the near-Earth domain of the magnetospheric field (at the auroral acceleration region at several 1000\,km altitude) to probe, along essentially the same magnetic field lines, the processes that throttle beginning magnetotail activity and that determine if that activity may develop into a magnetospheric (sub-)storm and associated intense GICs or not. The second recommendation for this priority emphasizes the need for coordinated ground-based and space-based networks of instrumentation, and the development of ``testbed'' locations where dense observational coverage provides information on key distinct environments for geomagnetic and ionospheric electron density variability (specifically for the auroral zone and the near-equatorial region). The two complementary sets of recommendations are described in detail below in Sections 6, 7b, and 7d-f.

In addition to these recommendations regarding magnetospheric and ionospheric processes driven from outside and within the M-I system, there are the processes that subject the ionosphere to forcing by the neutral terrestrial atmosphere for which we articulate this specific additional recommendation: 
Recommendation: Funding agencies should place the continuation of existing observational capabilities within the context of the upper atmosphere as well as the computational capability that leads to its understanding for the scientific and operational purposes at a high priority. It also recommended that the coverage of the ground-based instrumentation be coordinated in the development of new instrumentation, including integration of ground- and/or space-based data systems. Also, it is recommended to include that lower-atmospheric information into data-driven modeling capabilities for the near-real-time ionospheric state. Additionally, the scientific community should pursue a general consensus on the underlying physics with the processes associated with radio scintillation, in particular about the coupling between neutral atmospheric winds and waves, and development of the plasma turbulence that leads to the bubble formation at equatorial latitudes

\vskip 2mm
{\em 4)	Develop a comprehensive space environment specification:}

With robotic and manned spacecraft becoming ever more frequent and venturing into new orbits around Earth and the solar system, it is important that we establish in detail the conditions to be encountered by these spacecraft prior to their design so that they can be built to survive space weather storms and be resilient to recover from such storms. Similarly, design criteria for technological infrastructures used on Earth need to be set so that they can resist, survive, and recover from effects of space storms. This requires information on space weather conditions that may be encountered during the lifetime of such spacecraft and ground-based systems.

{\bf Recommendation:} We recommend for the near term that detailed specifications be compiled of all relevant space-weather phenomena (following the recommendation for an open-data policy formulated below) that may be encountered wherever human technologies are, or likely will be, deployed; this includes properties of particle storms, solar irradiance variability, and geomagnetic variability, from low-level generally present backgrounds to rare, extreme, impulsive events. We recommend for the near- and mid-term that research funding be allocated to this effort that includes studies of radionuclides in the terrestrial biosphere and cryosphere, as well as in lunar and terrestrial lithospheres; that analogies between Sun and Sun-like stars be explored; and that advanced modeling be deployed to understand extremes in geomagnetic, heliospheric, and solar conditions. 

\subsection{Teaming of research and users: coordinated collaborative environment}

\vskip 2mm
{\em I.	Quantify the vulnerability of technological infrastructure to space weather phenomena.}

As our ability to understand general properties of space storms advances, more refined forecasts will come within reach: ranges of forecast magnitudes may shrink, start times and durations of events may be better forecast, and characterization of variability within a storm will be increasingly feasible. But the value of such forecasts is defined by the impacts that these and other attributes of storms have on our evolving technologies. Therefore, studies of the impacts compared to observed properties should be performed. Such impact studies will also guide our understanding of impacts of high-impact, low-frequency extreme events by narrowing the range over which we need to extrapolate to model their impacts before they occur.

Knowledge and understanding of space weather impacts remains uncertain in all impact areas of space weather. For example, GICs cause potentially the largest space weather impacts on terrestrial infrastructure, but the disturbance and failure modes of power grids subjected to GICs remain poorly known. For example, work in South Africa, stimulated by the GIC problems experienced there in October 2003, shows the need for better management of transformer life cycles against all forms of stress and aging: strong GIC events can significantly age transformers, thus advancing the date on which they will fail  (e.g., Gaunt, 2013), unless replaced in good time.
Such quantification of the impacts of space weather events can be guided by the equivalent of epidemiological studies of grid performance subject to geomagnetic variability (e.g., Schrijver and Mitchell, 2013; Schrijver et al., 2014), combined with systematic engineering and impact studies. Epidemiological studies should be executed in partnerships between space scientists and power-grid experts, while detailed impact studies should be led by the power sector with involvement of the science community (ideally with international coordination). Such statistical and case studies are needed to understand the time sequence and structure of the magnetospheric field dynamics that are the most geo-effective in driving GICs. 

Another example is that there is a big gap between the knowledge of the space environment and the statistics of satellite anomalies. One aspect of this is that the occurrence of a satellite anomaly is strongly dependent on the material specification of a particular satellite. A complicating factor in the study of satellite anomalies due to space environment is that the correlation is often based on indirect evidence. Unfortunately, detailed information on satellite anomalies is generally not released by companies that design or operate satellites. 

A third example of insufficient information is related to the impacts of radiation on crews and passengers of airliners and on the aircraft avionics. The UK Royal Academy of Engineering study on space weather (Royal Academy of Engineering, 2013) stressed the importance of ground-testing of avionics against exposure to neutrons of energies typically found at 10-12\,km altitude; this report noted that appropriate test standards are being developed as are test facilities capable of testing whole avionic systems, e.g., the ChipIR facility in the UK. Other work in this area includes an aircrew dosimetry system operated by the French Institute for Radiological Protection and Nuclear Safety
footnote{https://www.sievert-system.org/?locale=en}.

Studies of this type do exist but are often restricted in their feasibility or in their distribution and availability to the general space-weather stakeholder community by concerns expressed as competition sensitive, proprietary information, or even national security. Removing or reducing these barriers is an important step to be taken towards understanding the true impacts of space weather phenomena. 

{\bf Recommendation:} We recommend for the near term that the space weather community engage with research agencies responsible for the study of economic and societal developments and encourage them to explore and support research to quantify the impact of space weather phenomena on societal technologies. Where these agencies are separate from space organizations, the community should encourage them to work with space organizations. In all cases, the support for research on impacts should enable and stimulate the trans-disciplinary research that is required, and should also encourage a data-sharing environment in which industries are given confidence that they can share information on space-weather impacts without concerns for their commercial competitiveness or their potential liabilities. To achieve this, we recommend that all research proposals on impact studies should be strongly encouraged to combine the research and user communities and their joint expertise and data. We also recommend that a trusted-broker environment be given shape by a group that involves stakeholders from government, academia, as well as industry. The community should also encourage international coordination of impact studies.

\vskip 2mm
{\em II. 	Build test beds in which coordinated observing supports model development.}

Some domains within the Sun-Earth system are covered better by observations than others. For example, there are sites for ionospheric research that are favorably located and well equipped with particularly valuable or state-of-the-art instrumentation. Coordination between the various space- and ground-based solar observatories in observing campaigns of particularly active space-weather source regions on the solar surface offer another opportunity for a ``laboratory'' with extensive empirical coverage of regions of interest. Test-bed data could be used to learn which types of observations are most suitable for data incorporation in modeling and which resolutions in space and time would be suitable compromises when considering both instrument maintenance costs and additional gain from data ingestion. 

{\bf Recommendation:} We recommend for the mid-term that international resources be optimally pooled, particularly in settings where significant investments have already been made: ground-based observatories of the ITM system in Jicamarca and around EISCAT appear well suited for focused further development; international coordination between solar observatories (in space and on the ground) should be mandated by funding agencies for a fraction of the total observing time available; data assimilation efforts with the continuously developing models should be supported by access to large-scale computer facilities (e.g., the IBM ExaScience Lab in Belgium) with national, EU, NSF or space agency funding. Results from test runs should be published in platforms that enable community based code development and testing (e.g., NASA CCMC and ESA virtual space weather modeling center) in order to harvest theoretical and computational skills from the entire SWx community. 

To make a step change in the international co-ordination, and delivery, of space weather observing systems we must increase the engagement of scientists working with this instrumentation with modelers, operators of space-based sensors, and other consumers of space weather data. That should be aimed to ensure the optimum interaction between the collected data and state-of-the-art international models, and the synthesis of these data and model results into, first, research tools and, later, into operational tools whose outputs can be made available to the applications and technology community. In order to improve the situation, the geospace observing system needs a more formal international structure to deliver its full scientific potential. 

{\bf Recommendation:} We recommend establishing a global program at inter-agency level for coordination of space weather observing systems, perhaps similar to coordination of space exploration activities.

\vskip 2mm
{\em III. Standardize (meta-)data and product metrics and harmonize access to data and model archives:}

Making efficient use of data, and even finding data that are in principle useful in advancing the science of space weather require adequate standardization in data formats and access protocols to on-line data archives. Knowledge of quality of data products and of space weather applications requires development of data product standards.

{\bf Recommendation:} We recommend that funding agencies require development of data archiving, data search, and data access plans as part of observatory/instrument operations support, retro-actively where needed; this should include standardization of data formats and of data retrieval methods (i.e., web and application interfaces). We recommend that funding agencies coordinate any needed reprocessing of historical datasets to make them generally available in a standard way. We recommend that a study be commissioned, possibly under the auspices of WMO, COSPAR, and/or ILWS, to address issues related to the long-term preservation of data archives. We also recommend that agencies resolve issues that hamper use of science data or products that are not at all available for commercial use (such as the Dst index) in an environment where tailored space weather surveces may increasingly be provided by commercial entities. Finally, we recommend that standard metrics are developed for data product quality and applications quality.

\vskip 2mm
{\em IV. Optimize observational coverage:}

The space from Sun to Earth is vast and poorly covered by observatories. Part of that problem is limited access to existing information - such as ionospheric information that is available to, but not effectively shared by, the GNSS navigational service sector - while another part is related to the cost of instrumentation, particularly in space. The high cost of space missions percolates into several concerns: high costs lead to low frequency of deployment which (a) causes instrumentation to be deployed one or two decades behind the technological frontier, and (b) slows or precludes filling of gaps in the vast space to be covered and in terms of observables that are accessible. The science of the Sun-Earth connections that underlie space weather is particularly affected by this problem because the system to be observed is vast and the number of instruments very small in comparison. But precisely because of that, it is a particularly suitable area for which this issue should be addressed because any new instrumentation is deployed as an augmentation to an existing Sun-Earth system observatory so that even relatively small new ground-based observatories or space missions can significantly strengthen the overall system observatory as it advances that into a scientifically much more capable state.

{\bf Recommendation:} We recommend that research organizations and industry sectors partner to optimize use of already available data related to space weather impacts; central government organizations (such as OSTP in the US) should play a leading role in fostering this sharing of resources. 
Recommendation: We recommend that space agencies urgently seek to lower the costs to achieve their science goals by  (a) emphasizing partnerships and effective use of small opportunities within the context of the Sun-Earth system observatory, (b) by developing infrastructures and rule sets that enable lower-cost access to space in the mid- and long-term, and (c) by implementing sustained diversification of their research fleets with small niche satellites and clusters of smallsats (perhaps even microsats surrounding a larger mother ship with central coordination and communication abilities), all to enable more frequent insertion of new observatories (standalone and as hosted payloads) that can better utilize state-of-the-art technologies, where possible in ``off-the-shelf"\" configurations. In working towards these goals, research agencies should take care to maintain the most important observational capabilities in the currently operating distributed Sun-Earth system observatory, ideally in coordination with, and in consultation with, the research community to most efficiently advance the system observatory's capabilities. 

As part of this recommendation, we note also that efforts should be made to increase the availability of ground-based data on the geomagnetic field and on the state of the ionosphere with high timeliness. This can be accomplished by: (i) considering the deployment of magnetometers and other low cost ionospheric instrumentation (radio wave based probing techniques) in regions with limited coverage; (ii) utilizing the WMO data infrastructure to disseminate data from existing instrumentation; and (iii) working with providers of proprietary data to allow their data to be used in space weather products. We also emphasize the opportunity to expand ground-based coverage by using relatively small facilities that enable emerging countries to become involved in space weather research.

\subsection{Collaboration between agencies and communities}

\vskip 2mm
{\em A.	Implement an open space-weather data and information policy:}

Open access to relevant space weather data will stimulate research by enabling innovative uses of observations of the Sun-Earth system, of modeling data relevant to that system, of impact data for space weather phenomena, and of societal consequences of such phenomena. An open data policy is already agreed upon in the G8 ``Open Data Charter'' (2013) between governments, but we argue for a wider adoption of that charter beyond the G8 nations, and to explicitly define ``open data'' to mean that data should be open to all users as soon as such data is archived for use any research group. Funding agencies supporting establishment and maintenance of SWx infrastructures should favor such initiatives that follow this open data policy. A general open data policy, supported by an infrastructure of virtual observatories, will stimulate research in general, in particular by enabling the utilization of ancillary and adjacent data to observations being analyzed by a research group. 
Recommendation: We recommend that observational data pertinent to space weather be made publicly accessible as soon as such data is calibrated adequately for scientific use. Data sets obtained with routine instrument setups or for monitoring or survey purposes should be mandatorily archived and opened up for immediate internet access without proprietary periods starting as soon as data are archived and provisionally calibrated and characterized by meta-data. We also recommend that data sharing be a formalized part of observatory planning, funding, and operating, including sharing between space-based and ground-based observatories and instrumentation.

{\bf Recommendation:}  Agencies that fund instrument operations should provide sufficient resources to enable efficient open data access - perhaps through a mix of direct funding to instrument operators to generate and store data products and their metadata, and funding for data infrastructures that catalog and provide access to these products.  Virtual observatories and/or general software (such as exists for SolarSoft IDL) should be sustained and developed to ease data identification and retrieval from the archives.
Recommendation: We recommend that a similar open-data policy for assimilation model data be developed and implemented where practical in the mid-term. 

\vskip 2mm
{\em B.	Identify develop, and provide access to quality education and information materials for all stakeholder groups:}

There is huge public interest in space weather and its impact on humankind, in part because it is a spectacular natural phenomenon, and in part because it is a risk that appeals to the human love of scare tales. As a result the internet provides access to a vast and ever-growing multitude of web sites and web pages related to the topic of space weather (see Figure 1). However, this information is of very variable quality and many web pages contain significant factual errors in both the science and the impact of space weather. These errors frequently propagate into other web sites, into the media, and even into the discussion of space weather by policy makers. Therefore, it is critically important to guide all stakeholders of space weather towards high-quality information on space weather; these stakeholders include policy-makers, regulators, educators, engineers, system operators, scientists, and many more. The wider scientific community is also a critical target, especially the closely related scientific areas such as plasma physics, radiation physics, planetary exploration, and astrophysics.
Furthermore, the increasing public interest on space weather is stimulating an increased focus on Sun-Earth connections as part of university level education. In this relatively young and specialized research discipline, the level and focus of education may vary between different universities, which can cause complications in student exchange or in recruitment of early career scientists.

{\bf Recommendation:} We recommend the development of an international structure that can coordinate a global effort to identify and/or develop high-quality information that addresses the needs of all types of stakeholders in space weather, and does so in a way that facilitates customization of information to local (national/regional) circumstances, especially targeting diversification towards culture and language. To deliver this we recommend a structure that engages the leading disseminators of space weather knowledge in each country, including space-weather science experts who have a good appreciation of how the science of space weather links into impacts on practical systems of local importance, and of how their national structures handle both the science and the impacts of space weather. A key goal is to link with the most suitable national structures, recognizing these will vary considerably from country to country. The structure should also encompass dissemination activities in international bodies, including ISES, WMO, UN/COPUOS, ESA, IUGG, IAU, COSPAR, EGU and AOGS.

In the very near-term we recommend that a working group (a) carry out a survey to identify and engage the national experts and structures most active in developing and disseminating high-quality information, and then (b) propose an international structure that can coordinate their efforts on the tasks shown below. This working group should also engage with relevant international bodies and ensure that the international structure can encompass their dissemination work. The working group must also consider how the international structure will develop the scientific authority needed to deliver on the tasks below. 

Once formed, the international structure should
\begin{enumerate}
\item	survey existing good practice in different countries and in international organizations, and encourage the exchange of ideas
\item	establish a procedure for validation of information, including a method for embodying that validation within the information, for example by a certification mark; 
\item	identify, validate and provide links to existing good information that can assist education and awareness-raising amongst all types of stakeholders in space weather;
\item	identify gaps in this information, promote efforts to fill those gaps and provide links to validated outputs;
\item	encourage adaptation of information to best suit the needs of particular groups and countries, including translation into local language and adaptation to local culture; and
\item	compile information on sources of space weather research data and of space weather applications and services, and provide links to these sources.
\end{enumerate}

To support the worldwide dissemination of space weather knowledge the international structure should establish an internet-based system that can point to a range of validated information on space weather covering the different needs of all types of stakeholders. Given the rapid evolution of internet technologies, we do not seek to prescribe how the system be implemented, rather we recommend that the international structure, once established, review the options available, having regard to the need for a resilient and sustainable system, and for ease of use by contributors and users around the world. 

Consideration should be given on whether the international structure should seek to validate space weather services and data access. However, it may be best to avoid this as it may open up significant legal issues, especially in respect of commercial services. 

\vskip 2mm
{\em C.	Execute an international, inter-agency assessment of the state of the field to evolve priorities subject to scientific, technological, and user-base developments to guide international coordination:}

With rapidly changing technologies and advancing understanding the focus priorities of the science of space weather are likely to evolve over time, so that a periodic evaluation of the state of the field will prove useful to provide guidance to the research organizations. Such roadmap studies should identify options for agencies to align domestic and international priorities, to identify projects most suitable for direct collaboration, and to identify where international activities might make domestic investments to address scientific priorities less urgent or superfluous.

{\bf Recommendation:} We recommend that a roadmap for the science of space weather be conducted at a 5-year cadence, ideally timed to provide information to national strategic assessments such as the Decadal Surveys in the US or to such multinational funding programs like EU H2020 program and its successors. These should be international projects, led by representatives from the scientific community, and involving at least representatives of all major space agencies as well as research organizations working on the ground segments of space weather. Moreover, agencies should regularly appoint inter-agency science definition study teams with representatives from agencies, the science community, and the space-weather user community to find optimal balance between scientific potential of new missions by themselves and as part of the Sun-Earth System observatory, where the potential value as element of the system observatory should weigh heavily in its design and evaluation. 

\vskip 2mm
{\em D.	Develop settings to transition research tools to operations.}

With the rapid advance of scientific knowledge and the continuous evolution of operational needs, it is essential that a constant cycle of improvement occur in space weather forecasting/specification capabilities. This cycle involves the development and utilization of new capabilities that improve services, and it involves the continuous feedback of the highest priority operational needs. Research tools comprise scientific models and research (measurement) infrastructure. The latter may require investment in order to be considered operational with respect to data availability, timeliness, quality and reliability. Operational tools need to be consistent in order to comply with any given Service Level Agreement and to be considered effectively operational.
Recommendation: We recommend that a roadmap for research-to-operations (R2O) transition be conducted at a 5-year cadence, timed to provide information to, and take information from, a parallel science roadmap. This R2O roadmap should focus on specific, high priority activities that can achieve measurable progress on service metrics.

In order for research models to be considered suitable for use in a forecast center, they should be confronted to a set of standardized evaluation criteria, such as: accuracy of output; confidence level and uncertainty estimation; stability; maintainability and scalability; support and documentation; autonomy and ease of use (forecasters and operators may not be specialists).

{\bf Recommendation:} We recommend that the international community establish standardized practices and procedures to evaluate research models, similar to the GEM/\-SHINE/\-CEDAR challenges. Internationally agreed metrics for model and forecast performance evaluation should be established.

It is vital to implement and support reliable archiving of data collections and to guarantee streamlined access to archives. Efficient dissemination of archived and near real time data is a prerequisite for maintaining reliable service operation.

{\bf Recommendation:} We recommend that a common set of space weather metadata be developed in order to facilitate harvesting and interpreting data. In addition, the development of standardized interfaces to data repositories is strongly encouraged, as well as common data analysis toolkits (standardized software packages). This should build on existing international programs that have addressed the provision of standard metadata for the Sun-Earth domain, including the NASA-led SPASE project, the ESPAS and HELIO projects in the EU FP7 programme, ESA's Cluster Active Archive, and the IUGONET project in Japan. 

\vskip 2mm
{\em E.	Partner with the weather and solid-Earth communities to share infrastructure and lessons learned:}

The uppermost troposphere and the thermosphere/meso\-sphere above it influence ionospheric processes. Moreover, space weather phenomena may couple into long-term regional weather and climate, as has been proposed for cosmic-rays modulation and solar spectral irradiance variations. 

Partnering with climate/weather stakeholders is beneficial also on another front: space weather scientists are only beginning to learn how to assimilate observations into data-driven ensemble models, and can learn much from the meteorological community. Space weather forecasting looks to the example of terrestrial weather forecasting as a field with similar goals and much longer history and experience. In particular, many of the advances in terrestrial weather forecasting have come about by the augmentation of physics-based models with advanced statistical methods, such as data assimilation and ensemble modeling.  The advancement of space weather forecasting likely requires similar endeavors.  However, we must recognize that for some regimes (such as in Earth's ionosphere) the meteorological techniques can be borrowed with little modification, while in other regimes (particularly where the data is sparse) new techniques will need to be developed.

Yet another reason for partnerships with weather and solid-Earth sciences could be the sharing of stations for geomagnetic, magnetotelluric, and ionospheric measurements in the much-needed expanded web of ground-based measurement locales. 

Engagement with the solid-Earth geophysics community is vital for space weather studies of GIC, since the geomagnetic induction of electric field inside the Earth is a critical element in the causal chain of physics leading to adverse GIC impacts on technological systems. The solid-Earth geophysics community has excellent expertise in this area, since geomagnetic induction is also a key technique for exploring the sub-surface structure of the Earth down to depths of at least several hundred kilometers. Thus their engagement in space weather studies is vital.

{\bf Recommendation:} We recommend intensification and formalization of   coordination between the ITM and climate/weather research and modeling communities, specifically (a) to develop, in the mid-term, whole atmosphere models (i.e. coupled models ranging from the troposphere into the high ITM domain), (b) to learn about data assimilation, ensemble modeling and other advanced statistical techniques in forecasting scenarios, (c) to partner in the multi-use of ground stations for both terrestrial and space weather, and (d) to analyze possible pathways by which space weather could impact Earth global or regional climate. In the US, NSF/UCAR is the natural focal point for this research, coordinated with NASA/SMD Heliophysics research in ITM physics, while organizations like WMO or IUGG may be engaged as an enabling coordinator and as an existing multi-national organization for intergovernmental communication. We also recommend strengthening ties to the solid-Earth community to support transformation of geomagnetic variability to electric fields involved in GICs.

\section{Research: observational, numerical, and theoretical recommendations}

The highest-priority scientific needs that were only very briefly mentioned in Section 5.1 are presented in more detail in this section, and complemented by specific instrumentation and model concepts in Section 7. In this Section, we identify the needs (a) to maintain existing observational capabilities, (b) to develop modeling capabilities, enable archival research, or improve the data infrastructure, and (c) to construct new space-based and ground-based instrumentation. By their nature, these three classes of investments can be implemented (a) immediately, (b) on a time scale of a year or two, or (c) on a time scale of about 4 to 10 years. 

The recommendations formulated in this section are taken from the impact tracings (discussed in Appendix B) that were initially produced looking separately at the solar-heliospheric domain and at the geospace domain (as summarized in Appendices D and E, respectively). In this section we merge all those requirements, sorting them into three general clusters of recommendations that we shall refer to as pathways of research.
Pathway I collects the needs for the full Sun-Earth domain that are related to the need to obtain forecasts more than 12\,hrs ahead of the magnetic structure of incoming CMEs to improve alerts for GICs and related ionospheric variability. Pathway II will need at least part of the new knowledge developed within Pathway I to forecast the particle environments of (aero-)space assets and to improve environmental specification and near-real time conditions. Pathway III is a particularly challenging one, aiming to enable pre-event forecasts of solar flares and CMEs, and related solar X-ray, EUV and energetic particle storms for the near-Earth satellites, astronauts, ionospheric conditions, and polar-route aviation, including all-clear conditions. 
From the viewpoint of solar physics, Pathway I contains those requirements that are most feasible to do within next five years. From the viewpoint of geospace research Pathway I contains the requirements linked with the advancements to be achieved from solar physics and support especially GIC/GMD. With the requirements of Pathway I fulfilled, we should be able to improve GIC/GMD forecast in all conditions. Also in GNSS and aero-space domains significant improvement will be achieved, but the issues involved will not be fully solved in Pathway I.

From the viewpoint of solar physics, Pathway I contains much the same things as Pathway II.  The geospace requirements are mostly separate in Pathway II because they address largely complementary physics in geospace, while they will gain significant benefit from the solar and heliospheric work on the variable solar wind supported by Pathway I. Aero-space assets are here the main impact area, but the models discussed in Pathway II will support the efforts to understand D-layer absorption (caused by RB energetic particle precipitation) and thus they will help also with HF communication. 

Recommendations in Pathway III are from the viewpoint of solar physics much more challenging than those of Pathways I and II because they concern forecasts before any substantial event has occurred on the Sun. 

\paragraph{Pathway I:}
To obtain forecasts more than 12\,hrs ahead of the magnetic structure of incoming coronal mass ejections to improve alerts for geomagnetic disturbances and strong GICs, related ionospheric variability, and geospace energetic particles:

\noindent Maintain existing essential capabilities: 
\begin{itemize}
\item	magnetic maps (GBO, SDO), X/EUV images at arcsec and few-second res. (SDO; Hinode), and solar spectral irradiance observations;
\item	solar coronagraphy, best from multiple perspectives (Earth's view and L1: GBO and SoHO; and well off Sun-Earth line: STEREO);
\item	in-situ measurements of solar wind and embedded magnetic field at, or upstream of, Sun-Earth L1 (ACE, SoHO; DSCOVR);
\item	for a few years, measure the interaction across the bowshock/magnetopause (as now with Cluster/\-ART\-EMIS/\-THEMIS; soon with MMS), to better understand wind-magneto\-sphere coupling;
\item	satellite measurements of magnetospheric magnetic and electric fields, plasma parameters, soft auroral and trapped energetic particle flu variations (e.g., Van Allen Probes, LANL satellites, GOES, ELEC\-TRO-L, POES, DMSP);
\item	ground-based sensors to complement satellite data for Sun, heliosphere, magnetosphere, and iono-/thermo-/mesosphere data to complement satellite data.
\end{itemize}

\noindent Archival research, develop data infrastructure, or modeling capabilities:
\begin{itemize}
\item	near-real time, observation-driven 3D solar active-region models of the magnetic field to assess destabilization and estimate energies; 
\item	data-driven models for the global solar surface-coronal field; 
\item	data-driven ensemble models of the solar wind including magnetic field;
\item	data assimilation techniques for the global ionosphere-magnetosphere system for nowcasts and near-term forecasts to optimally use and to coordinate ground and space based observations to meet user needs. Compare models and observations, ideally in select locations where laboratory-like test beds exist or can be developed at a few informative latitudes.
\item	coordinated system-level research into large-scale rapid morphological changes in the Earth's magnetotail and embedded energetic particle populations (using data from, among others, SuperDARN, SuperMAG,  AMPERE, etc.);
\item	system-level study of the mechanisms of the particle transport, acceleration, and losses driving currents and pressure profiles in the inner magnetosphere; 
\item	stimulate research to improve global geospace modeling beyond the MHD approximation (kinetic, hybrid, \ldots)
\item	develop the ability to use chromospheric and coronal polarimetry to guide full-Sun corona-to-helio\-sphere field models.     
\end{itemize}

\noindent Deployment of new/additional instrumentation:
\begin{itemize}
\item	binocular imaging of the solar corona at $\sim$1-arcsecond and at least 1-min. resolution, with $\sim$10$^\circ$-20$^\circ$ separation between perspectives;
\item	observe the solar vector field at and near the surface and the overlying corona at $<$200-km resolution to quantify ejection of compact and low-lying current systems from solar active regions;
\item	(define criteria for) expanded in-situ coverage of the auroral
  particle acceleration region and the dipole-tail field transition
  region (building on MMS) to determine the magnetospheric state in
  current (THEMIS, Cluster) and future high-apogee constellations, using hosted payloads and cubesats where appropriate;
\item	(define needs, then) increase ground- and space-based instrumentation to complement satellite data of magnetospheric and ionospheric variability to cover observations gaps (e.g., in latitude coverage or over oceans); 
\item	an observatory to expand solar-surface magnetography to all latitudes and off the Sun-Earth line [for which the Solar Orbiter provides valuable initial experimental views];
\item	large ground-based solar telescopes (incl.\ DKIST) to perform multi-wavelength spectro-polarimetry to probe magnetized structures at a range of heights in the solar atmosphere, and from sub-active-region to global-corona spatial scales;
\item	optical monitors to measure global particle precipitation [such as POLAR and IMAGE] to be used in data assimilation models for GMD and ionospheric variability.
\end{itemize}

\paragraph{Pathway II:}
For the particle environments of (aero)space assets, to improve environmental specification and near-real-time conditions. In addition to the remote-sensing and modeling requirements for Pathway 1:	

\noindent Maintain existing essential capabilities:
\begin{itemize}
\item	develop particle-environment nowcasts for LEO to GEO based on observations of electron and ion populations (hard/$\sim$MeV and soft/$\sim$keV; e.g., GOES, \ldots, striving for intercalibrated data sets with better background rejection, for at least a solar cycle), and of the magnetospheric field [see GMD/GIC recs.];
\item	maintain a complement of spacecraft with high resolution particle and field measurements and defined inter-spacecraft separations (e.g., the Van Allen Probes).
\end{itemize}

\noindent Archival research, develop data infrastructure, or modeling capabilities:
\begin{itemize}
\item	specify the frequency distributions for fluences of energetic particle populations [SEP, RB, GCR] for the specific environment under consideration, and maintain access to past conditions;
\item	develop, and experiment with, assimilative integrated models for RB particle populations towards forecast development including ionosphere, thermosphere and magnetosphere, including the coupling from lower-atmospheric domains, and validate these based on archival information.
\end{itemize}

\noindent Deployment of new/additional instrumentation:
\begin{itemize}
\item	increased deployment of high- and low-energy particle and electromagnetic field instruments to ensure dense spatial coverage from LEO to GEO and long term coverage of environment variability (including JAXA's ERG [Exploration and energization of Radiation in Geospace; launch in 2015]. Combine science-quality and monitoring instruments for (cross) calibrations, resolution of angular distributions, and coverage of energy range.
\end{itemize}

\paragraph{Pathway III:}
To enable pre-event forecasts of flares and solar energetic particle storms for near-Earth satellites, astronauts, ionospheric conditions, and polar-route aviation, including all-clear conditions. In addition to the remote-sensing and modeling requirements for Pathway I:	

\noindent Maintain existing essential capabilities:
\begin{itemize}
\item	solar X-ray observations (GOES);
\item	observe the inner heliosphere at radio wavelengths to study shocks and electron beams in the corona and inner heliosphere;
\item	maintain for some years multi-point in-situ observations of SEPs off Sun-Earth line throughout the inner heliosphere (e.g., L1, STEREO; including ground-based neutron monitors). 
\item	maintain measurements of heavy ion composition (L1: ACE; STEREO; near-future: GOES-R). 
\end{itemize}

\noindent Archival research, develop data infrastructure, or modeling capabilities:
\begin{itemize}
\item	develop data-driven predictive modeling capability for field eruptions from the Sun through the inner heliosphere;
\item	investigation of observed energetic particle energization and propagation within the inner-heliospheric field, aiming to develop at least probabilistic forecasting of SEP properties [see also Pathway-1 recommendations for heliospheric data-driven modeling];
\item	new capabilities for ensemble modeling of active regions subject to perturbations, to understand field instabilities and energy conversions, including bulk kinetic motion, SSI, and energetic particles. 
\end{itemize}

\noindent Deployment of new/additional instrumentation:
\begin{itemize}
\item	new multi-point in-situ observations of SEPs off Sun-Earth line throughout the inner heliosphere to improve models of the heliospheric field and understand population evolutions en route to Earth (e.g., Solar Orbiter, Solar Probe Plus).
\end{itemize}

\section{Concepts for highest-priority research and instrumentation}

In this section we present brief summaries of the rationales for the (new or additional) research projects and instrumentation investments given the highest priorities in the preceding section. These summaries, expanded upon in In 6, are meant to illustrate possible investments that address the identified scientific needs towards an integrated approach to the space-weather science. These investment concepts demonstrate possible approaches that we deem feasible in terms of technological requirements and budgetary envelopes, but we emphasize that science definition teams focusing on the top-level scientific needs may well be able to identify and shape other options. 

\subsection{Quantify active-region magnetic structure to model nascent CMEs}

We identified a very high-priority requirement to characterize the solar-wind magnetic field, and in particular the field involved in CMEs. To address that, in-situ measurements upstream of Earth at the Sun-Earth L1 sentinel point are insufficient. Moreover, technological means and budgetary resources are inadequate to position in situ sentinels on the Sun-Earth line sufficiently close to the Sun, or to launch a fleet of moving sentinels with always one near to that position. The variable solar-wind magnetic field associated with active-region eruptions must consequently be obtained from forward modeling of observed solar eruptions through the embedding corona and inner heliosphere, based on the inferred magnetic field of erupting structures. A key science goal in this roadmap is consequently the determination of the origins of the Sun's impulsive, eruptive activity that will eventually drive magnetospheric and ionospheric variability. Deriving that from surface (vector-) field measurements only has been shown to yield ambiguous results at best. 

One way to constrain the 3D active-region field makes use of novel modeling methods that can utilize the coronal loop geometry to constrain the model field, most successfully when 3D information on the corona is available. As described in Appendix F.1, binocular imaging of the active-region corona at moderate spatio-temporal resolution enables the 3D mapping of the solar active-region field structure prior to, and subsequent to, CMEs, thereby providing information on the erupted flux-rope structure. This could be achieved by a single spacecraft some 10 to 20 degrees off the Sun-Earth line if combined with, e.g., the existing SDO/AIA or other appropriate EUV imagers, or - if these are no longer available - by two identically equipped spacecraft off the Sun-Earth line with approximately 10-20 heliocentric degrees of separation.

A second new observational capability is also prioritized in this context to derive information on low-lying twisted field configurations in the deep interiors of unstable active regions before and after eruptions is key to determining what propagates towards Earth to drive space weather. The magnetic stresses in these regions that are involved in CMEs cannot be observed directly, but require modeling based on observations of the vector magnetic field at and immediately above the solar surface (in the solar chromosphere), guided by higher structures that are observed within the overlying atmosphere (differentiated by temperature from the low, cool, dense chromosphere to the high, hot, tenuous, field-dominated corona). With the help of these atmospheric structures that outline the magnetic field and the parallel electrical currents we can determine the fraction of the system's free energy that is converted to power eruptions. Binocular imaging is unlikely to succeed for very low-lying, compact magnetic structures within active regions. 
Moreover, chromospheric vector-mag\-netic measurements in addition to photospheric ones are expected to aid in the mapping of the 3D active region field as this provides observational access to electrical currents threading the solar surface, and observing the details of the low-lying  flux ropes before, during, and after eruptions to quantify the 3D field ejected into the heliosphere.

The information on the low-lying magnetic field and the electrical currents within can be obtained by a space mission (outlined in Appendix F.2) that obtains vector-mag\-netic measurements of active regions at the solar surface and within the overlying chromosphere, combined with optical and EUV imaging of the solar atmosphere from 10,000K up to at least 3MK, all at matching resolutions of order 0.2 arcseconds, to map all field structures that may carry substantial electrical currents. 

\subsection{Coupling of the solar wind to the magnetosphere and ionosphere, and strong GICs}

Like explained in Section 4, some key processes in the chain of magnetospheric energy release leading to large field-aligned currents and hence GICs remain poorly known: 
\begin{itemize}
\item	the dynamics nearer to Earth at the inner edge of the plasma sheet where bursty bulk flows brake and presumably give rise to a number of more or less effective plasma instabilities, 
\item	and the coupling of these processes into the ionosphere along magnetic field lines.
\end{itemize}
In particular, the processes that control the rate of energy transport and the partition between competing routes of dissipation in the coupled (M-I) system remain \referee{insufficiently} understood. We do not know the conditions which control the partitioning and exchange of energy between currents, waves, and particles that are believed to act as a gate for the ability of the tail to drive FACs through to closure in the ionosphere

To uncover the processes in this interface domain, a two-constellation satellite mission architecture is proposed (detailed in Appendix F.3), where the challenge lies not in advanced instrumentation but in the positioning of a sufficient number of spacecraft at the two key locations in space. The first constellation should focus on plasma instabilities and flow braking at the inner edge of the plasma sheet, in the transition region from dipolar to tail-like magnetic fields. It would probe the three-dimensional plasma and electrodynamic fields (E and B) in the transition region, from close to geosynchronous orbit to around 10-12\,$R_{\rm E}$ during the course of the mission. This constellation could consist of a central mother spacecraft accompanied by some 4 smaller spacecraft approximately 1\,$R_{\rm E}$ from the mother, providing coverage in the azimuthal and radial directions. 

The second, coordinated constellation is to focus on the (M-I) coupling in the auroral acceleration region on conjugate auroral field lines below. It would provide multi-point plasma and electrodynamic fields in the auroral acceleration region to determine the dynamical M-I coupling, at altitudes of around 4000\,km to 1\,Re, in a configuration designed to resolve spatio-temporal ambiguity of the filamentary FACs and to distinguish between alternative particle acceleration processes. 

These constellations should be supported by conjugate auroral imaging
from the ground, and if possible also from space, to probe both global
and small scale ($<$100\,km) structures.  LEO satellites to
additionally monitor the precipitating electrons as a measure of
conductivity changes below will provide valuable complementary
measurements. On the other end of the coupled system of the
magnetosphere measurements of incoming flows in the central plasma
sheet from the more distant tail are required to assess incoming flow
characteristics in the same meridian. These could be provided by
existing assets, such as Geotail, Cluster, THEMIS, or perhaps by MMS in an extended mission phase. An upstream solar-wind monitor is, as always, required as well.
The success of the constellations proposed here, as is true for the overall geospace observing system, is crucially dependent of measurements from rich networks of ground-based instruments, including (a) magnetometers to observe how electric currents in the ionosphere are modified by space weather, plus (b) a wide variety of radar and radio techniques to monitor changes in the density, motion and temperatures of ionospheric plasmas, as well as (c) optical techniques to measure thermospheric winds and temperatures. These data sources are all key inputs to aid the identification of the onset location and the resolution of the spatio-temporal ambiguity of the processes leading to large dB/dt. Ground-based data support also the development of improved models of the atmosphere and its response to space weather, increasingly so as we advance assimilative approaches. As outlined in Appendix F.4, we need to promote these ground-based networks as a global system for scientific progress on space weather, so that each instrument provider (and funder) sees how their contribution fits into the wider picture, i.e. that a local contribution builds and sustains local access to a global system.

\subsection{Global coronal field to drive models for the magnetized solar wind}

The Sun's surface magnetic field is a vital ingredient to any predictive model of the global magnetic field that defines the structure of the heliosphere, including the position of the heliospheric current sheet and the regions of fast and slow solar wind, and plays a key role in space weather at Earth: (1) the interaction of CMEs with the ambient field impacts their geo-effectiveness; (2) the connection of the heliospheric magnetic field to CME-related shocks and impulsive solar flares determines where solar energetic particles propagate, and (3) the partitioning of the solar wind into fast and slow streams is responsible for recurrent geomagnetic activity. In order to obtain the magnetic-field information that is needed to forecast the field arriving at Earth over a day in advance, the interaction of that field evolving from near the solar surface into the field moving through the corona-heliosphere boundary must be known.

Models for the global solar magnetic field typically use magnetic maps of the photospheric magnetic field built up over a solar rotation, available from ground-based and space-based solar observatories.  Two well-known problems arise from the use of these ``synoptic'' maps.  First, such maps are based on solar conditions that lie as much as 27 days in the past of an always-evolving surface field. Second, the field in the Sun's polar regions is poorly observed, and consequently the high-latitude fields in these maps are filled with a variety of interpolation and extrapolation techniques.  These two observational problems can strongly influence the global magnetic field model: poorly observed active regions near the limb (as viewed from Earth), unobserved regions beyond the limb, and inaccurate polar field estimates can introduce unacceptable errors in the high-coronal field on the Earth-facing side of the Sun.

To address these observational problems, we need to obtain photospheric magnetograms off of the Sun-Earth line, particularly of the east limb (just prior to becoming visible again on the Earth-facing side of the Sun that is the portion of the solar field with the oldest observations as viewed from Earth), to complement magnetograms obtained along the Sun-Earth line by SDO and ground-based observatories; Appendix F.5 describes a mission trailing Earth its orbit to provide the needed observations.  Moreover, we need to obtain magnetograms of the Sun's polar fields over a few years to understand the evolution of the Sun's polar magnetic flux. As described in Appendix F.5, the Solar Orbiter will provide some of the initial measurements needed by providing at least a good calibration of the highest-latitude solar field; but the orbit of the Solar Orbiter is such that magnetogram information of the solar far side as viewed from Earth will be provided only on occasion with long gaps between relatively short observing intervals and with long delays before the observations can be telemetered to Earth: magnetic field coverage at a cadence fast enough to follow active region evolution fully around the Sun is not possible with the Orbiter.

The improved magnetographic coverage of the solar surface is needed to improve the models for the corona-heliospheric interface and to feed heliospheric magnetohydrodynamic models: to yield meaningful predictions of CME magnetic structure, it is necessary to utilize an accurate representation of the time-evolving global solar coronal magnetic field as input to models of prediction, eruption and propagation of CMEs through the solar wind. Such models ultimately will need validation by sampling their results near Earth, but before that they need guidance as to the physical approximations made in the solar coronal field and plasma model. Such guidance comes from coronagraphic polarimetric observations, which we shall need to learn to absorb into global coronal-heliospheric models.

New observations and modeling techniques are needed (as summarized in Appendix F.6): we propose to start with simulated multi-wavelength coronal magnetometric observables for constraining the global magnetic field, to be incorporated into global MHD models. In order to reach this goal, we propose model testbeds (c.f., Section 5.2, Recommendation II) for synthetic polarimetric measurements related to the Zeeman and Hanle effects, gyroresonance and gyrosynchrotron radiation, thermal (free-free) emission, and coronal seismology.  These simulated observables can be tested as drivers for, and ultimately assimilation into, global MHD coronal models. Guided by the outcome of these experiments, observations need to be obtained, most likely requiring new instrumentation (see Appendix F.6).   

The MHD models for the global corona are the natural foundation for plasma and field models for the overall heliosphere. The latter are already being worked on, specifically for the background quiescent solar wind structure, but they are not yet driven by the full magnetic input from actual solar conditions simply because that information is yet to be derived as described above. Once that information is available, fully dynamic models for the entire inner heliosphere need to be strongly supported.  

\subsection{Quantify the state of the coupled magneto\-sphere-iono\-sphere system}

The response of magnetosphere to solar wind driving depends on the previous state of magnetosphere. Similar sequences in energy, momentum and mass transfer from the solar wind to magnetosphere can lead in some cases to events of sudden explosive energy release while in other cases the dissipation takes place as a slow semi-steady process. Comprehensive understanding on the factors that control the appearance of the different dissipation modes is still lacking, but obviously global monitoring of the magnetospheric state and system level approach in the data analysis would be essential to solve this puzzle. Continuous space-based imaging of the auroral oval would contribute to this kind of research in several ways. The size of polar cap gives valuable information about the amount of energy stored in the magnetic field of magnetotail lobes.  Comparison of the brightness of oval at different UV wavelengths yields an estimate about the energy flux and average energy of the particles that precipitate from magnetosphere to ionosphere. These estimates are not as accurate as those from particle instruments on board LEO satellites, but the additional value comes from the capability to observe all sectors of the oval simultaneously.  Such view is useful especially in the cases where the magnetosphere is prone to several subsequent activations in the solar wind.  The shape and size of the oval and intensity variations in its different sectors enable simultaneous monitoring of magnetosphere's recovery from previous activity while new energy comes into the system from a new event of dayside reconnection.

There is consequently a need to achieve continuously global UV or X-ray images to follow the morphology and dynamics of the auroral oval, at least in the Northern hemisphere, but occasionally also in the southern hemisphere. Imager data combined with ground-based networks similarly as suggested above in Section 7b allows solving the ionospheric Ohm's law globally which yields a picture of electric field, auroral currents and conductances with good accuracy and spatial resolution.  This would mean a leap forward in our attempts to understand M-I coupling, particularly the ways how ionospheric conditions control the linkage, e.g., by field-aligned currents (see Appendix F.7 for more discussion).

\subsection{Observation-based radiation environment modeling}

Trapped energetic particles in the inner magnetosphere form radiation belts. Solar wind variability modifies trapped particle transport, acceleration, and loss. The inner proton belt is relatively stable, but the outer radiation belt comprises a highly variable population of relativistic electrons. Relativistic electrons enhancements are an important space weather factor with a strong influence on satellite electronics. Around half of magnetic storms are followed by an enhancement of relativistic electron fluxes. It appears that the pre-existing magnetospheric state is a critical factor in radiation belt variability; for example, efficient production mechanisms appear to need a seed population of medium-energy electrons. Storm-time magnetospheric convection intensification, local particle acceleration due to substorm activity, and resonant wave-particle interaction are the main fundamental processes that cause particle energization and loss.

Current understanding of energetic electron and proton acceleration mechanisms within the radiation belts is insufficient to discriminate between the effectiveness of different energization, transport, and loss processes at different L-shells and local times. Quantitative assessment of the predictive properties of current and emerging models is thus essential. Given the current very limited understanding of geo-effectiveness, excellent monitoring of the real time environment (nowcasting) is the most useful product that can be provided to satellite operators. This, and well-characterized descriptions of historical events, can be used to interpret failures and improve the resilience of spacecraft design. We propose intensified support for comprehensive modeling of the behavior of particles within the radiation belts based on multi-point in-situ observations (see Appendix F.8 for more details). 

\subsection{Understand solar energetic particles throughout the Sun-Earth system}

SEPs present a major hazard to space-based assets.  Their production
is associated with large flares and fast CMEs in the low corona,
typically originating from complex active regions.  The prompt
response can arrive at Earth in less than an hour from the onset of
eruption, and some times in a few minutes after the onset of the flare in the case of a well-connected, relativistic particle event.   This is often followed by continued high SEP fluxes as a CME shock propagates out from the Sun and may exhibit a distinct peak, rising by as much as two orders of magnitude, as the CME shock passes over the point of measurement. The particles in this peak are often referred to as Energetic Storm Particles (ESPs) and are particles trapped close to the shock by plasma turbulence associated with the shock (e.g., Krauss-Varban, 2010).  They may provide most of the particle fluence within a gradual SEP event. While warning of events in progress is certainly important, many users require significantly longer warning, e.g., 24 hours (e.g., all-clear periods for Extra-Vehicular Activities for astronauts). Recent multi-point observations of SEP events off the Sun-Earth line show that prompt energetic particles have access to a wide longitudinal extent for some events. They also show that for a large proportion of SEP events, the prompt energetic particles do not propagate along the normal Parker spiral but in the magnetic field of a pre-existing CME. ESPs can also reach energies in excess of 500\,MeV and also pose major space weather hazard. The ESPs represents a ``delayed'' radiation hazard traveling with the CME shock, so that there is a good chance of making accurate predictions of the onset times of ESP events. Thus the key research challenge is how the intensity, duration, and arrival time of the prompt response depends on the SEP event near the Sun, presence of a type II or IV radio burst in the near-Sun interplanetary medium, and the source location of the flare/CME that drives the shock. 

In order to further the understanding of these SEP populations, we recommend that in addition to the recommendations for the solar and heliospheric domains above, there should be multi-point in-situ observations of SEPs off the Sun-Earth line and possibly closer to the Sun than the L1 distance. For this we recommend extensive use of any planetary and inner-heliospheric missions that may occur in the future (see Appendix F.9 for more details).  

\section{In conclusion}

Throughout the development of this roadmap, we were as much impressed by the observational coverage of the Sun-Earth system that is currently available as by the sparseness of that coverage given the size, complexity, and diversity of conditions of the system. Our main recommendations are readily condensed into a few simple phrases that reflect the team's philosophy: make good use of what we have now; bring considerable new resources to bear on those problem areas that have the largest leverage throughout the overall system science; and be efficient with the limited budgets that are available in the near future. Out of this philosophy come the recommendations that stress the paramount importance of system modeling using the system observatory, of collaboration and coordination on instrumentation and data systems, and of a shift from a few expensive, comprehensive satellite missions to investments in many efficient small satellites that target specific capability gaps in our current fleet.

The desire expressed in the charge to the team for demonstrable improvements in the provision of space weather information will need to be evaluated after our recommendations have started to be implemented. Our recommendations are formulated clearly with an expectation to meet that desire: knowing how to model the field in solar eruptions before its arrival at Earth, better understanding of the triggers and inhibitors of instabilities in the magnetospheric field, improved tracing of the sources and sinks of energetic particle populations from Sun to radiation belts, and greater modeling capabilities of the drivers of ionospheric variability, to name a few of the targets of this roadmap, appear all within reach if targeted investments are made.

One key challenge overall is the transformation into an effectively functioning global research community. Many of our recommendations are meant as catalysts to this process: open data, shared investments, code development as community effort, and coordinated instrumentation deployment are all part of this. 

In order to stimulate implementation of our recommendations as well as to assess their impact we recommend that the research community keep a finger on the pulse as active players and as stewards of the investments: this roadmap needs updates and course changes by the research community and their funding agencies every few years to remain relevant.

\appendix

\section{Roadmap team and process}

The roadmap team was appointed by the COSPAR leadership taking advice from the COSPAR Panel on Space Weather and the steering committee of the International Living With a Star (ILWS) program. The team organized its work to ultimately summarize its findings in three key areas: (1) observational, computational, and theoretical capabilities and needs; (2) coordinated collaborative research environment, and (3) collaboration between agencies and between research and customer communities. The team's deliberations proceeded hierarchically: a) key research questions to be addressed to make demonstrable advances in understanding the space environment and the space weather services derived from that; b) research methodologies required to effectively address those research questions, and c) specific crucial observables, models, data infrastructures, and collaborative environments that enable those research methodologies. Specific recommendations for observational, modeling, and data infrastructures are grouped into three pathways. The pathways reflect the assessed scientific urgency, the feasibility of implementations, and the likelihood of near-term success. Subsequent pathways need the recommendations of preceding ones implemented at least to some extent to achieve full success, but can be initiated in parallel. The roadmap team does not mean to imply priorities in the sense of scientific or societal importance in these pathways, but developed them to devise a pragmatic, feasible, affordable, international implementation plan to meet the roadmap's charge.

The roadmap team met for three face-to-face working meetings: twice in Paris at COSPAR headquarters (November 2013 and September 2014) and once in Boulder (April 2014, in conjunction with the US Space Weather Week). In between these work was coordinated in telecons every second week and email exchanges. Input from the community was sought through presentations at various meetings, notifications of an email input channel through community newsletters, and by personal discussions with the team membership. An initial draft of the Roadmap was presented in an interdisciplinary lecture at the COSPAR general assembly in Moscow in August 2014, followed by a dedicated community discussion session. Prior to the completion of the full written report, the principal findings were summarized in an Executive Summary (packaged as a brochure) and made available for final comments on the COSPAR web site\footnote{https://cosparhq.cnes.fr/sites/default/files/executivesummary$\_$\-compressed.pdf}, and handed out at the US LWS meeting and the European Space Weather Week, both in November 2014.

The roadmap team was supported in its activities by an oversight group from agencies and research organizations: Philippe Escoubet (ESA, ILWS Steering Committee Chair, ESA), Madhulika Guhathakurta (NASA, ILWS), Jerome Lafeuille (WMO, ICTSW, ILWS), Juha-Pekka Luntama (ESA-SSA Applications Programme, ESA), Anatoli Petrukovich (RFSA, ILWS vice-chair), Ronald Van der Linden (International Space Environment Service), and Wu Ji (Chinese National Space Science Centre, COSPAR Deputy Chair). 

For a partial compilation of space weather resources, we refer the reader to web sites such as at WMO/OS\-CAR\footnote{http://www.wmo-sat.info/oscar/spacecabapilities} (Observing Systems Capability Analysis and Review Tool), the WMO Space Weather Portal\footnote{http://www.wmo-sat.info/product-access-guide/theme/space-weather}, and the online Space Weather Catalogue\footnote{http://www.spaceweathercatalogue.org}. 

\begin{figure}
\includegraphics[width=8.8cm]{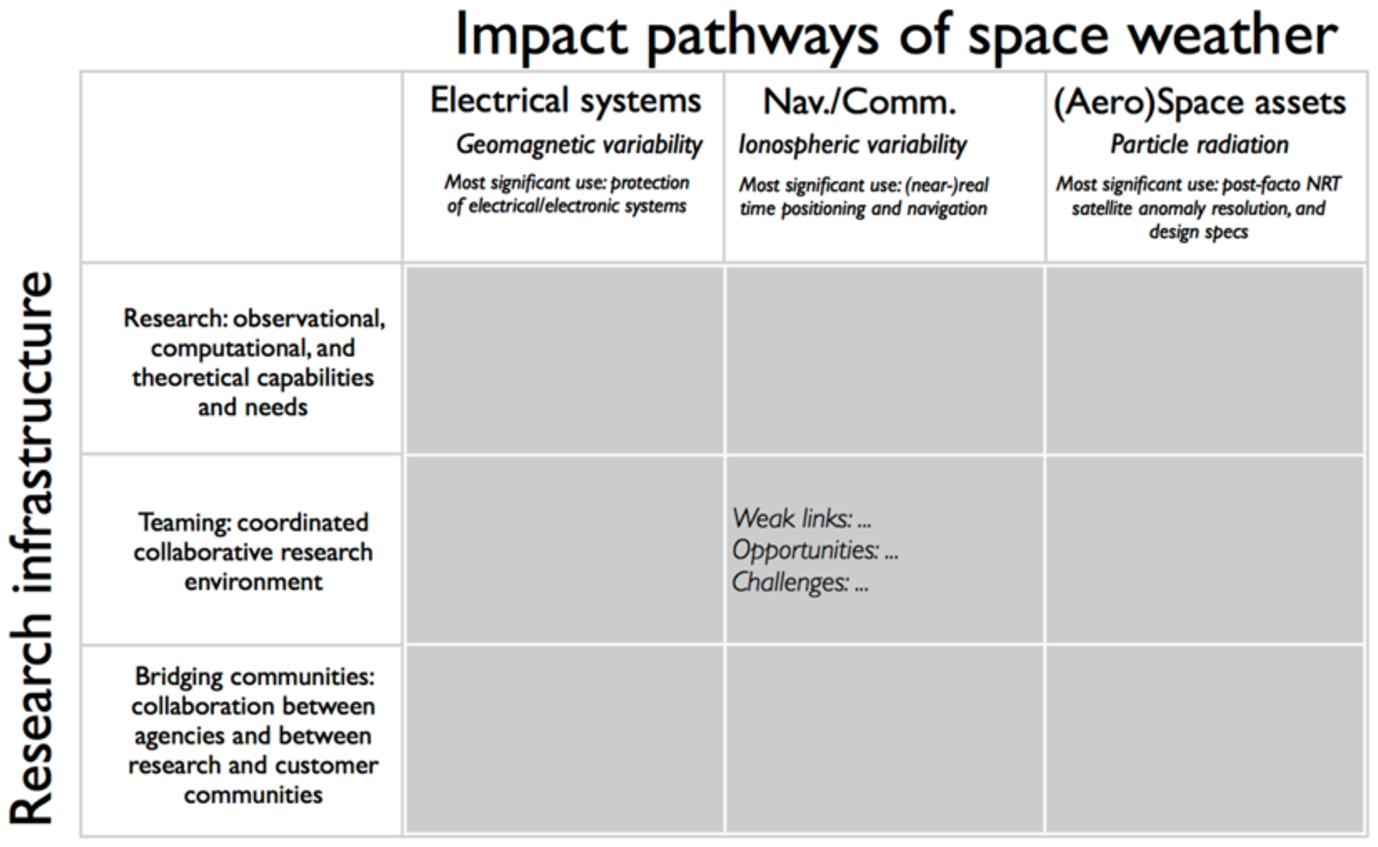}
\caption{Logical structure of the roadmap: After tracing impact pathways, requirements are prioritized in three major aspects of the research infrastructure, after which opportunities and challenges are identified in each to reach a set of prioritized recommendations.\label{fig4}}
\end{figure}
\section{Roadmap methodology: tracing sample impact chains}

The complexity of the interconnected system of physical processes involved in space weather precludes a single, comprehensive, yet understandable and concise exploration of the network as a whole. To effectively and efficiently address its charge the roadmap team therefore developed a strategy that focuses on three largely distinct space weather phenomena with largely complementary impact pathways on societal technologies: A) electrical systems via geomagnetic disturbances, B) navigation and communication impacts from ionospheric variability, and C) (aero-)space assets and human health via energetic particles. These impact pathways collectively probe the physics of the Sun-Earth system whose variability is quantified, for example, in the G (geomagnetic), R (radio blackout), and S (radiation) storm scales used by NOAA's Space Weather Prediction Center. The physical processes involved in the probed pathways often map across these storm scales that are but one way to quantify some of the attributes of space weather. No three pathways or scales can characterize the rich spectrum of space weather phenomena, just as terrestrial weather forecasts typically do not characterize the multitude of phenomena of weather, but in both cases a carefully selected set will encompass much of the dynamics and will help users understand the threat of space and terrestrial weather in the coming hours and days.

Subsequent to this impact tracing, the roadmap team integrated these parallel assessments into a comprehensive view with associated requirements and prioritized actions to be taken to advance our understanding of space weather and to improve the quality of space weather services. We do so by looking from the Sun, through the heliosphere, into geospace which we use to describe the domain encompassing the coupling region between solar wind and the geomagnetic field, throughout the geotail, and down into the high layers of Earth's atmosphere. A summary of the organizational structure is shown in Fig. A.4.

\section{State of the art in the science of space weather}

\subsection{Achievements}
The research activities in ``space weather'' are increasing rapidly: for example, based on entries in the NASA Astrophysics Data System, over a recent 15-year period (1997-2012) the number of publications per year has increased by a factor of 12 (refereed publications) to 16 (all publications) to 925 publications per year in total on the topic for 2012.  This rapid growth in interest in, and implied support for, space weather research (see Fig. 1) has advanced our insights in and understanding of space weather tremendously. That advance largely coincided with the growth of space-based observational capabilities, whose discoveries included the direct confirmation of the existence of the solar wind in the early 1960s (with undeniable confirmation during a six-month flight of the Mariner II spacecraft en route to Venus) and of the phenomenon of the coronal mass ejection first recognized as the highly dynamic process originating from the low corona in the early 1970s (with the coronagraphs flown on OSO-7 and shortly after that on Skylab). 

Since those early observations, recent achievements in the space-weather science arena are supported by an increased availability, accessibility, and sharing of both archival and near-real-time space- and ground-based data. These involve, for example, the US ``virtual observatories'' for solar and for geospace data, and the Information System of the World Meteorological Organization [WMO] through its Interprogramme Coordination Team on Space Weather [ICTSW] and the International Space Environment Service [ISES], and the ICSU World Data System and World Data Center [WDS/WDC], and e-infrastructrue programs in the EU (e.g., ESPAS, SEPServer, HELIO) and Japan (IUGONET), as well as by the development of end-to-end modeling frameworks (including NASA's CCMC), and various space agency programs (e.g., ESA SSA; NASA iSWA).

On the solar side, we now have available continuous full-Sun observations at multiple wavelengths, from multiple perspectives, from below the surface into the heliosphere. This observational coverage includes far-side observations and stereoscopic imaging by STEREO, SDO, and SoHO combined with other observatories, and helioseismic probing of the interior and even the far side of the Sun. Unfortunately, this full-sphere coverage does not include the (near-)surface magnetic field  which is measured only from Earth's perspective (continuously by SDO, and by some ground-based observatories). This full-Sun coverage supports the study of conditions leading to variable space weather, including the evolution leading up to and following solar storms, and the coupling to adjacent flux systems. But the space to cover is vast, and the observables range in size scale, energy, wavelength, and other physical attributes over many orders of magnitude, requiring multiple types of observatories and necessarily leaving gaps in our knowledge.

Years of increasing coverage and open-data policies have led to a growing open-access event archival database of well-observed flares, CMEs, and particle events that enables event comparisons and inter-domain coupling studies. Search and visualization tools are developing to increase the utility of and access to observational data. On the numerical front, leaps forward in the realism of numerical modeling that enable us to explore, for example, the generation, rise, and emergence of magnetic flux bundles into the solar atmosphere and the subsequent initial phases of a CME, the formation of sunspots upon field emergence onto the solar surface, and the analysis of chromospheric processes in a realm with radiative transfer, partial ionization, and comparable field and plasma forces. These numerical experimental settings are complemented by initial developments of data-driven models of the solar coronal field and of the Sun-heliosphere coupling in the nascent solar wind.

Within the heliosphere, we now have multi-perspective observations of propagating disturbances from Sun to Earth and beyond, often as combinations of in-situ and remote-sensing observations.  These include long-term, continuous, in-situ monitoring of solar wind and SEP properties at the Sun-Earth L1 point, one million miles upwind from Earth, and multi-point, multi-scale measurements in the near-Earth solar wind. These observations often provide information on the structure of CMEs, including the shocks between CME and the preceding solar wind (itself often containing earlier CMEs) that are important to the generation of solar energetic particles (SEPs). It has been demonstrated that approaching CMEs can be detected through interplanetary scintillation and that these show signatures in cosmic-ray muon intensities that may be used as proximity proxies. Sets of spacecraft around the inner heliosphere have demonstrated the prevalence of wide-angle SEP events. Exploratory models exist of CME propagation, including CME-coronal-hole and CME-CME interactions, from near the Sun to beyond Earth orbit.

Understanding of the magnetospheric dynamics greatly benefits for the
availability of continuous remote sensing of solar activity and
in-situ L1 monitoring of solar wind and SEP properties. Many key
advances in the research of energy storage and release processes are
being made owing to the availability of multi-point, multi-scale
measurements throughout the magnetosphere (Cluster, THEMIS) as well as
specialist coverage of the magnetotail (ARTEMIS) and the radiation
belts (Van Allen Probes), and improved coverage by ground-based
observatories (radar and magnetometer networks, e.g., SuperDARN and
SuperMAG). Here, as elsewhere in the science of Sun-Earth connections,
significant advances in numerical modeling are made, including the
coupling between CMEs and the geomagnetic field as well as magnetotail
dynamics, based on developments that utilize different approaches
(magneto-hydrodynamic [MHD], hybrid, kinetic) and that include
energetic particles. Regional studies have improved understanding of
GIC effects at auroral latitudes and ionospheric impacts in the Southern Atlantic
(magnetic) anomaly (SA(M)A-region). Although still challenging areas
of research, advances are being made in the characterization of
substorm dynamics (specifically their field evolution and system
couplings), understanding of the ring current system, and analyzing
the importance of ionospheric feedback for geomagnetic field dynamics.

In the ionosphere-thermo\-sphere-mesosphere (ITM) domain we see growing observational coverage and increasing regional resolution of ITM dynamics (including GNSS and ionosonde data), enabling near real-time maps of total electron content (TEC) and critical-frequency maps for radio communication, and advances in 3D reconstruction (by, e.g., GNSS radio-occultation and Beacon receivers, AMISR, and forthcoming EISCAT$\_3$D+ instrumentation and the Swarm mission). There is improved understanding of the contribution of the plasmasphere in the ionospheric total electron content (TEC), especially on the night side and during extreme events (e.g., plasmaspheric plumes). Insight has been deepened in sub-auroral effects (e.g., SAPS), in vertical coupling between troposphere and ionosphere (tidal modes, gravity waves), and in solar wind momentum storage and release in the thermosphere. There is improved access to data-product repositories, which in part is due to empirical model development based on long-term databases. All this supports the development of physics-based models and understanding of the processes involved, including data-ingestion into assimilative models, some of which are approaching transition to real-time operations.

The rapid growth of observational and numerical resources is advancing our understanding of the Sun-Earth system as a coupled whole, but many major issues remain to be addressed on the path towards improved scientific understanding and desired specificity in forecasts for the various sectors of society that are impacted by space weather.  

\subsection{Prospects for future work}
With regards to solar drivers of space weather, we have to advance the diagnostic capability to forecast times, magnitudes, and directionalities of flares and eruptions; at present diagnostics with the best skill scores perform quite poorly with often an unguided ``no flare'' forecast performing best, and with trained observers unable to differentiate forecasts within 24-h increments and flare magnitudes uncertain to one or more orders of magnitude, and with only the first advances to forecast the magnetic content of eruptions into the heliosphere being made in recent years. New magnetic field intruding into pre-existing active regions (a phenomenon referred to as ``active-region nests'' that one could view as pre-conditioning) is often involved in major flaring, making even ``all clear'' forecasts difficult on time scale more than about a quarter of a day. Long-range couplings and large-scale forces affect how eruptions propagate into the heliosphere, and we face the problem of limited coverage of solar magnetic field and coronal structure observations to incorporate within our growing modeling capabilities: only little more than a quarter of the solar-surface is accessible for high-resolution, well-calibrated magnetography, with vector-field data routinely made only for the active regions of their relatively young early-decay products. 

Another major challenge is establishing what determines whether a solar event inserts SEPs into the heliosphere and how such SEPs propagate through the heliosphere. Advancing on these fronts requires the development of physics-based, data-based models for the research community, including in particular a general-purpose observ\-ation-driv\-en magnetic-field-modeling tool. Quantifying the properties of extreme flares, CMEs, and SEP events requires investments in analyzing natural archives (e.g., biosphere, ice deposits and rocks - including those from the lunar surface; e.g., Kovaltsov and Usoskin, 2013), the use of cool-star analogs to our Sun to capture a sufficient number of rare extreme events to quantify the probability distribution for such extreme events powered by our present-day Sun. From a climatic point of view, we need to advance our understanding of solar spectral irradiance variations and its modeling based on magnetograms and models of the solar magnetic atmosphere. 

Once solar disturbances enter the heliosphere, we have yet to understand what establishes the characteristics (width, direction, velocity), deflection, and propagation of CMEs, including the interaction with the background solar wind. That problem requires the development of data-driven MHD modeling through general-access community models for the background solar wind and for interplanetary CME evolution from Sun to Earth (such as increasingly available and supported at NASA's CCMC). We have yet to establish the properties of the 3D magnetic structure of CMEs starting with CME initiation and following their propagation through and interaction with the magnetic field of the solar wind, required for forecasting CME arrival times and CME magnetic properties. Another major issue is the understanding of shock propagation through the pre-existing solar wind and the resulting SEP production and propagation (distinct from the prompt relativistic particles). 

Remote-sensing and in-situ observations are needed to validate heliospheric models, including continued in-situ L1 monitoring of solar wind and SEP properties. We need to acknowledge, moreover, that the evolution of CMEs, and therefore their geo-effectiveness, is affected by the solar wind ahead of them: we need to include pre-conditioning of the heliosphere into account, and thus think in terms of continuous evolution rather than in terms of isolated single events. On larger geometric and temporal scales, we need improved understanding of how heliospheric magnetic field and solar wind modulate galactic cosmic rays, as well as improved understanding of solar-wind patterns that can significantly enhance the energy of magnetospheric electrons.
For the terrestrial magnetosphere, we need improvement of knowledge of conditions in the incoming solar wind and incoming SEP populations, both by continued and improved upstream monitoring and modeling/forecasting. At the ground-level ``base'' of the magnetospheric state we need to ensure the continuity of global measurements of electric field and currents by ground-based radars and magnetometers and to find accurate methods to measure also particle precipitation and consequent ionospheric conductance variations in global scales. The details of the solar-wind/magnetosphere coupling remain elusive, so that understanding of the geo-effectiveness of CMEs has to include how the CME phenomena evolve through the bow shock. An additional challenge comes from magnetospheric pre-conditioning that can significantly modulate the impact of solar activity in the geosphere. Here, once more, we need physical data-driven modeling (MHD and kinetic) through general-access community models reaching across dipole-like and tail-like field domains, with ability to include regional impact modeling and the forecasting of substorms, of auroral currents, of the ring current, and of particle populations and their precipitation. We need multi-scale multi-component measurements in shocks, reconnection, and turbulence, and understanding of cross-scale coupling from micro-scale physics to meso- and macro-scale. 

For particle populations in the radiation belts we need better understanding of energetic ion and electron injection into radiation belts in storms and substorms, and also better forecasting of solar-energetic particle radiation storms at LEO and aircraft altitudes. Coupling across domains is also important, requiring space-based multi-point exploration of the MIT interfaces to understand, e.g., the origin of outflow ions and their roles in the magnetic storm development and the characteristics of the ionospheric-thermospheric phenomena caused by magnetic storms.  Lacking natural records, the specification of worst-case magnetospheric processes leading to strong GICs will require physics-based models, to be somehow adequately validated for conditions not normally encountered in geospace.

A key goal, perhaps the key goal, in mitigating space weather impacts mediated by the ionosphere is to develop 3D global models of the morphology of the ionosphere. Given such a model it will then be straightforward to derive many products needed by users, e.g., nowcasts and forecasts of TEC for GNSS and space radars, of usable frequencies for HF communications (e.g., foF2, MUF(3000)F2) and of where there are strong spatial gradients that will lead to ionospheric scintillation. Such models must be physics-based in order to encompass all the processes and phenomena that can affect the morphology of the ionosphere, especially in highly disturbed conditions. They must therefore include coupling to all the systems associated with the ionosphere, especially the co-located neutral atmosphere (thermosphere) as well the magnetically-coupled regions of the plasmasphere and magnetosphere. The key processes to be studied include the coupling of solar-wind-magnetospheric variability, ion-neutral coupling, the development and evolution of ionospheric storms, traveling ionospheric disturbances (TIDs), small-scale irregularities and bubbles, the mechanisms behind equatorial F-region radio scattering, and the role of small-scale structure in the mid-latitude ionosphere. An adequate understanding of all this necessary physics requires extending coverage of these coupled domains by globe-spanning ground-based networks for continuous observation (including mid- and low-latitude electron and neutral density, electric field, and neutral wind variations during geomagnetic events). Coordination in the development of these networks is also needed, including integration of ground- and space-based data systems, to develop data-driven modeling capability for the near-real-time ionospheric morphology. In this domain, too, we identify how physical data-driven modeling feeds into general-access community models for, e.g., regional perturbation knowledge, including scintillation, and near-term TEC forecasting. ITM studies will also feed into improved predictions of satellite drag for the purpose of orbit calculations (including collision avoidance, scheduling of critical operations and re-entry predictions) via better modeling of the thermospheric neutral density, temperature and composition.

Finally, we note that there is much synergy with radio astronomy. Many ionospheric processes add noise to radio astronomy observations, so there is scope for sharing of data and knowledge on the basis of one scientist's noise is another's signal. In addition, the technologies being developed for advanced radio telescopes (e.g., LOFAR and the Square Kilometer Array SKA) have great potential for spin-out into ionospheric work, e.g., as demonstrated by the Kilpisj{\"a}rvi Atmospheric Imaging Receiver Array (KAIRA) in Finland. Thus the ITM space weather the community needs to strengthen the links with the radio astronomy community in this exciting new era.
 
\section{Research needs for the solar-heliospheric domain}
In this Appendix D, and in the subsequent Appendix E, we provide a series of needs derived from detailed formulations of the top-level goals. In specifying the needs, we differentiate between data that we require to enable the basic process and data that we desire to significantly improve our ability or to accelerate our progress. We also differentiate between observables that are already being obtained and which we need to maintain, and observables for which we need to improve instrumentation or diagnostic capability, or observables for which we need to develop new capabilities. Priorities in this Appendix are given for the science of the solar-heliospheric domain; they are merged and renumbered with the priorities from the geospace domain from Appendix E in the main text of this roadmap.

\vskip 2mm
{\em Goal SH-1:  Time-dependent description of the coronal magnetic field, both for the space-weather source regions (i.e., quiet-Sun filament environments and active regions) and for the embedding global field.}

The cornerstone of any understanding or prediction of solar activity lies in the magnetic field. We note that
\begin{itemize}
\item	the three-dimensional magnetic field of an erupting region and of the surrounding corona are essential for specifying the characteristics of the transient solar activity that causes space weather impacts;
\item	the magnetic structure of the solar ejecta and its interaction with the embedding coronal and heliospheric fields result in the strength and orientation of the changing magnetic field that will reach the Earth; and that 
\item	the coronal magnetic field and its extension into the heliosphere affect SEP creation and transport.
\end{itemize}

\vskip 2mm
{\em Goal SH-1a: Specify magnetic structure of space-weather sources associated with active regions}
\begin{itemize}
\item REQUIRE: High-resolution vector photospheric field and plasma flow boundary condition. MAINTAIN the ability (SDO/HMI) to observe the full disk at about 1 arcsec resolution. IMPROVE for source regions by an order of magnitude spatially to at least 0.3 arcsec resolution as combined 2nd priority by new solar observations (as envisioned, for example, for Solar C).
\item REQUIRE: Vector boundary condition at a force-free layer, e.g., the top of the solar chromosphere. IMPROVE on the current ground-based experimental ability at NSO/SOLIS and MLSO ChroMag; NEW ability to be developed for space-based observations at about 0.2 arcsec resolution as combined 2nd priority by new solar observations (such as planned for Solar C), along with extensive theoretical/modeling developments.
\item REQUIRE: Development and validation of algorithms for magnetic field models using, e.g., coronal polarimetry and imaging data, including consideration of connections between active region magnetic fields and the surrounding global background coronal field. IMPROVE on current modeling capabilities in MHD and nonlinear force-free codes using currently available (e.g., SDO) observations [MAINTAIN] at about 1 arcsec resolution. NEW: combine with information from radio gyroresonance measurements (FASR, CSRH).
\item REQUIRE: Off-Sun-Earth line observations of the coronal plasma structures for on-disk source regions of solar eruptions. Develop a NEW capability to obtain and ingest binocular observations from near-Earth perspective and at about 10-20 degrees off Sun-Earth line as 1st priority in solar observations with a new mission concept.
\end{itemize}

\vskip 2mm
{\em Goal SH-1b: Specify the evolving global background coronal magnetic field:}
\begin{itemize}
\item REQUIRE:  Full-disk, front-side photospheric boundary condition. MAINTAIN the ability to obtain full-disk magnetograms as now possible with SDO/HMI and some ground-based observatories (GBOs). 
\item REQUIRE:  Models for the global solar field reaching into the heliosphere are crucially limited by magnetograph data being available only for the Earth-facing side of the Sun.  More comprehensive boundary data for a significantly larger portion of the solar surface, especially near the east limb of the Sun (where data obtained from Earth-based observations is oldest as regions evolve on the far side) is necessary.  This may occur through incorporation of information from the backside of the Sun as seen from Earth, via helioseismology/time-shifted data and/or by direct observations off the Sun-Earth line of the photospheric field, specifically at least about 50 heliocentric degrees trailing Earth (from around an L5 perspective, or - at least for some time - with the Solar Orbiter mission). IMPROVE modeling combined with NSO/GONG and SDO/HMI helioseismological data.  NEW: 3rd priority capability of solar magnetography off the Sun-Earth line, at first with the Solar Orbiter, while planning for sustained such observation for more extensive coverage in the more distant future. 
\item REQUIRE: Calibration of the high-latitude magnetic field (using, e.g., measurements from above and below the ecliptic plane) to validate surface flux transport models for the largest-scale field that determines the overall heliospheric structure. IMPROVE modeling combined with high-resolution observations from, e.g., Hinode and GBOs; obtain observational coverage of the high-latitude regions by the out-of-ecliptic observatory Solar Orbiter as 3rd priority in NEW solar observations.  
\item DESIRE: Full-disk, chromospheric magnetic boundary to validate and constrain magnetic models; the first step should be a study to establish the potential value of full-disk chromospheric vector magnetographs as addition to surface field measurements; NEW measurements under development by MLSO COSMO and planned for Solar C. 
\item DESIRE: Coronal polarimetry  (radio, infrared, visible, UV) to validate and constrain magnetic models. A first step here is to IMPROVE our abilities to use present-day GBO data to guide and validate continuous MHD field modeling of the global solar corona.
\item REQUIRE: Coronal imaging to validate and constrain magnetic models. MAINTAIN the ability to obtain X/EUV imaging of the solar corona with space-based observatories at about 1 arcsec resolution as now possible with SDO, supported by STEREO. 
\end{itemize}

\vskip 2mm
{\em Goal SH-1c: Evolving description of the ambient inner-heliospheric magnetic field and plasma flow in the solar wind through which CMEs propagate and with which they interact, that is important for the generation and propagation of energetic particles, and that drives recurrent geomagnetic activity.  The foundation of such models is derived from Goal SH-1b.}
\begin{itemize}
\item REQUIRE: Data-driven modeling of the solar wind throughout the inner heliosphere, improved with better descriptions of underlying physics, and routinely incorporating magnetic field to go beyond present-day hydrodynamic modeling.  MAINTAIN the ability to feed coronagraphic observations into heliospheric models, but substantially IMPROVE the modeling by including magnetic information from the corona and advancing it towards geospace and beyond. 
\item REQUIRE: Development of techniques to adjust models based on heliospheric observations, including available heliospheric images, and the development of NEW tools, such as radio-based interplanetary scintillation (IPS) methods, to detect and map perturbations approaching Earth. 
\item DESIRE: Solar and interplanetary observations from above the ecliptic plane to validate models or to guide models once such capabilities are developed. Such NEW observations could for some time be provided by the Solar Orbiter, the 3rd priority solar instrumentation, to develop and evaluate the utility of such observations.
\item DESIRE: Continuation of multi-view, multiple EUV emission line, full-disk photospheric vector magnetographs, multi-view coronagraphs and heliospheric imaging, and L1 solar wind measurements.  These capabilities should be MAINTAINed to drive modeling.
\end{itemize}

\vskip 2mm
{\em Goal SH-1d: Specify magnetic structure of space-weather sources associated with long-lived (quiescent) filaments}
\begin{itemize}
\item REQUIRE:  Observations of the photospheric line-of-sight
  boundary condition. MAINTAIN the ability to obtain full-disk
  magnetograms, such as now observed by SDO/HMI and GBOs. NEW: a capability to cover more of the solar surface for magnetic measurements as 3rd priority.
\item REQUIRE:  Characterization of coronal currents using non-potential magnetic models, and/or coronal evolution models, that are validated/fit to data such as coronal polarimetric (radio, infrared, visible, UV) and imaging observations. IMPROVE our abilities to used space and ground-based imaging and GBO polarimetric observations to guide and validate magnetic modeling capabilities based on MAINTAINed surface field maps. IMPROVED and NEW coronal polarimetric measurements under development for GBOs in the U.S. and China, and potentially off Sun-Earth line.
\item DESIRE: Chromospheric magnetic field (boundary condition and prominence) measurements to validate and constrain magnetic models. IMPROVED/NEW GBO observations and data assimilation techniques.
\item DESIRE: off-Sun-Earth line observations of coronal plasma; this NEW ability, the 1st priority for solar observations, is to be combined with NEW data assimilation techniques.
\end{itemize}

\vskip 2mm
{\em Goal SH-2: Description of CME/flux-rope evolution throughout the heliosphere, enabling prediction of CME arrival time, kinematics, and magnetic field strength and direction as function of time.}

Once the plasma and magnetic properties of the ejecta from the solar source regions are specified, we are in a position to determine its heliospheric evolution. We note that
\begin{itemize}
\item	quantitative knowledge of the properties of the erupting field and the field into which it erupts enable modeling of the transport and evolution of that eruption throughout the heliosphere en route to Earth;
\item	specifying the solar wind conditions incident upon the magnetosphere is critical for determining the space weather response in the geospace system; and that,
\item	in particular, high-impact hazards of space weather strongly depend on the strength and orientation of magnetic fields and the density and velocity at solar wind at the Earth.  
\item REQUIRE: NEW data-assimilative models coupling coronal models to solar wind propagation models (in near-real-time), explicitly incorporating magnetic flux rope structure established from Goal SH-1a and 1d into background wind model (Goal SH-1c).
\item REQUIRE: MAINTAIN/IMPROVE on validation of solar-wind propagation models with in situ observations at Earth/L1. 
\item DESIRE: MAINTAIN/IMPROVE validation of solar-wind propagation models with off-Sun-Earth-line in-situ observations (STEREO) and NEW off-Sun-Earth line remote-sensing observations (e.g., L5) and above-ecliptic (Orbiter, SPP).
\item DESIRE: MAINTAIN/IMPROVE on validation of solar-wind propagation models with remote imaging: e.g., Heliospheric imagers (structure), muon detection (orientation via pitch angle anisotropy), radio (Faraday rotation measurements constrain magnetic field strength), IPS. 
\end{itemize}

\vskip 2mm
{\em Goal SH-3: Develop the capability to predict occurrence of transient solar activity and the consequences in the heliosphere}

Given the coronal magnetic field, we are in a position to characterize and predict solar drivers of space weather. We note that
\begin{itemize}
\item	transient solar activity leads to a conversion of magnetic energy into space-weather-driving phenomena such as flares, coronal mass ejections, and energetic particles;
\item	interactions between the space-weather source and the global background corona affect the properties of these phenomena; and that
\item	the ability to predict transient solar activity in advance is generally desired, and required for warning of prompt solar energetic particles
\end{itemize}

\vskip 2mm
{\em Goal SH-3a: Predict transient solar activity in advance}

\begin{itemize}
\item REQUIRE: MAINTAIN observations that may be used for precursor warnings, e.g., active-region complexity, neutral-line shear, filament activation, helioseismology obtained from a range space and ground-based observatories, and IMPROVE upon theoretical and empirical justifications and implementation strategies for their use.
\item REQUIRE: Develop NEW capabilities for ensemble modeling in which probability of eruption is obtained based on perturbing coronal magnetic structure determined as in Goal SH-1, through e.g., flux emergence/flows, with consideration of sympathetic eruptive likelihood.
\item DESIRE: IMPROVE on theoretical understanding and modeling capability for predicting instability by analyzing the topological properties of coronal magnetic structures determined as in Goal SH-1.
\item DESIRE: MAINTAIN STEREO observations that may be used for precursor warnings using off-Sun-Earth-line observations; NEW off-Sun-Earth line observations.
\end{itemize}

\vskip 2mm
{\em Goal SH-3b: Specify the consequences of transient solar activity into the heliosphere}

\begin{itemize}
\item REQUIRE: MAINTAIN space and ground-based multi-wavelength imaging (including coronagraph) during eruption including Sun-Earth-line (SoHO, SDO), off-Sun-Earth-line (STEREO), and NEW off-Sun-Earth line observations.
\item REQUIRE: IMPROVE on existing techniques to quantify kinematic properties of ejecta including velocity and trajectory
\item REQUIRE: IMPROVE on eruptive models starting from coronal magnetic structure determined as in Goal SH-1; NEW incorporate interactions between the source and the background during eruption, and constrain/validate by data - leading to quantification of magnetic structure of ejecta.
\end{itemize}

\vskip 2mm
{\em Goal SH-4: Prediction of particle intensities, including all-clear forecasts, flare-driven acceleration near Sun, and shock acceleration at CME fronts.}

Energetic particles present a major hazard to space-based assets as
well as possible consequences for aircraft and aircrews and aircraft
passengers, particularly in polar routes.  The production of SEPs is
associated with large flares and fast CMEs in the low corona,
typically originating from complex active regions.  The
prompt-response particles can arrive at Earth in less than an hour
(some times as rapidly as a few minutes) after the onset of an eruption. These prompt SEPs may be followed by a longer lasting flux of particles originating at the CME shock as it propagates through the heliosphere, and a short, sharp, very high flux of energetic storm particles (ESPs) as the CME shock passes the point of measurement. Many of the needs to improve understanding of the prompt SEPs and ESPs are contained in the needs for solar and inner-heliospheric phenomena; the purpose of this short section focusing on SEPs and ESPs is to explicitly highlight their significance.

The strong increases in fluxes of high-energy protons, alpha particles and heavier ions in SEP events can cause or contribute to a number of effects, including:
\begin{itemize}
\item	Increases in astronaut radiation doses resulting in increased long term cancer risk and short term incapacitation
\item	Single Event Effects (SEEs) in micro-electronics
\item	Solar cell degradation
\item	Radiation damage in science instruments and interference with operations
\item	Effects in aviation include increases in radiation dose levels and interference with avionics through SEEs. 
\end{itemize}

While warning of events in progress is certainly important, many user needs (e.g., all-clear periods for EVAs for astronauts) require significantly longer warning, e.g., 24 hours.

\vskip 2mm
{\em Prompt energetic particles vs.\ energetic storm particles.}\null

\vskip 2mm
{\em Prompt particles:}  At the onset of major (M or X-class) flares and CMEs, SEPs can arrive in minutes after the start of the event.  These events are typically associated with large, magnetically complex active regions.  The energy source of the particles is still under debate: it may be entirely due to shocks generated low in the corona, or it may be that reconnection in solar flares plays a vital role, or both.  The prompt SEPs represent the most difficult forecasting problem, as they require
\begin{itemize}
\item 	The prediction of major solar flares and eruptions prior to their occurrence. 
\item	The efficiency of the flare/CME  to produce energetic particles
\item	The escape conditions of the particles from the coronal acceleration site
\item	The angular extent for injection of particles in the interplanetary medium (STEREO observations show that prompt energetic particles can be measured on a wide longitudinal extent). 
\item	The conditions of propagation of the prompt particles in the interplanetary medium (a large proportion of SEP events are found to propagate not along the normal Parker spiral but in the magnetic field of a CME). In addition to the conditions of diffusion in the medium, this evidently affects the delay between production of energetic particles and arrival at the earth. 
\end{itemize}

\vskip 2mm
{\em Energetic Storm Particles (ESPs):}  Energetic storm particles are particles detected in situ when shocks pass by spacecraft. These are particles trapped by turbulence just ahead of the shock front, and can be prominent events with the particle flux jumping by 1-2 orders of magnitude. Occasionally, ESPs can reach $>$500\,MeV and hence pose major space weather hazard. Therefore, the ESPs represent a ``delayed'' radiation hazard because the ESP event forecast is essentially a shock arrival forecast.   The duration of the ESP event generally depends on the shock geometry: At quasi-perpendicular shocks (where the shock normal vector is perpendicular to the local magnetic field), generally a narrow spike is observed, while a broad profile is observed in a quasi-parallel shock.

A good indicator of an impending ESP event is a radio-loud shock, i.e. a shock that produces type II radio burst near the Sun.  Stronger shocks generally produce more intense ESP events.  Radio-quiet shocks can also produce ESP events. Shock travel time ranges from about 18 hours to a few days, so there is a good chance of making prediction of ESP events. 

ESP events events precipitate in the polar regions and thus they are hazardous to satellites in polar orbits. They lead to radio fadeouts in the polar region and may cause communication problems for airplanes in polar routes.

Understanding and predicting particle acceleration relies on all preceding SH goals. We note that
\begin{itemize}
\item	Given a heliospheric field description, connectivity of Sun to Earth can be determined enabling all-clear forecasts if no active region is connected to the Earth (thus, even if flaring occurs, prompt particles on time scales of 10s of minutes are unlikely to occur). However, there are unsolved problems related to unusual spread of SEPs in longitude.
\item	Given prediction of a flare/CME, warning may be given of prompt particles and generally of SEPs on time-scales of 1-2 days; information about connectivity combined with models/observations may give warning of their likely geoeffectiveness.
\item	CME driven shocks continuously accelerate particles from the corona to 1 AU and beyond, so once a CME occurs, models that propagate them through the background heliosphere may predict shock-related SEPs, and in combination with observations of shocks can allow prediction of geoeffectiveness.
\end{itemize}

\vskip 2mm
{\em Goal SH-4a: All clear (ongoing)}

\begin{itemize}
\item REQUIRE: IMPROVE/NEW Evolving heliospheric model driven by 3D coronal field model (Goals SH-1b and c).  Modeling/analysis to understand unusual longitude spread and prompt rise of some SEP events.
\item DESIRE:  IMPROVE/NEW Predictive models of eruption, analyzed in the context of global field, to establish likelihood of connectivity change and sympathetic eruptions (Goal SH-3a).
\end{itemize}

\vskip 2mm
{\em Goal 4b: Predict Rapidly Arriving SEPs (time scale minutes to hours)}
\begin{itemize}
\item REQUIRE: MAINTAIN/IMPROVE nowcast observations to predict geoeffectiveness of incoming SEPs: relativistic and mildly relativistic particles (neutron monitors and GOES) to establish connectivity and forewarning of future, less energetic particles; measurement of type III radio bursts to establish that the flare/eruption site has access to open magnetic fields.
\item REQUIRE for 1-2 day forecast: IMPROVE/NEW ability to predict flares/CMEs (Goal SH-3a).
\item DESIRE: IMPROVE/NEW Evolving heliosphere model driven by 3D coronal field model (Goal SH-3), coupled with particle acceleration models and probabilistic/stochastic methods to characterize likelihood of strong SEPs from an active region.
\end{itemize}

\vskip 2mm
{\em Goal 4c: Predict SEPs from shock acceleration/transport (time scale hour to days)}
\begin{itemize}
\item REQUIRE: MAINTAIN radio observations of TYPE II bursts indicating shocks in front of CMEs; GBO radio observations to provide information on shocks close to the Sun at high frequency; L1-observations of shocks driven in interplanetary space that must be measured from space because they are below ionospheric cutoff 
\item REQUIRE: MAINTAIN in situ measurements of SEPs at L1.
\item DESIRE: IMPROVE/NEW evolving heliosphere model driven by 3D coronal field model with treatment of response to propagation of CME (Goal SH-1b and c), coupled with particle acceleration and transport models. 
\item DESIRE: MAINTAIN/NEW multi-point in situ measurements of SEPs off Sun-Earth line (STEREO, Solar Probe Plus, Solar Orbiter).
\end{itemize}
 
\section{Research needs for the geospace domain}

Following our impact tracings for the three different technologies presented earlier in this roadmap, we categorize the scientific activities as follows:\
\begin{itemize}
\item	Magnetospheric field variability and geomagnetically-induced currents;
\item	Energetic particles, in order of impact importance:
  \begin{itemize}
  \item	Solar energetic particles,
  \item	Radiation-belt energetic particles, and
  \item	Galactic cosmic rays;
  \end{itemize}
\item	Ionospheric variability. 
\end{itemize}

\subsection{Magnetospheric field variability and geo\-magn\-etically-induced currents}

\vskip 2mm
{\em Goal GM1:  Understand the dynamical response of the coupled geospace system to solar wind forcing to improve GIC forecast capabilities}

The dynamical Geospace Environment, in the form of the coupled MIT system, responds to energy input from the solar wind through processes of energy input, storage, release and dissipation (in severe GIC events the energy input from solar spectral irradiance plays at most a minor role). Rapid reconfigurations of this coupled system can be sufficiently fast, so that inductive magnetic fields produce large and unwanted GICs which present a threat to power grid infrastructures on the ground, which can be very serious during severe space storms.

The examples of GIC events with significant social and economic impact in recent history (GIC storms in 1989 and 2003) motivate research to specify and forecast the geo-electric field imposed on power grid and transmission line infrastructure in order to facilitate better preparedness for such impacts. One of the major challenges is that the fundamental physical processes which control especially the fast release of stored energy in this coupled system remain poorly understood. 

Under some conditions energy that is stored is released gradually with little GIC impacts; conversely under other conditions stored energy is released very rapidly from prior quasi-stable configurations. The dependence of this rapid energy release on the current and prior state of the coupled geospace system is not well-known, and present-day prediction capabilities are thus severely hampered, both by the poor understanding of the impacts of system-level coupling, and through the lack of knowledge of the severity of the effects arising from the prior state of the system through pre-conditioning. For example, the response of the coupled magnetosphere-ionosphere system to a given solar wind input depends upon the sequence of these drivers, and the time integrated response within the coupled geospace system. Whether there is a pre-existing ionization state in the ionosphere, a well-developed ring current, a heated atmosphere, or a dense tail current etc. can all change the rate at which the geospace system dissipates stored energy, and exactly when violent dissipation might begin. In the absence of a proper understanding of these impacts, the inductive electric fields arising from dB/dt from this coupled system have both a severity and indeed an onset timing which cannot be accurately forecast with present observations and understanding.
 
In order to improve the GIC forecasting capabilities, the required advances in the underlying physics will necessitate further discovery-based research. This must incorporate both existing and new measurements of the global state of the system and its dynamics, including data from extensive networks of ground-based instruments supported by constellation class in-situ satellite infrastructure. Such measurements, together with targeted improvements in model development (including data assimilation) will be key.

\vskip 2mm
{\em Goal GM-1a: Understand energy storage and release in the MIT system in driving large dB/dt at ground.}

\begin{itemize}
\item REQUIRE: Goal oriented research for improved physical
  understanding of the onset of rapid and large scale magnetotail
  morphology changes, including substorm onset, for improved
  physics-based forecasting (5th priority in Pathway I recommendations
  for modeling). At the moment we only see that some times the
  magnetosphere decides to deliver the excess energy to the ionosphere
  in a steady flow or small parcels (e.g., in the form of BBFs in the
  tail), and some times in one big parcel (substorm), driving GICs. We need to find out which characteristic of the coupling or which characteristic of the driver determines this choice. In addition, we need improved understanding of the role of preconditioning in determining the magnetospheric response to a time sequence of solar wind drivers (cf., the largest ever historical super-storms have occurred as isolated events which can (often) be during solar minimum periods).
\item REQUIRE: Systematic observations characterizing the global state of the MIT system (3rd priority in Pathway I recommendations for new instrumentation). Such observations require a maintained fleet of spacecraft in crucial locations in the tail and close to the ionospheric end of the conjugated field lines as described in section 7c. In addition, in order to solve the remaining key physics questions, we will require 2 constellation type missions at locations at the inner edge of the plasma sheet (from 7-12\,$R_{\rm E}$), and within the so-called auroral acceleration region at several thousand km altitude. None of these satellites needs radically new and expensive instrumentation though. The challenge lies in populating the key locations with spacecraft housing standard plasma instrumentation suites, and designing orbits which optimize the presence of s/c at these locations often enough, if not continuous. Again, the THEMIS mission provides some useful perspectives for the planning of the necessary configurations of future fleets or constellations. It would be valuable to have simultaneous measurements from both inside and outside of the onset region (at it's Earth side and further in the tail) in order to follow energy conversion processes and associated plasma flows towards the Earth and away from the magnetotail in the form of plasmoids. In addition a set of 4-6 probes at geostationary distances and distributed evenly in different MLT sectors would help in understanding how a major partition of this released energy gets stored into the ring current and radiation belts. An important asset supporting these missions would be a set of LEO satellites with magnetic field and particle instruments in order to get a better handle on the magnetosphere-ionosphere coupling (energy carried by currents and precipitating particles). A particular challenge is to measure the intensity of field-aligned currents that directly control the horizontal currents in the auroral ionosphere and consequently also the dB/dt values as measured on the ground.  NEW

MAINTAIN: When planning satellite fleets or constellations for probing the MIT system it is important to keep in mind the support that other existing satellite missions can provide. A nice example towards this direction is the AMPERE project which maintains an impressive set of magnetometers hosted by the Iridium satellites on LEO orbits. Another example of fruitful collaboration with long heritage of support to other space plasma missions is the DMSP program hosted by the United States Department of Defense. Ensuring continuity of this kind of missions will be an important factor in the attempts to keep the costs of multipoint MIT-monitoring on a tolerable level.   

\item	REQUIRE:  With the help of observations described above we need to develop magnetospheric modeling capacity taking upstream solar wind input and deriving forecast dB/dt at the Earth`s surface, including transformation and coupling at the bow shock and the magnetopause, coupling between hot and cold magnetospheric plasma populations, and coupling to the ionosphere including ionospheric conductivity modules and feedback (5th priority in Pathway I recommendations for modeling). We have still little understanding on the relative importance of such pre-states in the prime coupled elements of our system, nor the time constants of their potential impacts, but they are in order of distance from the Earth:
  \begin{itemize}
  \item	{\em Atmosphere:} temperature, altitude/density and winds - long lived (a day or so)
  \item	{\em Ionosphere:} conductivity and convection (strength and location) - short lived (from seconds to hours)
  \item	{\em Plasmasphere:} location and density - long lived (days)
  \item	{\em Ring current:} strength, composition, energy of particles, shape and/location - long lived (days)
  \item	{\em Tail current:} strength, mass composition, temperature, shape and/or location - short lived (hours)
  \item	{\em Lobe field:} pressure from previous energy coupling to solar wind, i.e. the energy reservoir from previous events is not completely emptied - no intrinsic time constant, influenced by other system input and output.
  \end{itemize}

\item REQUIRE: Coordinated ground-based measurement can provide crucial support for satellite missions (6th priority in Pathway I recommendations for maintaining existing capabilities). We propose to establish a formal basis for collaboration and data sharing between heliospheric and magnetospheric missions and ground-based assets. Informal (e.g., Cluster Ground-based Working Group) and formal (e.g., THEMIS mission) agreements have been made before, but usually missions do not involve a requirement for GB support in their level-1 science requirements for ground-based support (Van Allen Probes, MMS, Solar Probe, etc). This must be rectified. Satellite missions require support from a global network of
  \begin{itemize} 
  \item	MAINTAIN/IMPROVE: ground-based magnetometers, for the global and local current patterns. Instruments operated in the auroral zone and polar cap (magnetic latitudes above about 65 degrees) yield often the most interesting data, but during storm periods also sub-auroral stations (magnetic latitudes of some 55-65 degrees) are needed for grasping the entire picture.  According to the recommendations of the  WMO Interprogramme Coordination Team on Space Weather (WMO/ICTSW) the threshold for spatial resolution in ground-based magnetic field recording is 500\,km for SWx monitoring while for breakthrough science it would be $\sim$100\,km\footnote{http://www.wmo-sat.info/oscar/applicationareas/view/25}.
  \item	MAINTAIN/IMPROVE: radars for the electric field and convection patterns on meso-scale. On global scale such measurements can provide information on the energy state of the magnetosphere from oval size and dynamics. Similarly as for magnetometers, the interesting magnetic latitudes are mostly above 55 degrees. WMO/ICTSW recommendations for spatial resolutions are 300\,km and 10\,km for monitoring and break-through science, respectively.  
  \item	NEW: optical monitoring from space or ground for the precipitation and conductivity state (7th priority in Pathway I recommendations for new instrumentation).  While ground-based instruments (all-sky cameras) are useful in the research of meso-scale physics, again as in the case of radars for system level science global images by space-born imagers will be of key importance at least for two reasons: i) Space-based observations do not suffer from cloudiness problems and ii) global scale instantaneous images of the auroral oval provide valuable information on the magnetospheric energy storage and release processes during storm and substorm periods.  
  \end{itemize}

  \item DESIRE, IMPROVE/NEW:  Improved magnetosphere-ionosphere coupling models, including the impact of feedback between field-aligned current structures and energetic particle precipitation on ionospheric conductivity - which is an important aspect of the creation of ionospheric currents and hence ground dB/dt, resulting in an electric induction field. 

  \item DESIRE, IMPROVE/NEW: Improved data assimilation method and models applied to the coupled magnetospheric system including magnetic fields, current systems, and hot and cold plasma populations. 

  \item DESIRE, IMPROVE: Capacity to specify the global state of the magnetosphere will ultimately be required to advance our knowledge of space weather to the true point of predictability.  Even with currently operating multi-satellite missions in the so-called Heliophysics Great Observatory, geospace remains extremely sparsely sampled in-situ. Global networks of ground-based magnetometers, radars, and optical imaging (both ground and space) provide the only currently credible approach to specifying the global state of the coupled magnetosphere ionosphere system. Even supplemented by current satellite missions, the state remains sparsely specified in the critical regions of in-situ geospace. 
\end{itemize}

Future efforts to develop nano- and micro-satellite infrastructure could offer a cost-realistic route to a true constellation class mission (perhaps 50-100 satellites) needed to advance the system level science beyond its current typical case study methodology. Hosted payloads may offer an additional attractive route to securing constellation class specification of the state and dynamics of the in-situ geospace environment. Note, however, that this would be only an additional and not alternative route, as those instruments would most likely have to rely on simple measurements of plasma characteristics and relatively coarse and ``dirty'' field measurements due to host instrumentation interference. Nevertheless to monitor simple arrival times of changes in the plasma environment these would still be very valuable for science, but more so for monitoring. Suitable missions, which could host SWx instrumentation are, e.g., Galileo satellites, meteorological satellites on polar and geostationary orbits, and satellites testing new space technologies (e.g., the PROBA missions). Keeping these opportunities in mind, new SWx instrumentation should be based on generic solutions with are suitable for payloads hosted by several different missions. NEW

\vskip 2mm
{\em Goal GM-1b: Understand the solar wind - bow shock - magnetosphere-interactions}

\begin{itemize}
\item REQUIRE, MAINTAIN/IMPROVE/NEW:  Ongoing, reliable and continuous upstream monitoring of incoming solar wind conditions and propagation to 1AU such as from L1, or from other locations closer to the sun (4th priority in the Pathway I recommendations for maintaining existing capabilities). 
 
A suitable solar wind monitoring capability will be required. Data from L1 typically gives the general idea and this location is definitely suited for the primary mission for SWx forecasts, but in the near future we will also need satellites closer to the subsolar bowshock/magnetopause to understand what really arrives at Earth. For such remaining science purposes during the years 2007-2009 a good constellation was formed by the Cluster and THEMIS multi-satellite missions. Systematically during spring and autumn times one of these missions was monitoring the energy feed from solar wind and the other was probing the consequences in the magnetosphere. Ensuring such coordination between different missions will be vital for better physical understanding of energy transfer processes between solar wind and magnetosphere. In longer run the research conducted with Cluster-THEMIS (or in the future MMS-Cluster-ARTEMIS) observations should pave the way to design future cost-efficient solutions for continuous monitoring of upwind conditions with optimally configured satellite constellations. 

Both Cluster and THEMIS have been designed for science purposes. For SWx monitoring a less comprehensive suite of instrumentation would most likely be sufficient. A basic set of instrumentation could include, e.g., magnetic and electric field (at least AC in the case of E-field) instruments and a plasma instrument(s) with similar specification as in Cluster (e.g., electron and ion spectrometers, energy ranges 1\,eV - 30\,keV). The combined THEMIS and Cluster constellation provides a good first basis for examining the baseline for future constellations, for observing the time evolution of solar wind structures as they cross the bow shock, magnetosheath and magnetopause regions (mainly for case study purposes). 

\item DESIRE, IMPROVE/NEW: Improved global kinetic and/or hybrid models of the solar wind-bow shock magnetosphere interaction, including
  \begin{enumerate} 
  \item	downstream impacts of kinetic bow shock and magnetosphere processing of upstream solar wind, 
  \item	development and improved coupling of kinetic modules into global magnetospheric models especially plasmasheet-ring current-magnetosphere-ionosphere system. 
  \end{enumerate}

\item DESIRE, NEW: Research in order to achieve a consolidated view whether a satellite (or satellite constellation) located closer to the Sun-Earth line than currently at L1 is crucial to get better GIC predictions or not. Previous studies presented in the literature give a controversial view about this question. As a consequence it is at present not clear:
  \begin{itemize}
  \item	how close to the Sun-Earth line does a monitor need to be to get good upstream conditions for space weather/GIC input (in the literature the correlation length is different for the magnetic field, for the plasma velocity, and for electric field/potential (Burke et al., 1999); this knowledge determines whether an L1 orbit (perhaps constrained to within 60\,$R_{\rm E}$ only) is good enough or whether observations directly upstream from, and close to, the Earth on Earth orbiting spacecraft or constellation of spacecraft is needed.
  \item	whether we can specify what we need in terms of SW parameters with a single satellite and some assumptions about field orientations to constrain it (Weimer and Kind 2008); or whether we would rather need multi point measurements in the upstream solar wind for improved predictions.
  \end{itemize}
\end{itemize}

\vskip 2mm
{\em Goal GM2:  Uncover patterns in geomagnetic activity associated with disturbances in, and failure modes of the power grid infrastructure.}
\begin{itemize}
\item	GICs have potentially the largest space weather impacts on terrestrial infrastructure, with potentially long lasting and very severe impact and the modes of impact and consequent failure risk need to be understood.
\item	Models of the impacts of imposed dB/dt from the coupled geospace system in the form of geoelectric fields on power grids need to be developed and improved, and the effects arising from underlying solid earth physics such as sub-surface and sea-water conductivity should be included.
\end{itemize}
 
To date our understanding of the temporal and geographical patterns of geo-electric fields that are the most effective in driving GIC impact on the hardware in, or the operation of, electric power grids is very limited. 

{\em Priority: Develop the tools and data needed to enable engineering and impact studies to understand the failure modes of power grids in dependence of GMD driver characteristics.}

\vskip 2mm
{\em Goal GM-2a Understand the factors controlling the geo-electric field in the regions of primarily critical power grids, i.e serving dense populations at latitudes typically suffering from large GMDs.}

\begin{itemize}
\item REQUIRE:  MAINTAIN/IMPROVE/NEW: Deployment of a network of ground-based magnetometers and magnetotelluric instrumentation for geo-electric field measurements with infrastructure for data collection and distribution in near-real time. The critical density of real-time magnetometer and magnetotelluric networks needed for the regions of critical power grids under the risk of SWx disturbances should be defined on the basis of knowledge about underlying gradients in ground conductivity in the relevant regions, as these may cause very localized features in the geo-electric field. 

\item REQUIRE, IMPROVE/NEW: Systematic studies on ground-conductivity in the regions of high GIC risk. Combining results from ground-based, air-borne and satellite (Swarm in particular) measurement campaigns including both solid-earth and sea-water conductivity effects.  
\end{itemize}

\vskip 2mm
{\em Goal GM-2b: Assess and access space weather GIC risk and impact data.}

\begin{itemize}
\item REQUIRE, IMPROVE/NEW: Make power grid GIC (current and electric field) data available to the wider space weather research community (c.f. General recommendations for collaboration between agencies and communities, point i). Data of actual GICs effects in power grids are often collected by grid operators, but such data are for a variety of reasons often not (easily) made available for the scientific and technical study of risks, except internally with the power grid companies. This kind of data will be essential in the efforts to gain improved understanding of the different modes of malfunction (non-linear behavior, extra harmonics, and heating, overload etc) of transformers and power grids under GIC forcing. For space weather purposes this is particularly important as the scientists do not always know which type of disturbance is actually leading to potentially damaging impacts on various critical parts of the power grid infrastructure. This requirement includes therefore also determining the characteristics of the most geo-effective time sequences of dynamic fields (one large event, or a fluence of many smaller repetitive or pulsating events or persistent moderate forcing) and their impact on infrastructure spanning from individual single transformer failure and lifetimes to catastrophic network collapse. Knowledge of such impact factors could both improve the relevance of space weather predictions (not only large dB/dt warning required) and assist in the industries own assessments of approaches to critical infrastructure protection (CIP) through design of resilient power networks avoiding design with single point failure under the expectation that some elements of the network will likely fail.  

\item REQUIRE, IMPROVE/NEW: Nations should consider undertaking a vulnerability and risk assessment for space weather impacts on their power grids and other infrastructure, recording the resulting risks in national Risk Registers and adopting appropriate policy and taking appropriate action to mitigate such risks. 
\end{itemize}

\subsection{Magnetospheric field variability and particle environment}

The energetic particle environment in the magnetosphere is an important factor of space weather. The main external sources of the magnetospheric energetic particle population are SEPs and GCRs, as well as solar wind plasma particles. The last ones, being accelerated by magnetospheric processes, are responsible for radiation belts, auroral FACs and particle precipitation into the upper atmosphere. The dynamics of the magnetospheric magnetic field under solar wind driving controls particle distribution, acceleration and losses. We note that:
\begin{itemize}
\item	Solar activity is the main controlling factor for the particle environment in the Earth's magnetosphere. 
\item	Active processes on the Sun and in the heliosphere reveal themselves in the particle distribution inside the magnetosphere. Omitting the internal magnetosphere processes like plasma instabilities and wave activities, which are currently not predictable, MHD or hybrid models of solar wind - magnetosphere coupling are at present the prevailing means to provide space weather related forecasting of the magnetospheric particle fluxes for operational purposes.
\item	Magnetospheric magnetic field variations determine the particle dynamics, and disturbances in this field caused by solar wind variations can drastically change the near-Earth's particle environment.
\item	GCRs and SEPs penetrate deep into the Earth magnetosphere. The resulting particle population depends on the rigidities of the primary particles as well as on the magnetospheric magnetic field structure and dynamics.
\item	Enhancements of trapped electrons accelerated by inner magnetospheric processes are usually the consequence of interplanetary shocks, CME or high-speed streams arrivals.
\item	Low-energy particle precipitations in auroral regions are usually manifestations of magnetospheric processes taking place during geomagnetic disturbances.
\end{itemize}

It is essential for satellite operators to know the past, current and future condition of the space environment (particles and fields) around their own satellite, especially in case of actual satellite anomaly or before critical operation periods. Measurements and modeling of high and low energy particle fluxes, and observational data assimilation are of key importance for operative space weather predictions in near-Earth space. 

\vskip 2mm
{\em Goal MEP-1: Description of the magnetospheric state}

In order to move the predictability of particle fluxes forward, we first need a complete science based description of the magnetospheric particle populations, which give rise to space weather effects (1st and 2nd priorities in Pathway II recommendations for maintaining existing capabilities). Solar wind conditions are the controlling factors for the near Earth energetic particle population, but the previous state of the system is also critical. The details of the resulting population always depend on the present driving condition and the prehistory of the event (timescales to be considered are of the order of several days to a week).

Trapped energetic particles in the inner magnetosphere form the radiation belts. The inner proton belt is relatively stable, but the outer radiation belt comprises a highly variable population of relativistic electrons. Solar wind changes produce changes to trapped particle transport, acceleration, and losses. Current understanding of the ways in which the magnetosphere responds to solar inputs, indicate that the effectiveness of the input is highly dependent on the initial state. For example, magnetospheric reconnection rates are believed to be dependent on the oxygen content of the tail in the reconnection region, so that substorm triggering may be modified during storm time, when oxygen concentrations may be greatly elevated. Moreover, the production of relativistic electrons, is believed to depend on a seed population of high energy electrons as a result of earlier substorm activity. There are numerous other examples for such preconditioning.

Storm-time magnetospheric convection intensification, local particle acceleration due to substorm activity and resonant wave-particle interaction are the main fundamental processes that cause particle fluxes energization and loss. However, the details of the mechanisms that may be able to contribute to these processes remain a subject for active research. While progress has been made in the theoretical understanding of these competing processes, there is as yet no clear consensus on which of these will be significant in particular situations, and no real predictive methods which can give precise fluxes at different local and universal times. There is therefore a pressing need to confront theoretical models with detailed measurements, in order to resolve these shortfalls.

Finally, there are short-term populations of ring current particles produced by substorms and enhanced convection. Currently our understanding of substorm onset is evolving rapidly, in particular with respect to local dipolarization structures. However, we are still a long way from an ability to predict precise events, and a new physics understanding is almost certainly required. Current spacecraft measurement configurations are probably not sufficient to make accurate predictions. Since these effects are the major contributor to spacecraft failures due to charging, it is essential that current investigations are maintained and extended. 

\vskip 2mm
{\em Goal MEP-1a  Magnetic field and solar wind description:}
\begin{itemize}
\item REQUIRE: MAINTAIN current continuous solar and solar-wind observations as a basis for the prediction of magnetospheric disturbances (1st, 2nd, and 3rd priorities in Pathway I for maintaining existing capabilities). 
\item REQUIRE: Advanced magnetospheric modeling is needed to take into account local plasma processes affecting the magnetic field variations producing particle accelerations, losses etc.  (4th priority in Pathway I for modeling).

\item REQUIRE: In order to provide the diverse system level data set essential for evolving predictive models of energetic particles, and to provide nowcasting of energetic particle fluxes, a required space weather product, we should MAINTAIN similar space and ground-based infrastructure as required for the prediction of GMD and GIC (see above)
\begin{itemize}
\item 	Simultaneous GEO, LEO, MEO, GTO near real-time magnetic field and particle data collection and processing. (1st and 2nd priorities in Pathway II for maintaining existing capabilities)
\item	Measurements of current solar wind conditions and solar UV monitoring for solar wind prediction.
\item	Monitor magnetospheric activity and dynamics by means of networks of ground based magnetometers, radars and auroral imagers, including space born global auroral imagery. (6th priority in Pathway I for maintaining existing capabilities).
\end{itemize}

\item	REQUIRE: Improved magnetic field models (empirical, numerical) with particle tracing codes to form a basis for nowcast/forecast particle distributions, (1st priority in Pathway II for maintaining existing capabilities).

\item REQUIRE: Improved models for solar wind propagation from the Sun.
(1st, 2nd, and 3rd priorities in Pathway I for modeling).
\item DESIRE: Accurate Magnetic field models to predict Dst and forecast particle distributions. We suggest this as a particularly fruitful area for modeling.
\item DESIRE:  Visualization tools for magnetic fields and low and high-energy particle distribution in geospace.
\end{itemize}

\vskip 2mm
{\em Goal MEP-2a: Specify extreme condition of soft particle (keV) environment in Geospace - Surface charging}

Dynamical variations of keV plasma, such as substorm injection, auroral particle precipitation, and field-aligned currents are a critical factor in surface charging of satellites at LEO and GEO. Thus soft plasma descriptions are an essential precursor for the mitigation of surface charging problems from extreme conditions in the soft particle (keV) environment; this information is essential for satellite design. 

\begin{itemize}
\item REQUIRE: Monitoring soft particle fluxes by LEO satellites (e.g., DMSP, POES) (1st Priority in Pathway II recommendations for maintaining existing capabilities; 5th priority in Pathway I recommendations for maintaining existing essential capabilities)
\item DESIRE: permanent auroral oval monitoring from spacecraft in UV (like Polar) (7th priority in Pathway I recommendations for new instruments)
\item REQUIRE: Large databases of current and historical data obtained from in-situ particle measurements onboard spacecraft should be preserved (and compiled with cross-calibration) as we need to analyze the occurrence frequency distribution of extreme conditions of soft particles;   Understanding the physical mechanism of particle injection and precipitation and relating solar wind conditions with extremes in the geospace environment is an important aspect in order to make progress in predicting extreme conditions in the soft particle environment (1st priority in the Pathway II recommendations for modeling). 
\end{itemize}

\vskip 2mm
{\em Goal MEP-2b: Nowcast and forecast of soft particle (keV) environment in Geospace - surface charging.}

The soft particle environment in geospace is dynamically changing depending on solar wind conditions. Understanding the current and future soft particle environment (spatial distribution, time variation, energy spectra, etc.) is important to assess the risk of space assets and to examine on-going satellite anomalies in geospace (1st priority in the Pathway II recommendation for maintaining existing capabilities; 1st priority in the Pathway II recommendation for new instrumentation).

\begin{itemize}
\item DESIRE: Ability to reconstruct three dimensional particle distributions using limited numbers of satellite observations based on the data assimilation method with magnetospheric models.
\item DESIRE: IMPROVE our understanding of the relationship between the solar wind parameters and the variations of injecting/precipitating soft particles for correct predicting the space environment around GEO (2nd priority in Pathway II recommendations for modeling). Development of soft particle model in geospace including supply and loss of soft particle variations are important for the prediction.
\end{itemize}

\vskip 2mm
{\em Goal MEP-3a: Magnetospheric energetic particle measurements.}

Trapped energetic electrons ($\sim$MeV) can penetrate spacecraft shielding leaving their energy and charge embedded in devices. This energy deposit contributes to total ionizing dose and deep dielectric charging; the deposited charge can build up leading to electrostatic discharge when a breakdown voltage is reached. Dose and damage lead to catastrophic failure or progressive degradation of the performance of solid-state devices, including electronic components, solar cells, and focal planes. Electrostatic discharge can physically damage spacecraft materials, create short circuits, or manifest itself as phantom commands through electromagnetic or radio frequency interference. Trapped energetic protons and heavier nuclei (keV to GeV) cause dose and single event effects (SEE). This is especially important near the SAA.

At the onset of most geomagnetic storms the outer zone of the
radiation belts may be depleted in several hours. Subsequently, the
outer zone will some times build up to levels higher than before the
magnetospheric activity began.  A variety of processes can apply in
different conditions (see MEP1), with more extreme events some times causing prompt effects. There is no clear correlation between ring current enhancement, as measured by Dst, and the effects in terms of relativistic electron penetration deeper into the outer zone, slot, and even inner zone. In the past it was believed that the higher the solar wind speed, the more likely is the event to lead to elevated post-event electron fluxes; however this has recently been called into question. Monitoring such events at GEO is routinely undertaken. Given the very limited understanding currently available of the relative geoeffectiveness of various solar wind disturbances (see above), the best practice consensus is that excellent monitoring of the real time environment (nowcasting) is the most useful product that can be provided to satellite operators. This, and well characterized descriptions of historical events, can be used to interpret failures and improve the resilience of spacecraft design.

\begin{itemize}
\item We REQUIRE high quality measurements of the energetic particle environment in Geospace, particularly LEO and GEO.  We need to MAINTAIN the existing GOES satellite measurement, and ENSURE continued access to LANL GEO data, even if not in real time. We NEED better energy resolution at GEO. 
\item REQUIRE: maintain (and enhance) the availability of high-energy, widely distributed, multiple local time, GEO energetic particle data (5th priority in Pathway I recommendations for maintaining existing capabilities; 1st priority in Pathway II for maintaining existing capabilities): 
\item REQUIRE: We need to preserve access to data sets, current and historical, ground and space based (e.g., preservation of LANL, Ground magnetometers, HF radars), in standard archival formats, in order to enable detailed validation of physics models (1st priority in Pathway II for archival research). 
\item REQUIRE We need a standardization of instrument documentation and documentation on data processing; a minimum level of documentation required for user community should be identified. (General recommendation for Teaming, point g)
\item REQUIRE: Need to establish methodologies and metrics for validation and calibration models and of data.  (General recommendation for Teaming, point g)
\item DESIRE: hosted radiation monitor payloads at least at LEO and L1, Perhaps in the long run widespread miniaturised radiation monitors can populate all other key regions of geospace from field aligned acceleration regions to the ring current and magnetotail.
\item REQUIRE: To improve our understanding of processes leading to better understanding of the radiation belt dynamics, we need continuing support of the Van Allen Probes (RBSP) or a similar mission (2nd priority in Pathway II for maintaining existing capabilities). 
\end{itemize}

\vskip 2mm
{\em Goal MEP-4: Model of magnetospheric particle acceleration.}

It is essential for satellite operators to know if relativistic electrons can be expected to show a major increase. Thus processes of acceleration, transport, and loss of energetic electrons should be the basis for predicting the dynamics of trapped energetic particles.
However, as remarked previously, understanding geoeffectiveness is a major shortfall in current theoretical understanding. Currently understanding of Energetic Electron and Proton acceleration mechanisms is insufficient to discriminate between the effectiveness of different energization, transport and loss processes at different L-shells and local times. Quantitative assessment of the predictive properties of current and emerging models is thus essential. Hence detailed testing against fine-grained observations is needed.
\begin{itemize}
\item REQUIRE: We need to preserve access to data sets, current and historical, ground and space based (e.g., preservation of LANL and ground-based magnetometers), in standard archival formats, in order to enable detailed validation of physics models. In the future cubesats should be used in detailed campaigns for this purpose also (1st priority in Pathway II for archival research).
\item REQUIRE: Need to establish methodologies and metrics for validation and calibration models and of data (General recommendation for Teaming, point g). We suggest the establishment of an energetic particle index for MEO high-energy flux. Multipoint in-situ measurements from MEO satellites can provide the needed information about trapped particle population.
\end{itemize}

The end goal should be a fully predictive dynamical model of radiation belt particle populations, at different times, L-shells and local times, related to geomagnetic activity indices (especially Dst) and solar wind conditions (velocity, field orientation and pressure). Such a model will certainly also need to encompass an accurate description both of the current conditioning of the system and the current and predicted inputs. It should be capable of predicting extreme events with an accuracy that increases as the event realistic approaches, until realistic warnings can be given of major disruptions. Moreover, such a model (or models) will have provided a realistic basis for improved engineering designs and procedures that mitigate potential impacts.

\vskip 2mm
{\em Goal MEP-5: Description of SEP and GCR penetration into the magnetosphere.}

SEP and GCR originate from solar system or the Galaxy, respectively, and propagate to the Earth's orbit. Energetic particle can cause dose and SEE either through direct ionization events, or through the nuclear showers they create. The terrestrial magnetic field topology in general effects the penetration of these particles into the near Earth environment. While GCR particles of specific energies can enter the Earth's magnetic field, solar particles of relatively low energies cannot penetrate deep into the Earth's magnetosphere. Magnetospheric structure changes during geomagnetic disturbances can provide expansion of the region populated by energetic particles coming from heliosphere. Knowledge of the geomagnetic cutoff is thus important for satellite and aviation operators, as it controls particle fluxes at low altitudes. The latitude of the penetration region for particles of higher energies depends on their rigidity and the geomagnetic activity level. On occasion SEP penetrate deep into the atmosphere where they produce the secondary particles measured by neutron monitors. Independent on-ground measurements give important information on SEP spectra, while SEP variations are also observed by satellites in GEO and LEO. The slowly-varying GCR component is a persistent threat, while SEP are the main source of variable radiation effects on commercial aircraft, particularly at high latitudes, and represent a critical hazard to astronauts, as well as having effects on spacecraft.

Energetic particles originating from SEP and GCR can cause dose and damage in similar ways to trapped electrons, but also cause SEE. We need observations and models to tackle these factors (3rd priority in Pathway III for maintaining existing capabilities). SEP forecasting based on solar observations is of key importance.  This opportunity is described in detail in the SH sections.
\begin{itemize}
\item MAINTAIN energetic particle monitoring to specify solar particle access at different latitudes (1st and 2nd priorities in Pathway II for maintaining existing capabilities). 
\item MAINTAIN measurements onboard polar LEO: POES, MetOP, Meteor M, in proton channels 10\,MeV and above.  (5th priority in Pathway I for maintaining existing capabilities).
\item MAINTAIN supplemented monitoring at L1 (ACE) and GEO (GOES, LANL, Electro-L) to establish free space flux.
\item MAINTAIN solar observations in UV (See SH section).
\item REQUIRE: IMPROVE models of the effective vertical cut-off rigidities dependent on local (or universal) time, and geodetic coordinates (altitude, latitude and longitude), as well as on the conditions of geomagnetic disturbances described by the Kp/Dst-indices, and possibly also auroral electrojet indices, AE/AU/AL which can potentially modify access.
\end{itemize}

\subsection{Ionospheric variability}
The users of ionospheric space weather services have differential needs depending on the application areas. GNSS users in precise and safety of life applications need nowcasts and short-term forecasts of disturbed periods, so that they are aware of potential for disruption of GNSS signals, leading to higher uncertainties in measurements, and in severe cases, to loss of service. In surveying, forecasts on electron density variations up to 1-2 days and nowcasts and 1-day forecasts on scintillation would be desirable for planning purposes. For aviation aircraft and ground-systems supporting them we need both short-term alerts and forecasts of disturbed conditions with 0.5-1 day lead times (e.g., at least six hours before take-off) in order to be prepared for potential disruptions in satcom and HF signals. Operators of space radars require nowcasts and short-term forecasts of TEC in order to correct for range and bearing errors in radar tracking of space objects in LEO.

We note that statistical ionosphere models tuned with data from ground-based networks or LEO satellites can in many cases provide relatively good results for nowcasts or short-term forecasts in regional or global scales. However, with increasing lead times more comprehensive physics-based modeling with data assimilation is required.

\vskip 2mm
{\em Goal IO-1: Understand/quantify the benefits of data assimilation in ionospheric modeling.}

In the work to get longer lead times in predictions with adequate reliability space weather research can benefit from synergies with meteorology (c.f. General recommendations for Collaboration between agencies and communities, point m). Data assimilation is today the standard approach in numerical weather prediction. In space weather assimilation is utilized in some research projects, but comprehensive understanding on how assimilation could in optimal way support ionospheric SWx forecasts is still missing. 

\vskip 2mm
{\em Goal IO-1a: Upgrading MLTI models with assimilation capability}

Variability in the ionospheric electron density is controlled by
several processes that are coupled with solar activity, and with the Earth's magnetosphere and neutral atmosphere (thermosphere). In thermosphere global circulation patterns, thermal conditions and chemistry all have their impact to the evolution of ionospheric conditions. Therefore, priority in the upgrading work should be given to such models that have the capability to take into account the various coupling processes.  

We recommend the following:
\begin{itemize}
\item 	REQUIRE, IMPROVE/NEW: Advance projects in which the research
  community can investigate the opportunity to use research models for
  Magnetosphere-Lower Thermo\-sphere-Iono\-sphere-(MLTI) system in
  extensive use with assimilation capability (4th priority in Pathway
  I recommendations for modeling).  The CCMC service maintained by
  NASA is an example of a platform that could be upgraded to become a
  forum for centralized assimilation code development as a joint
  community effort. Especially in the development phases, solutions
  should be favored and further developed that have flexible
  interfaces for adopting several types of observations, e.g., solar (spectral) irradiance, ionospheric plasma convection, 2D and 3D views of electron density, and neutral wind properties (chemical composition, density, temperature, wind). 
\end{itemize}

\vskip 2mm
{\em Goal IO-1b: Defining optimal observation capabilities for ionospheric SWx services.}

For the mission to utilize both observations and models in ionospheric SWx forecasts it will be valuable to know which instruments provide the best support for improved model results. The optimal use of data requires improving/developing actual/new measurement and modeling techniques taking into account current and future customer requirements. In particular, the development of data-driven, physics-based models and associated assimilation techniques must be supported by funding agencies.  Furthermore, it has to be investigated which temporal and which spatial resolution is needed to provide the optimum cost-benefit ratio in operational use. Dedicated research projects are needed to answer these questions.

We recommend the following: 
\begin{itemize}
\item	REQUIRE, IMPROVE/NEW: During the coming five years national
  funding agencies, space agencies, EU and other stake holders should
  support research projects which investigate the observational needs
  for advanced ionospheric space weather modeling and forecasts in
  relation to existing and planned ground and space based monitoring
  capabilities. Funded projects should include sensitivity analyses
  that show how much different measurements with variable time and
  space resolutions contribute to improving our ability to
  forecast. Potentially new metrics need to be developed in order to
  quantify in an objective way the benefits from various input data
  sets (General recommendations for Teaming, point g). The studies
  should demonstrate how a complementary mix of measurements can be
  used effectively in operational forecast systems. Suggestions of optimal cost-benefit solutions should be one outcome from these research projects. 
\item	REQUIRE, MAINTAIN/IMPROVE: To support the work described above the research groups and agencies maintaining SW instrumentation should establish test beds for evaluating new observation methods, techniques and data products and for performance validation of ionospheric forecast codes at high, middle and low latitudes (General recommendation for Teaming, point f). Such test beds can be identified from the existing ground and space based ionospheric monitoring capabilities, many of which can provide long data records from versatile instrumentation. Considering ground based measurement facilities only, a test bed area should be covered by a dense network of GNSS receivers accompanied by a wide variety of other relevant instrumentation (ionosondes, ISR radar, magnetometers, riometers, HF-radars, Beacon receivers). Preference should be given to such set-ups that can provide also neutral atmospheric measurements (wind, density and temperature, in particular at low and mid-latitudes) and are supported frequently with relevant space-based observations. Suitable, already existing candidates for test-bed areas are often located in the surroundings of ISR systems. For example, the measurements of the Resolute Bay and Poker Flat ISRs in northern US and Canada, of the EISCAT ISRs in Fennoscandia and Svalbard, of the Millstone Hill ISR in US Massachusetts and of the Jicamarca radar in Peru are supported by networks of relevant instrumentation.
\item	DESIRE, MAINTAIN/IMPROVE: Science organizations involved in SW (e.g., COSPAR, ILWS, WMO, ISES) should do overarching coordination and review work in order to compose unified view from the studies described above (i.e., re-do the Roadmap task that is presented here, but with the input from the various sensitivity studies; see General recommendations for collaboration between agencies and communities). Recommendations from this work should i) help in identifying from existing assets those which deserve prioritization in maintenance ii) and support the planning work for future investments in SW instrumentation both in national and international level. 
\end{itemize}

\vskip 2mm
{\em Goal IO-2: Improved understanding on physics of ionospheric processes disturbing trans-ionospheric radio wave propagation.}

In ionospheric research there are still several topics that are relevant for SWx prediction but are lacking for general consensus on the underlying physics. Examples of such topics are processes associated with radio scintillation (in particular plasma turbulence at high latitudes and bubble formation at equatorial latitudes) and the coupling between neutral atmospheric dynamics (wind and waves) and ionospheric disturbances. Systematics in the appearance of ionospheric storms in different latitudes, local times and during the different storm phases has not been studied comprehensively yet, which complicates the efforts even for short-time global forecasts.  

\vskip 2mm
{\em Goal IO-2a: Conducting ionospheric research with enhanced observations.}

The spatial resolution of measurement stations providing ionospheric data has great variability. Some regions are populated with dense GNSS receiver networks while no ground-based receivers are available in the ocean areas. To fill this gap that is typically for ground based measurements, space based monitoring techniques such as radio occultation, and satellite altimetry data can be used.  While ionospheric plasma can be probed with a wide variety of different cost effective and robust instrumentation,  monitoring of thermospheric properties (chemical composition, neutral wind, density and temperature) is challenging. Achieving reliable continuous estimates of the global magnetospheric energy input as Joule heating or as energetic particle precipitation is also difficult. 

We recommend the following
\begin{itemize}
\item	REQUIRE: Advance combined use of ground-based, space based and airborne instrumentation in ionospheric modeling and forecasts (6th recommendation in Pathway I for maintaining existing capabilities; 2nd recommendation in Pathway III for new instrumentation).  The already existing ground-based networks (magnetometers, ionosondes, riometers, HF-radars, Beacon and GNSS receivers) should be maintained and their data dissemination systems should be homogenized and streamlined. Solar SSI measurements, Beacon transmitters, Radio occultation receivers and altimeter radars on-board LEO satellites are examples of data sources which can complement the specific view of ground-based measurements to jointly lead to a comprehensive image of the physical processes. Efforts for open data policy should be supported (MAINTAIN/IMPROVE, MAINTAIN refers to the existing instrumentation, IMPROVE refers to measurement in ocean areas and to upgrades in data dissemination and to open data policy, c.f. General recommendations for collaboration between agencies and communities, point i). In particular, efforts to get access for science community to the data archives of Real Time Kinematic (RTK) positioning networks should be supported. Commercial enterprises in several countries maintain RTK systems (dense GNSS receiver networks) that can provide information on ionospheric electron content with much better time and space resolution than typically available in scientific missions.
\item	REQUIRE, IMPR\-OVE/\-NEW: Establish and maintain collaboration with research groups and agencies which conduct thermospheric and magnetospheric research in order to search new ways to get neutral atmospheric or Joule heating data as input for ionospheric model testing and forecasts. Advancements in Fabry-Perot interferometry should be harvested, especially where advanced instrument technologies will open up daytime observations.  Opportunities provided by some forthcoming satellite missions should also be investigated (e.g., the NASA missions ICON and GOLD for thermospheric measurements, the US-Taiwanese atmosphere-ionosphere mission COSMIC II and the ESA mission Swarm for MLTI research). 
\item	DESIRE: Advance the search of new data sources to support SWx forecasts. Examples of new potential sources are GNSS reflectometry, SAR imaging, and multi-satellite radio occultation sounding by COSMIC II satellites.
\end{itemize}

\vskip 2mm
{\em Goal IO-2b: Enhancing system level modeling of the ionosphere.}

Models available today provide a mosaic on processes from different parts of the system (with separate pieces for high latitudes and low latitudes, for magnetosphere-ionosphere coupling and ionosphere-thermo\-sphere coupling, for global processes and micro-scale processes etc.) but knowledge on the linkage between the different pieces is missing. Today ionospheric models are typically coupled with neutral atmosphere models (e.g., MSIS) that describe the climatology quite well, but they fail to describe the processes in lower atmosphere which affect ionospheric conditions (tides, gravity waves, planetary waves).  For this reason, the models are not able to reproduce properly e.g., the background and seeding conditions for bubble formation. It is important to have a model where ionosphere is properly coupled with such neutral atmosphere and magnetosphere models that are capable to describe properly at least those processes that control ionospheric electron content.

We recommend the following:
\begin{itemize}
\item	REQUIRE, IMPROVE/NEW: Advance efforts for three-dimensional imaging and modeling of ionospheric phenomena with the goal to gain better understanding on non-equilibrium plasma processes and on cause and consequence relationships between processes in different scale sizes (particularly between regional and microscales in order to gain better understanding on processes causing scintillation).
\item DESIRE, NEW: Advance efforts for the development of a framework where the whole atmosphere (up to $\sim$400\,km altitudes )is treated as one system. That should give a better means to understand vertical coupling, and also stimulate stronger links between solar-terrestrial research community and the meteorological community.
\end{itemize}
 
\section{Concepts for highest-priority instrumentation}

\subsection{Binocular vision for the corona to quantify incoming CMEs}

{\em Rationale:} Knowledge of the magnetic structure of the solar wind, and in particular that of coronal mass ejections, that will interact with the magnetospheric field that in turn drives the underlying ITM is needed at least a day prior to reaching geospace, i.e., well before reaching the solar-wind sentinel(s) positioned a million miles (or about an hour of wind travel time) upstream of Earth at the Sun-Earth L1 point. This knowledge, in particular of the magnetic direction and strength of the leading edge of coronal mass ejections [CME]s) can be obtained by forward modeling an observed solar eruption through the embedding corona and inner heliosphere, provided the magnetic structure of the erupting structure is known. Deriving the magnetic configuration of an erupting solar active region based on surface (vector-) field measurements alone yields ambiguous results at best; these are insufficient for the purpose of MHD modeling of CMEs. New field modeling methods have been developed that can utilize the coronal loop geometry to constrain the model field, particularly when 3D information on the corona is available (e.g., Malanushenko et al., 2014). Binocular imaging of the active-region corona at moderate spatio-temporal resolution enables the 3D mapping of the solar active-region field structure prior to, and subsequent to, CMEs, thereby providing information on the erupted flux-rope structure. Combined with full-disk coronal imaging and ¬- if feasible within the mission parameters - with coronagraphic imaging provides valuable information on the direction taken by the nascent CME en route to the inner heliosphere through the high corona.

{\em Objectives:} Obtain EUV images of solar active regions, of the solar global corona, and (if not provided by other instrumentation) of the inner heliosphere from a perspective off the Sun-Earth line, to complement EUV images from existing instruments on the Sun-Earth line such as SDO/AIA or, if not available, by a second identically-equipped spacecraft.

{\em Key requirements:} EUV images of active regions at (at least) two wavelengths (characteristic of 1-2MK and 2-3MK plasma) at $\sim$1.5-arcsec resolution and 1-min. cadence, and full-disk observations at $\sim$3\,arcsec resolution and 30-sec. cadence with low noise to detect signals out to at least 1.5 solar radii, observed from a perspective between 5 and 15 degrees from the complementary second imager. If not provided by other resource: coronagraphic images at 2-min. cadence from about 1.5 to 5 solar radii.

{\em Implementation concept:} A single two-channel full-disk EUV imager a 1 arcsec resolution with on-board data processing to (selected) high-resolution active region images and binned full-coronal images (to reduce overall telemetry rates). A compact coronagraph if needed. In a slightly elliptical solar-centric (``horseshoe'') orbit drifting away from Earth at a rate of no more than about two degrees per year for the first three years after reaching 5-degrees of separation from the Sun-Earth line. If combined with SDO/AIA a single spacecraft suffices; if a standalone mission, two similar spacecraft are needed to drift apart by the required separation for a period of at least one year. Each spacecraft could be scoped within, e.g., the NASA SMall EXplorer (SMEX) envelope.

{\em Status:} Proposed in this Roadmap. Whereas the STEREO mission has given us a temporary glimpse of stereoscopic coronal EUV imaging, the field modeling capabilities at the time did not yet exist and the image resolution was insufficient on the STEREO and SoHO spacecraft.  

{\em Supporting observations:} New observational technologies can support the goals of determining the 3D structure of the magnetic field in source regions of solar activity, particularly when combined with the above direct observations and modeling techniques. This includes radio observations such as, for example, with proposed instrumentation for FASR, etc., particularly when multi-frequency ultra-wideband radio array(s) are available over frequency range of 50 MHz to 20 GHz, with full Sun images at all wavelengths taken once per second (partly completed in China, US development efforts continuing). 

\subsection{3D mapping of solar field involved in eruptions}

{\em Rationale:} A key science goal in this roadmap is to determine the origins of the Sun's activity to help predict variations in the space environment. Knowing the low-lying twisted filament-like field configurations and their embedding field in the deep interiors of unstable active regions before and after eruptions is key to determining what propagates towards Earth to drive space weather, which, in turn, is needed to forecast the dynamics and coupling of the Earth's magnetosphere-ionosphere-atmosphere system driven by the incoming CMEs. Magnetic stresses involved in coronal mass ejections cannot be observed directly, but require modeling based on observations of the vector magnetic field at and above the solar surface, guided by, and compared to, observations of the solar atmosphere in which structures from chromospheric to coronal temperatures are carriers of the electrical currents that reflect the system's free energy converted to power eruptions. In the context of Pathway 1, the primary goal is to observe electrical currents threading the solar surface, and observing the details of the low-lying configurations known as filaments and their embedding flux ropes before, during, and after eruptions to quantify the 3D field ejected into the heliosphere. In the context of Pathway 3, the same instrumentation provides observations of the small-scale processes involved in the triggering of flares and eruptions, needed for short-term SEP forecasts, and for CMEs forecasts of more than 2-4 days and SEP all-clear forecasts in the coming hours.

{\em Objectives:} Obtain vector-magnetic measurements of active regions at the solar surface and within the chromosphere to measure electrical currents. Image the solar atmosphere from 10,000K up to at least 3MK at matching resolutions to observe all field structures that may carry electrical currents. Provide observations before, during, and after eruptions to derive CME field structure to drive heliospheric models, to study how the nascent CME is restructured as it propagates through the active-region magnetic field.

{\em Key requirements:} High-resolution imaging (at matching resolutions of 0.2 arcsec or better) is needed throughout active-region atmospheres, spanning the entire active region footprint, with observations at temperatures characteristic of photosphere, chromosphere, and corona. Spectro-polarimetric observations for photospheric and chromospheric magnetic field measurements. Imaging cadence of approximately 10 seconds, or better. (Near-)Continuous solar viewing.

{\em Implementation concept:} Geo-synchronous or low-Earth orbiter in high-inclination orbit with UV-optical telescope with polarimetric imaging capabilities, enabling photospheric and chromospheric imaging and polarimetry. Soft X-ray and EUV imagers. Substantial ground-based network to enable large effective telemetry rates, and/or onboard image selection from large memory.

{\em Status:} Considered as a multi-agency mission between JAXA, ESA, and NASA to share cost and expertise, in support of international space-weather research.

\subsection{Strong GICs driven by rapid reconfigurations of the magnetotail}

{\em Rationale:} The processes of energy input from the solar wind and storage in the magnetotail are now reasonably well-understood, with further recent discoveries from the THEMIS and Cluster missions beginning to reveal the tantalizing physics of the development and penetration of Earthward propagating bursty bulk flows and localized dipolarizing flux tubes which transport flux and plasma Earthwards. However, the nearer Earth dynamics which couple these flows to the inner edge of the plasmasheet and how exactly the flow braking region couples to the ionosphere and produces large field-aligned currents and hence GICs are poorly understood. A two-constellation satellite mission architecture is proposed. The first will reveal the key plasma physical processes associated with plasma instabilities and flow braking at the inner edge of the plasmasheet, in the transition region from dipolar to tail-like magnetic fields; the second will reveal the nature of magnetosphere-ionosphere (M-I) coupling in the auroral acceleration region on field lines conjugate to the inner edge of the plasma sheet. From the GIC perspective, the processes that control the rate of energy transport and the actual partition between competing routes of dissipation in the coupled M-I system remain insufficiently understood. In terms of space weather impacts, the conditions leading to and the physical processes responsible for enabling large field aligned currents to reach the ionosphere and drive large GICs are not known.

{\em Objective:} Determine the M-I processes controlling the destabilisation of the near-Earth magnetotail, which will lead to the establishment of large field-aligned currents resulting in extreme GICs. 

{\em Key requirements:} To meet this objective we need the flight of two coordinated satellite constellations in the inner edge of the plasmasheet, which marks the transition region between tail-like and dipole fields, and on the conjugate auroral field lines below. It should be noted that the challenge of these constellation missions lies in the population of magnetospheric key regions and not in particularly fancy or expensive instrumentation.

{\sc Transition Region Explorers:} Three-dimensional plasma and electrodynamic fields (E (at least AC) and B) in the transition region between dipole-like and tail-like fields, which is the originator for large field-aligned currents. Coverage is required from close to geosynchronous orbit to around 10-12 Re. This could be accomplished by at least one classical well-instrumented spinning spacecraft with magnetic field, electric field, plasma measurements for pitch angle resolved electrons and ions in the energy range from around 10's eV (as low as possible without ASPOC) to several hundred keV, including species resolution. This likely requires a standard suite of particle instruments including an electrostatic analyser, solid state detector, and ion composition spectrometer. Potentially this satellite should carry sufficient fuel to change apogee altitude between 8-12\,$R_{\rm E}$ during the course of the mission. 

This should be supplemented by a swarm of around 4 smaller spacecraft approximately 1\,$R_{\rm E}$ from the mother, providing coverage in the azimuthal and radial directions. The smaller daughters could carry a more limited basic plasma payload of a magnetometer, miniaturised electrostatic analyser, and  Langmuir probes for total (including cold) plasma density and temperature.

{\sc Field-Aligned Current Explorers:} Multi-point plasma and electrodynamic fields in the auroral acceleration region in order to determine the dynamical coupling between the magnetosphere and the ionosphere including the partitioning and exchange of energy between currents, waves and particles which are believed to act as a gate for the ability of the tail to drive FAC through to closure and eventual energy dissipation in the ionosphere. Operational altitudes should encompass the range of around 4000\,km to 1\,Re, utilising conjugate measurements from different altitudes on the same field line, and multiple along track satellites providing a capability to resolve spatio-temporal ambiguity related to the filamentary nature of FAC and examine and distinguish between the dynamics of Alfv{\'e}nic and inverted-V auroral acceleration processes. Optionally an additional single 3-axis stabilized satellite for in-situ auroral imaging.

Baseline of two or three spinning spacecraft providing electric (via wire booms) and magnetic fields and waves monitoring, as well as plasma electrons and ions from energies of around 10\,eV to 30\,keV, likely from an electrostatic analyzer. Options to add higher energy coverage from an solid state detector, to resolve the populations up to some 50\,keV should also be studied. Again, Langmuir probes would provide significant information about very low energy populations below the energy range of the particle instrument. Payload for the 3-axis stabilized satellite for in-situ auroral imaging is TBD.

{\sc Auroral Imaging and Supporting Ground Networks and LEO Satellite Constellations:} The constellation missions should be complemented by conjugate auroral imaging from the ground, as well as supporting networks of ground-based magnetometers, HF radars, riometers etc, to aid the identification of the onset location and the resolution of the spatio-temporal ambiguity of the processes leading to large dB/dt. Existing or newly provided constellations of low-Earth orbiting satellites which can additionally monitor the precipitating electrons as a measure of ionospheric conductivity changes will provide valuable complementary measurements; global measurements of the background large scale FAC distributions, such as available from AMPERE, provide the capability to identify onset locations with respect to the nightside convection pattern.

{\sc Incoming Tail Flows and Upstream Solar Wind Monitor:} Measurements of incoming flows in the central plasmasheet of the more distant tail are also required to assess incoming flows in the same meridian. These could be potentially be provided by pre-existing assets, such as Geotail, Cluster, THEMIS, or perhaps MMS in an extended mission phase. An upstream solar wind monitor is of course required as always

{\em Implementation concept:}
The challenge for mission implementation is not in the instrumentation, which is readily available and should thus not be a cost driver. Rather, the challenge lies in the positioning of a sufficient number of spacecraft at the two key locations in space, and thus in both the possible orbit configuration and the number of spacecraft. Likely this requires detailed future study of at least the potential orbits which can launch and deliver the satellites into the appropriate operative orbits at modest cost. We recommend that the formulation of an international study with representatives of national space agencies is considered, perhaps in the context of ILWS. 

{\em Status:} Two satellite-constellation concept proposed in this roadmap. 

{\em Supporting observations:} by existing mid-tail and upstream solar wind satellites, as well as existing complementary multi-instrument ground-based networks and existing LEO satellite constellations. 

\subsection{Coordinated networks for geomagnetic and ionospheric variability}

{\em Rationale:} Our understanding of space weather impacts on the upper atmosphere is crucially dependent on measurements from rich networks of ground-based instruments, including (a) magnetometers to observe how electric currents in the ionosphere are modified by space weather, plus (b) a wide variety of radar and radio techniques  to monitor changes in the density, motion and temperatures of ionospheric plasmas, as well as (c) optical techniques to measure thermospheric winds and temperatures. These data sources are all key inputs into the development of improved models of the atmosphere and its response to space weather. This will become even more important in future as we focus on assimilative approaches to modeling. As in meteorology, these approaches will advance our science (e.g., through use of reanalysis techniques to reveal new systematic features in the data) and will also provide an efficient practical basis for future applications of our science. These assimilative approaches are significantly enhanced by the availability of diverse datasets from spatially dense networks as these data then provide strong constraints on the assimilation.  We need to promote these ground-based networks as a global system for scientific progress on space weather, so that each individual (and often independently funded) instrument provider (and responsible funding agency) sees how their contribution fits into the wider picture, i.e. that a local contribution builds and sustains local access to a global system.

{\em Objectives:} To make a step change in the international co-ordination, and delivery, of ground-based space weather observing systems to optimize our ability to observe space weather processes. This must include increased engagement with modelers, operators of space-based sensors, and other consumers of space weather data, in order to ensure the optimum interaction between the collected data and state-of-the-art international models, and the synthesis of these data and model results into operational tools whose outputs can be made available to the applications and technology community. This would be a major advance on the current situation where ground-based instruments are mostly established and supported by individual national programs, with only a few projects formally constituted as multi-national programs (EISCAT being the notable example). There are a good number of projects that operate internationally through working level agreements in the scientific community (e.g., SuperDARN) but experience shows that these are difficult to sustain in modern conditions; they rather reflect an older (make-shift) way of working that dates as far back as the IGY in 1957/58. Given modern approaches to funding and governance, these observing systems need a more formal international structure that can give them the continuous support and stimulus that they need to deliver their full scientific potential for a future operative space weather system.

{\em Key requirements:} Maintenance and extension of the SuperDARN network to measure electric fields in the high- and mid-latitude regions at least in the north hemisphere, preferably both (these measurements are a crucial factor in modeling the ionospheric, magnetospheric and radiation belt response to space weather); high-resolution volumetric measurements of ionospheric properties by incoherent scatter radars (ISR) at several locations (to resolve the detailed ion chemistry and plasma physics at work in the ionosphere); rich networks of magnetometers, GNSS receivers, ionosondes and riometers to provide regional and global maps of key ionospheric properties including ionospheric current systems (auroral, equator and Sq), total electron content and ionospheric scintillation, ionospheric critical frequencies and layer heights and D region absorption; improved operation of Fabry-Perot interferometers (FPI) to enable daytime as well as nighttime measurements of thermospheric temperature winds. It is highly recommended to ensure a concentration of networked instruments, including FPI, around major facilities such as incoherent scatter radars, as these provide vital context for that technique. These concentrations of instruments will allow us to use their locations as a scientific test bed where we can explore the detailed response of the ionosphere to space weather.

{\em Implementation concept:} Establish a global program at inter-agency level for coordination of space weather observing systems, perhaps similar to coordination of space exploration activities. The involvement of agencies is crucial as it is vital to involve funding bodies to develop a sustainable system that is subject to periodic review and where there is a proper emphasis on, and awareness of, the global nature of the program. The program would establish working groups with strong expert membership to carry out its technical tasks, including detailed review of measurement requirements, review of advances in instrument technology, coordination with space-based measurements, recommendations on standards, exploitation of secondary data sources especially radio astronomy (much of their ``noise'' is actually ionospheric signals that we wish to exploit), etc. The crucial aspect of this program is to build a framework where individual agencies can see that a modest contribution enables global science and, in particular, is the logical way to enable first-class participation by the scientific community that they support.

{\em Status:} Proposed in this Roadmap.

\subsection{Mapping the global solar field}

{\em Rationale:}  The global solar magnetic field extends out into the heliosphere. It defines the structure of the heliosphere, including the position of the heliospheric current sheet and the regions of fast and slow solar wind, and plays a key role in space weather at Earth: (1) The interaction of CMEs with the ambient field impacts their geoeffectiveness.  (2) The connection of the heliospheric magnetic field to CME-related shocks and impulsive solar flares determines where solar energetic particles propagate.  (3) The partitioning of the solar wind into fast and slow streams is responsible for recurrent geomagnetic activity.  The Sun's surface magnetic field is a vital ingredient to any predictive model of the global magnetic field, as it is used to derive boundary conditions.  Global magnetic field models (both the simpler potential-field source-surface (PFSS) models, and the more sophisticated MHD models) have shown significant success in describing coronal and heliospheric structure.  These models typically use magnetic maps of the photospheric magnetic field built up over a solar rotation, available from a ground-based and space-based solar observatories.  Two well-known problems arise from the use of these ``synoptic'' maps.  First, the maps contain data that is as much as 27 days old.  The Sun's magnetic flux is always evolving, and these changes in the flux affect coronal and heliospheric structure.  Second, the line-of-sight (LOS) field at the Sun's poles is poorly observed, and the polar fields in these maps are filled with a variety of interpolation/extrapolation techniques.  Unfortunately, these observational gaps can strongly influence the solution for the global magnetic field.  In particular, poorly or unobserved active regions at the limbs (as viewed from Earth) as well as inaccurate polar field estimates can introduce unacceptable errors in the field on the Earth-facing side of the Sun.

{\em Objective:} Model the evolving global solar magnetic field.  This requires the near simultaneous observation of the Sun's magnetic field over a larger portion of the sun's surface than is available from the Earth view alone.  Obtaining photospheric magnetograms off of the Sun-Earth line off of the east limb (portion of the Sun with the oldest observations as viewed from Earth), to complement magnetograms obtained along the Sun-Earth line by SDO and ground-based observatories, is the most crucial component.  Obtaining magnetograms of the Sun's polar fields over a few years is required to understand the evolution of the Sun's polar magnetic flux.

{\em Key requirements:} Ideally, several spacecraft would observe the Sun's magnetic field continuously, including the polar fields, but such a plan is unlikely to be economically feasible in the foreseeable future.   The processes by which the magnetic flux on the Sun evolves have been studied for many years, and resulted in the construction of flux transport models capable of predicting the evolution of the field.  The incorporation of magnetograms away from the Sun-Earth line would be used to augment existing, Earth-view magnetograms to capture a significantly larger portion of the Sun's evolving flux.  LOS magnetograms with MDI resolution and approximate cadence are likely to be adequate, although vector magnetograms with HMI resolution and cadence are desirable.  Flux transport models also predict the evolution of the Sun's polar fields, but are largely uncalibrated there.  Observing polar flux evolution with MDI resolution over a few years would significantly constrain these models. EUV or X-ray imaging (STEREO cadence/resolution) to capture coronal holes and evolving structures for model validation from all of these views are desirable.  

{\em Implementation concept:} Primary instrument:  Full-disk magnetograph with MDI-like spatial resolution and hourly time resolution in an ecliptic orbiting spacecraft reaching at least 45 degrees off the Sun-Earth line.  Orbits going beyond this point are desirable.  Given such a spacecraft, a heliospheric imager would augment the goals of (1) and (4) by imaging earth-directed CMEs.  If feasible, X-ray or EUV imaging (1 channel) at STEREO spatial resolution and cadence are desirable.  A separate, high-latitude (30 degrees or more above the ecliptic) spacecraft mission of a few years with the same instrumentation is desirable.

{\em Status:}  Images from Solar Orbiter may be sufficient to provide testing of concept of far side imaging, but are not adequate to fulfill the goal of more continuous monitoring of a larger portion of the Sun's magnetic flux.  Solar Orbiter may partially fulfill high latitude mission requirements at the latter stage of the mission.

{\em Supporting observations:} Direct imaging data of the solar atmosphere (such as possible with the STEREO spacecraft) or indirect information derived from far-side helioseismology (such as with SDO/HMI and GONG) provide useful constraints, but are no substitute for direct magnetography because these methods do not provide adequate information on the magnetic field. 

\subsection{Determination of the foundation of the heliospheric field}

{\em Rationale:}  The global solar magnetic field plays a crucial role in space weather at Earth.  It influences the internal magnetic structure of interplanetary coronal mass ejections; the connectivity of the magnetic field determines where solar energetic particles propagate, and structure of the field determines whether fast solar wind streams will cross the Earth's location.   An accurate representation of the time-evolving global solar coronal magnetic field is a required input to models of prediction, eruption and propagation of CMEs through the solar wind.

{\em Objectives:} Determine and obtain a critical set of multi-wavelength coronal magnetometric observables for constraining the global magnetic field.  Develop methods for incorporating these observations into global MHD models of the solar corona.  

{\em Key requirements:}  Testbeds of synthetic polarimetric measurements at multiple wavelengths to provide diagnostics related to the Zeeman and Hanle (saturated and unsaturated) effects and coronal seismology.  Techniques for efficiently modifying global MHD models of the solar corona to match data, synthetic or observed.  Ultimately, full-sun synoptic observations sufficient to enable a data-assimilative, real-time updated representation of the global coronal magnetic field.

{\em Implementation concept:} Different wavelengths diagnose different aspects of the solar coronal magnetic field - strong vs. weak field, disk vs. limb, eruptive vs. non-eruptive - and are weighted differently helping to remove line-of-sight ambiguity.  By utilizing testbeds of synthetic data at all wavelengths (from radio to extreme ultraviolet), effective measurement and optimization strategies can be developed which will set priorities and for future observational development.  

{\em Status:}  DKIST will provide opportunities for polarimetry and testing of observational techniques with high resolution and sensitivity but in a small (5 arc minute) field of view.  Proposed new observations include large (1.5 meter) ground-based coronagraph(s) with narrow-band filter polarimeter and spectropolarimeter to observe the full Sun corona at the limb in the visible and infrared (currently undergoing engineering design and preliminary design review as US-China collaboration).  Space-based missions would provide better duty cycle and continuity of measurements than ground-based, and also would allow measurement at short wavelengths otherwise blocked by the Earth's atmosphere.  Mission concepts have been proposed (ESA) with instruments including spectropolarimetric coronagraphs in the EUV and IR for off-limb observations, and spectropolarimeters to observe the solar disk at heights from the corona down into the chromosphere.

\subsection{Auroral imaging to map magnetospheric activity and to study coupling}

{\em Rationale:} The response of magnetosphere to solar wind driving depends on the previous state of magnetosphere. Similar sequences in energy, momentum and mass transfer from the solar wind to magnetosphere can lead in some cases to events of sudden explosive energy release while in other cases the dissipation takes place as a slow semi-steady process. Comprehensive understanding on the factors that control the appearance of the different dissipation modes is still lacking, but obviously global monitoring of the magnetospheric state and system level approach in the data analysis would be essential to solve this puzzle. Continuous space-based imaging of the auroral oval would contribute to this kind of research in several ways. The size of polar cap gives valuable information about the amount of energy stored in the magnetic field of magnetotail lobes.  Comparison of the brightness of oval at different UV wavelengths yields an estimate about the energy flux and average energy of the particles, which precipitate from the magnetosphere to the ionosphere. These estimates are not as accurate as those from particle instruments onboard LEO satellites, but the additional value comes from the capability to observe all sectors of the oval simultaneously and continuously. Such view is useful especially in the cases where the magnetosphere is prone to several subsequent activations in the solar wind.  The shape and size of the oval and intensity variations in its different sectors enable simultaneous monitoring of, e.g., nightside magnetospheric recovery from previous activity, while new energy already enters the system from a new event of dayside reconnection.

{\em Objectives:} To achieve continuously global UV-images to follow the morphology and dynamics of the auroral oval, at least in the Northern hemisphere, but occasionally also in the southern hemisphere. Imager data combined with ground-based SuperDARN and SuperMAG networks allows solving the ionospheric Ohm's law globally, which yields a picture of electric field, auroral currents and conductances with good accuracy and sufficient spatial resolution.  This would mean a leap forward in our attempts to understand M-I coupling, particularly the ways how ionospheric conditions control the linkage to the magnetosphere by, e.g., by field-aligned currents.

{\em Key requirements:} An imager which can observe the electron auroral emissions in the Lyman-Birge-Hopfield Nitrogen waveband, discriminating between the LBH-long and LBH-short bands. This set-up provides information about precipitating electrons in the range 1-20\,keV.  For solving the ionospheric electrodynamics (by estimation of ionospheric conductances) also an imager for Bremsstrahlung radiation (more energetic electrons, 20-150\,keV) would be necessary. For proton precipitation the mission would need an imager capable to capture Doppler-shifted Lyman-α emission from charge-exchanging precipitating protons. Time resolution of the images should be better than 60 sec and spatial resolution should reach $\sim$50\,km (at perigee) 

{\em Implementation:} The objective of continuous monitoring can be achieved with a constellation of two identically-instrumented spacecraft in identical highly-elliptical polar orbits (apogees close to 7 RE above the northern pole and perigees near 2 RE.). The orbits of the two spacecraft should be phased so that one spacecraft is at perigee while the other is at apogee and the imagers onboard should be able to observe the auroras from both positions.

{\em Status:} Undergoing more detailed definition within international science teams

{\em Supporting observations:} Global ground-based networks, existing LEO and GEO satellites, existing mid-tail constellation missions, and - as always - upstream solar wind monitor.

\subsection{Observation-based radiation environment modeling}

{\em Rationale:} The radiation belts are key domains in the Earth's magnetosphere, which cause spacecraft anomalies. It is essential for satellite operators to know if relativistic electrons can be expected to show a major increase, which is related to the high risk of spacecraft anomalies. Thus processes of acceleration, transport, and loss of energetic electrons should be the basis for predicting the dynamics of trapped energetic particles.
However understanding radiation-belt dynamics is a major shortfall in current theoretical understanding. Currently understanding of Energetic Electron and Proton acceleration mechanisms is insufficient to discriminate between the effectiveness of different energisation, transport and loss processes at different L-shells and local times.
 
Trapped energetic particles in the inner magnetosphere form radiation belts. The inner proton belt is stable, but the outer radiation belt comprises a highly variable population of relativistic electrons. Solar wind changes produce changes to trapped particle transport, acceleration, and loss. However the pre-existing magnetospheric state is also a critical factor. For example efficient production mechanisms appear to need a seed population of energetic electrons. Relativistic electrons enhancements are an important space weather factor with a strong influence on satellite electronics. Around 50\%\ of magnetic storms are followed by a corresponding enhancement of relativistic electron fluxes.

Storm-time magnetospheric convection intensification, local particle acceleration due to substorm activity, resonant wave-particle interaction are the main fundamental processes that cause particle fluxes energisation and loss. However, the details of the mechanisms that may be able to contribute to these processes remain a subject for active research. While progress has been made in the theoretical understanding of these competing processes, there is as yet no clear consensus on which of these will be significant in particular situations, and no real predictive methods which can give precise fluxes at different local and universal times. There is therefore a pressing need to confront theoretical models with detailed measurements, in order to resolve these shortfalls. Quantitative assessment of the predictive properties of current and emerging models is thus essential. Hence detailed testing against fine-grained observations is needed.

At the onset of most geomagnetic storms the outer zone of the
radiation belts may be depleted in several hours. Subsequently, the
outer zone will some times build up to levels higher than before the
magnetospheric activity began.  A variety of processes can apply in
different conditions (see MEP1), with more extreme events some times causing prompt effects. There is no clear correlation between ring current enhancement, as measured by Dst, and the effects in terms of relativistic electron penetration deeper into the outer zone, slot, and even inner zone. In the past it was believed that the higher the solar wind speed, the more likely is the event to lead to elevated post-event electron fluxes; however this has recently been called into question. Monitoring such events at GEO is routinely undertaken. Given the very limited understanding of radiation-belt dynamics currently available (see above), the current consensus is that excellent monitoring of the real time environment (nowcasting) is the most useful product that can be provided to satellite operators. This, and well characterized descriptions of historical events, can be used to interpret failures and improve the resilience of spacecraft design.

{\em Objective:} In-situ multipoint measurements of relativistic and sub-relativistic electrons in the inner magnetosphere, at least at GEO to control radiation belt particle populations, at different times, L-shells and local times.  Continuous control of the geomagnetic activity indices (especially Dst) and solar wind conditions (velocity, field orientation and pressure) to predict the possible relativistic electron fluxes variations. Realistic models capable of predicting the radiation belts dynamics can be constructed after detailed testing against fine-grained observations. Such a model (or models) will have provided a realistic basis for improved engineering designs and procedures, which mitigate potential impacts.

{\em Key requirements:} Multi-point in-situ observations and real-time analysis of energetic particle fluxes (mostly, 0.1 - 10\,MeV for electrons) in the inner magnetosphere, of the current geomagnetic conditions (mostly, Dst and AL indices) and solar wind / IMF conditions at L1 point. 
	
{\em Implementation concepts:} To realize our objectives, maintaining the current observation facilities related to radiation belt dynamics (particle and electromagnetic field measurements in the inner magnetosphere is essential). To fill the gap of observational data, hosting radiation monitor payloads (and/or cubesat missions) at LEO, MEO, GEO, and constructing new ground-based observatories for sparsely covered region is also encouraged. Models (empirical, theoretical, numerical) which are based on these observational data will improve our understanding of transport, acceleration, and loss processes in the radiation belt. Establishing methodologies and metrics for validation and calibration of models and data is another important issue.

{\em Status:} Continuing observations of energetic particles from LANL, GOES, ELECTRO-L, POES, Meteor-M. Geomagnetic indices from WDCs. Solar wind parameters from L1 (ACE).

{\em Supporting observations:} The current constellation of Cluster, Van Allen Probes, THEMIS, with the upcoming MMS, along with data from the geostationary satellites, and ground-based observation networks by magnetometers and HF radars represent a perfect opportunity to achieve a major step forward. Theoretical studies and modeling is underway. What is needed is investment in a few minor expansions and in particular international coordination to push the program forward.

\subsection{Solar energetic particles in the inner heliosphere}

{\em Rationale:} SEPs present a major hazard to space-based assets.  The strong increases in fluxes of high-energy protons, alpha particles and heavier ions can cause or contribute to a number of effects, including tissue damage for astronauts, increases in radiation doses, single event effects (SEEs) in micro-electronics, solar cell degradation, radiation damage in science instruments and interference with operations. In addition, large SEP events can also affect aviation, through increases in radiation dose levels and interference with avionics through SEEs.

The production of SEPs is associated with large flares and fast CMEs
in the low corona, typically originating from complex active regions.
The prompt response can arrive at Earth in less than an hour from the
onset of eruption, and some times in a few minutes after that onset in the case of a well-connected, relativistic particle event.   This is followed by a longer lasting, often rising, flux of particles originating at the CME shock as it propagates through the heliosphere, and a short sharp very high flux of energetic storm particles (ESPs) as the CME shock passes the point of measurement.  While warning of events in progress is certainly important, many users require significantly longer warning, e.g., 24 hours (e.g., for (e.g., all-clear periods for EVAs for astronauts).

At the onset of major (M or X) class flares and fast CMEs, relativistic SEPs (GeV protons) can arrive minutes after the start of the event in the case of a well-connected event. Less energetic protons will arrive minutes to hours later depending on the energy but also on the propagation in the interplanetary medium. Recent multi-point observations of SEP events off the Sun-Earth line (STEREO observations) furthermore show that prompt energetic particles have access to a wide longitudinal extent for some events. Recent studies also show that for a large proportion of SEP events, the prompt energetic particles do not propagate along the normal Parker spiral but in the magnetic field of a pre-existing CME.  All this represents an additional difficulty for the forecasting of the arrival of SEPs at Earth, since in addition to the conditions of diffusion in the medium, this affects the delay between production of energetic particles and arrival at the earth. On the other hand, shocks produced by fast CMEs start to be identified in coronagraph images.

Energetic storm particles (ESPs) can also reach $>$500\,MeV and also pose major space weather hazard.  They represent a ``delayed'' radiation hazard and the ESP event forecast is essentially a shock arrival forecast. A good indicator of an ESP event is thus a radio-loud shock, i.e. a shock that produces type II radio burst near the Sun and the interplanetary medium.  Shock travel time ranges from about 18 hours to a few days, so there is a good chance of making prediction of ESP events. However, the main question is how the intensity, duration, and arrival time depends on the SEP event near the Sun, presence of a type II radio burst in the near-Sun interplanetary medium, and the source location of the CME that drives the shock. ESP events are generally of low energy, so they are expected to precipitate in the polar region. Thus, they are hazardous to satellites in polar orbits. They lead to radio fadeouts in the polar region and may cause communication problems for airplanes in polar routes.

{\em Objectives:} In addition to the objectives for solar observations of active regions and of the solar global corona and of the inner heliosphere from a perspective off the Sun-Earth line (7.1) obtain multi-point in-situ observations of SEPs off the Sun-Earth line and possibly closer than the L1 distance.  

{\em Key requirements:} Multi-point in situ observations of SEPs off the Sun-Earth line and possibly closer than the L1 distance. 

{\em Implementation concept:} Suite of sensors measuring electrons, protons, and ions from helium to iron in the keV to over 100\,MeV per nucleon range.

{\em Status:} Continuing observations of energetic particles from STEREO, ACE, and other platforms. Upcoming Solar Orbiter and Solar Probe Plus missions will provide key measurements of SEPs close to the acceleration region.  Different concepts of missions at L5 proposed (NASA solar and space physics road map, ESA/CAS small mission opportunity, ...) Development of solar sails would open up new research opportunities for energetic particle science and monitoring.

{\em Supporting observations:} Continuing operations of ground-level neutron monitors and near real-time access to data of these observatories. They provide information on the arrival of the most energetic protons from flares.   Radio observations (ground-based and satellite) of electron beams and shocks propagating in the interplanetary medium. Also: continued analysis of radio-nuclide data in biosphere, ice cores, and in terrestrial and lunar rocks, as these provide information on pre-historical extreme events that cannot otherwise be obtained. Use of multiple data sources and radio-nuclides helps to provide some information on particle energy spectra that are needed to better constrain fluences and to specify environmental conditions.
 
\section{Acronyms}

{\parindent=0pt
- ACE	Advanced Composition Observatory

- AIA       SDO's Atmospheric Imaging Assembly

- AMPERE   Active Magnetosphere and Planetary Electrodynamics Response Experiment

- AOGS   Asia Oceanea Geosciences Society

- ARTEMIS Acceleration, Reconnection, Turbulence and Electrodynamics of the Moon's Interaction with the Sun

- ASPOC  Active Spacecraft Potential Control Experiment (on board Cluster)

- AU        astronomical unit (Sun-Earth distance)

- BBF	bursty bulk flow

- CAS      Chinese Academy of Sciences

- CCMC	Community Coordinated Modeling Center

- CEDAR  coupling, energetics, and dynamics of atmospheric regions program

- CIP	critical infrastructure protection

- CME	coronal mass ejection

- COPUOS  UN Committee on the Peaceful Uses of Outer Space

- COSMIC  Constellation Observing Sysgtem for Meteorology, Ionosphere, and Climage

- COSMO  Coronal Solar Magnetism Observatory (in MLSO) 

- COSPAR  ICSU's Committee on Space Research

- CSRH    Chinese Spectral Radio Heliograph

- DHS	Department of Homeland Security

- DKIST   Daniel K. Inouye Solar Telescope

- DMSP	defense meteorological satellite program

- DSCOVR deep-space climate observatory

- Dst	disturbance storm time index

- EGNOS  European Geostationary Navigation Overlay Service

- EGU      European Geophysical Union

- EISCAT  European Incoherent Scatter Scientific Association

- ERG      Exploration and energization of Radiation in Geospace

- ESA	European Space Agency

- ESP       energetic storm particle(s)

- ESPAS	European strategy and policy analysis system

- EU	European Union

- EUV	extreme ultra violet

- eV        electron-Volt

- FAC	field-aligned current

- FASR    Frequency-Agile Solar Radio telescope

- FPI       Fabry-Perot interferometer(s)

- FR       Faraday rotation

- GB	      ground-based

- GBO	ground-based observatory

- GCR	galactic cosmic ray(s)

- GEM	geospace environment modeling program

- GEO   geostatioinary orbit

- GeV    gigaelectron-Volt

- GIC	geomagnetically induced current

- GLE     ground-level enhancement

- GMD	geomagnetic disturbance

- GNSS	global navigation satellite system

- GOES	Geostationary Operations Environmental Satellite

- GOLD	Global-scale Observations of the Limb and Disk

- GONG	The Global Oscillation Network Group in NSO

- GPS   Global Positioning System

- GTO   geostationary transfer orbit

- H2020	Horizon 2020, funding program of EU for research and innovation

- HELIO	heliophysics integrated observatory

- HF	high frequency

- IAU	International Astronomical Union

- ICON	The Ionospheric Connection mission

- ICSU	International Council for Science

- ICTSW	Interprogramme Coordination Team on Space Weather of WMO

- IDL	Interactive Data Language

- IGY	International Geophysical Year 1957-58

- ILWS	International Living With a Star program

- IPS   interplanetary scintillation

- IR	infra-red

- ISES	International Space Environment Service

- ISR	incoherent scatter radar

- iSWA	Integrated Space Weather Analysis System (NASA)

- ITM	ionosphere-thermosphere-mesosphere

- IUGG	International Union of Geodesy and Geophysics

- IUGONET Inter-university upper atmosphere global observation network

- JAXA	Japan Aerospace eXploration Agency

- JCSDA  Joint Center for Satellite Data Assimilation

- KAIRA	Kilpisj{\"a}rvi Atmospheric Imaging Receiver Array

- keV	kilo electron volt

- L1	Lagrangian point 1

- L5	Lagrangian point 5

- LANL  Los Alamos National Laboratories

- LASCO	Large Angle and Spectrometric Coronagraph

- LBH	Lyman-Birge-Hopfield Nitrogen waveband

- LEO	low-Earth orbit

- LF	low frequency

- LOFAR	Low-Frequency Array for Radio astronomy

- LORAN	Long Range Navigation

- LOS	line-of-sight 

- LWS   NASA/SMD Living With a Star program

- MeV	megaelectron-Volt

- MEO	medium Earth orbit

- MDI	Michelson Doppler Imager (on SoHO)

- MHD	magneto-hydro-dynamic

- MI	magnetosphere-ionosphere

- MHD	magnetohydrodynamic

- MIT	magnetosphere-ionosphere-thermosphere

- MMS	magnetospheric multi-scale mission

- MLSO Mauna Loa Solar Observatory

- MLTI	magnetosphere-lower thermosphere-ionosphere

- MMS	magnetospheric multi-scale

- MSAS	MTSAT satellite based augmentation system 

- MSIS	Mass Spectrometer and Incoherent Scatter Radar (empirical atmosphere model)

- MTSAT	Multi-functional Transport Satellite

- MUF	maximum usable frequency

- NASA	National Air and Space Administration

- NERC  US National Energy Regulatory Commission

- NRT	near real time

- NOAA  US National Oceanographic and Atmospheric Administration

- NSF	National Science Foundation

- NSO   US National Solar Observatory

- OSCAR	Observing Systems Capability Analysis and Review Tool of WMO

- OSTP	Office of Science and Technology Policy

- PCW   Polar Communications and Weather satellite system

- POES	polar operational environmental satellite

- PFSS  potential-field source-surface model

- PSW   ICSU/COSPAR panel on space weather

- R2O	research-to-operations

- RB    radiation belt

- RBSP	Radiation Belt Storms Probes

- R$_{\rm E}$	Earth radius of 6371\,km

- REP	relativistic electron precipitation

- RTK	Real Time Kinematic

- SAA	South Atlantic anomaly

- SAMA	South Atlantic magnetic anomaly

- SAPS	sub-auroral polarization streams

- SAR	synthetic aperture radar

- SBAS  space-based augmentation system

- SCW	substorm current wedge

- SDO	Solar Dynamics Observatory

- SEE	single event effect(s)

- SEP	solar energetic particle(s)

- SEU	single-event upset

- SMD	NASA's science mission directorate

- SHINE	solar, heliosphere, and interplanetary environment program

- SKA	Square Kilometer Array

- SMEX	NASA SMall EXplorer 

- SOAP	Simple Object Access Protocol

- SoHO	Solar and Heliospheric Observatory

- SOLIS	Synoptic Optical Long-term Investigations of the Sun (NSO)

- SPASE	space physics archive search and extract

- SPE  solar particle event

- SPP	Solar Probe Plus (NASA mission)

- SSA	space situational awareness

- SSI	solar spectral irradiance

- STEREO	Solar-Terrestrial Relations Observatory

- SuperDARN	Super Dual Auroral Radar Network

- SWPC  US/NOAA Space Weather Prediction Service

- SWx	space weather

- TEC	total electron content

- THEMIS  time history of events and macroscale interactions during substorms

- TID	traveling ionospheric disturbance(s)

- UCAR	university corporation for atmospheric research

- UN    United Nations

- WAAS	wide area augmentation system

- WDC	World Data Center of ICSU

- WDS	World Data System of ICSU

- WMO 	World Meteorological Organization
}

\nocite{iea2013}
\nocite{sia2013}
\nocite{prb2013}
\nocite{nic2013}
\nocite{rac2013}
\nocite{jrc2011}
\nocite{ams2011}
\nocite{ndp2011}
\nocite{ames2012}
\nocite{orma2011}
\nocite{jason2011}
\nocite{severeswx2008}
\nocite{2014SpWea..12..487S}
\nocite{2013JSWSC...3A..19S}
\nocite{meteosat1998}
\nocite{2004SpWea...2.9002B}
\nocite{boteler+etal98}
\nocite{swximpactlloyds2011}
\nocite{lloyds2013}
\nocite{wef2013}
\nocite{nerc1989}
\nocite{gaunt2013}
\nocite{schulte2014}
\nocite{2011AdSpR..47.2059H}
\nocite{uncopuos2013}
\nocite{uncopuos2014}
\nocite{oecd2011}
\nocite{schrijverrabanal2013}
\nocite{g8_2013}
\nocite{1955RSPSA.228..238H}
\nocite{1964Natur.203.1214H}
\nocite{2010ApJ...715L.104B}
\nocite{trichtchenkoetal2013}
\nocite{newelletal2007}
\nocite{hpIII-11}
\nocite{hpII-8}
\nocite{2014EOSTr..95Q.201S}
\nocite{ds2012}
\nocite{weather2010}
\nocite{2014ApJ...783..102M}
\nocite{2013SoPh..tmp..182K}
\nocite{derosa+etal2008}
\nocite{2013ApJ...770L..28B}

\nocite{nerc2012}
\nocite{nerc2013a}
\nocite{nerc2013b}
\nocite{ferc2013}


\end{document}